\documentclass[
 reprint,
nofootinbib,
 amsmath,amssymb,
 aps,
pra,
]{revtex4-2}

\usepackage{graphicx}
\usepackage[dvipsnames]{xcolor}
\usepackage{aligned-overset}
\usepackage{hyperref}
\usepackage{orcidlink}
\usepackage{twemojis}
\usepackage[caption=false,font=footnotesize]{subfig}
\usepackage{placeins}
\usepackage{dcolumn}
\usepackage{bm}
\usepackage{tikz}
\usepackage{bbm}
\usepackage{dsfont}
\usepackage{mathrsfs}
\usepackage{amsmath}
\usepackage{romannum}
\usepackage{braket}
\usepackage{ragged2e}
\usepackage{float}
\usepackage[american, siunitx]{circuitikz}
\usepackage[protrusion=true, expansion=true]{microtype}
\usepackage{siunitx}
 \ctikzset{RF/scale=0.8}
\usepackage{circuitikz}
\begin{document}

\preprint{APS/123-QED}

\title{ Gate-Based Microwave Quantum Repeater Via Grid-State Encoding }

\author{Hany Khalifa \orcidlink{0000-0002-1276-5428}}
\thanks{Corresponding author}
\email[]{hany.khalifa@oulu.fi}

 \affiliation{Nano and Molecular Systems Research Unit, University of Oulu, P.O. Box 3000, FI-90014 Oulu, Finland}

 \author{Matti Silveri \orcidlink{0000-0002-6319-2789}}
 \affiliation{Nano and Molecular Systems Research Unit, University of Oulu, P.O. Box 3000, FI-90014 Oulu, Finland}
\begin{abstract}
In \textit{autonomous quantum error correction} the lifetime of a logical bosonic qubit can be extended beyond its physical constituents without feedback measurements. Leveraging autonomous error correction, we propose a gate-based microwave quantum repeater (GBMQR) with encoded bosonic grid states. Each repeater station comprises a transmon and two bosonic resonators: one resonator serving as a stationary quantum memory utilizing autonomous error correction, and the other as an information bus for entanglement generation. Entanglement is generated sequentially through the successful absorption of a microwave photon wavepacket. This method enables deterministic entanglement generation, in contrast to a probabilistic mixing of two heralding signals on a balanced beamsplitter. Furthermore, our GBMQR employs an all-bosonic entanglement swapping Bell-state measurement. This is implemented via a bosonic controlled-Z gate and two separate X-basis projective homodyne measurements on the stationary stored codewords. Our approach circumvents mode-mismatch losses associated with routing and interfering of heralding modes on a beamsplitter, and confines losses to those arising from stationary storage. We evaluate the performance of the proposed quantum repeater by calculating its secret key rate under realistic lab environments. Moreover, we explicitly demonstrate that at stationary damping rate of $\kappa^{-1}_{\text{damp}}=$~\SI{40}{\milli\second}, GBMQR can achieve entanglement generation and swapping success probabilities approx.~$0.75$, and $0.58$ respectively, surpassing the hallmark success probability of $1/2$ set by ideal linear beamsplitter-based Bell-state measurements. 
The proposed device can be implemented using currently available superconducting microwave technology and is suited for secure chip-to-chip communication and distributed quantum computing. 
\end{abstract}
\maketitle
\section{Introduction}

The fascinating phenomenon of quantum entanglement now has the potential to enable a wide range of applications—such as \textit{quantum key distribution} \cite{ekert1991quantum}, \textit{quantum teleportation} \cite{bennett1993teleporting}, \textit{quantum sensing} \cite{degen2017quantum}, and many others. The efficient realization of these protocols paves the way towards achieving the long-sought goals of quantum technologies: the quantum internet~\cite{kimble2008quantum} and {distributed quantum computing}~\cite{van2016path}. 

The non-ideal transmissivity of a transmission medium degrades propagating quantum signals, limiting the distance over which entanglement can be reliably shared. Quantum repeaters \cite{RevModPhys.83.33} have evolved to overcome this limitation. The essence of a quantum repeater is to divide the original transmission distance into two segments using a middle station. In each half-channel, a pair of entangled states is generated, such that one part belongs to one of the communicating parties, while the other belongs to the middle station. After that, entanglement swapping\cite{pan1998experimental} takes place by performing a \textit{Bell-state} measurement on the two states received at the middle station. When the protocol succeeds, a final entangled state is shared between the two remote parties. 
This has motivated the development of three generations of quantum repeater networks. The first generation exploits quantum memories \cite{duan2001long}. The second makes use of a highly entangled cluster state in order to eliminate the need for quantum memories \cite{azuma2015all}. The third uses encoding and  error correction to achieve the same goal \cite{jiang2009quantum}. It is worth noting that these technologies were not developed sequentially to improve upon each-other. Rather, they represent fundamentally different approaches, each optimized for different operational setting and constraints. 

Quantum repeater technology has been developed primarily for optical-domain networking. With the emergence of highly resilient microwave \textit{circuit quantum electrodynamics} (cQED) platforms, it is essential to adapt quantum repeater technology to the microwave domain. Although losses in microwave transmission lines are substantially higher compared to optical fibers, a microwave quantum repeater is still particularly advantageous for small-scale microwave networks, where the scalability of superconducting cQED constrains the processing power of prospective superconducting quantum computers \cite{van2016path}. Recent proposals have advocated a hybrid implementation via microwave-to-optical transduction to fully exploit the versatility of fiber-optic transmission lines \cite{chelluri2025bosonic}. However, highly efficient microwave-to-optical transduction at the single-photon level has yet to be demonstrated experimentally. Furthermore, the hybrid approach assumes an ideal two-photon interference Bell-state measurement realized through a balanced beamsplitter and ideal single-photon counters—devices that are currently unavailable in the microwave domain and exhibit only finite efficiency in the optical domain.  

Leveraging bosonic encoding and error correction, this article proposes a \textit{gate-based microwave quantum repeater} (GBMQR) with autonomous quantum error correction. Each repeater node comprises a transmon and two bosonic resonators, one serving as a stationary quantum memory, and the other facilitating information transfer during entanglement generation. The state of each quantum memory is encoded in a \textit{Gottesman–Kitaev–Preskill} (\texttt{GKP}) grid state entangled with the transmon. This particular choice of bosonic encoding is motivated by the fact that logical operations in the codespace are performed on the resonator quadratures, making the protocol backward-compatible with legacy microwave systems. Moreover, the versatility of \texttt{GKP} codewords against the main sources of decoherence in the microwave domain—namely noise and losses—has been recently demonstrated \cite{noh2020encoding}. In addition to that, it has been shown experimentally that the lifetime of an error-corrected \texttt{GKP} codeword can exceed that of an un-encoded discrete qubit state \cite{sivak2023real, brock2025quantum}. The proposed repeater architecture is a hybrid design, combining quantum memories from second‑generation repeaters with quantum error correction from third‑generation repeaters. It also exploits the high controllability and deterministic gate operations available in cQED platforms \cite{krantz2019quantum}. While the ideas presented here are abstract, the proposed device model can likewise be implemented in the optical domain using equivalent tools. 

Autonomous quantum error correction \cite{royer2020stabilization, sivak2023real, brock2025quantum} is performed synchronously on all bosonic modes immediately after each memory is loaded with its \texttt{GKP} grid state. In this error correction method, dissipative engineering of the storage cavity implements a Hamiltonian proportional to the stabilizers of the \texttt{GKP} basis states, confining the codewords to the logical codespace without measuring quadratures modulo $\sqrt{\pi}$. This eliminates ancilla-induced errors associated with modular quadrature measurements.

In the entanglement generation step, the protocol utilizes an un-encoded microwave wavepacket to create entanglement \textit{sequentially} \cite{campagne2018deterministic}. Contrary to \textit{concurrent} entanglement generation \cite{narla2016robust}, where both a transmitter and receiver send their signals to interfere on a path-erasure balanced beamsplitter, sequential entanglement requires only one of the two communicating nodes to act as a sender. The other node implements a phase shift on the received signal, then reflects it back towards the sender to be absorbed. This approach eliminates the further damping and mode-mismatch errors associated with signal routing towards a beamsplitter.  Furthermore, successful concurrent entanglement requires ideal microwave single-photon counters registering a two-mode interference-signature. Currently, efficient microwave photon number resolving detectors are not available because of the small energies in the microwave domain \cite{casariego2023propagating}. On the other hand, sequential entanglement is successfully achieved by implementing a deterministic absorption event. In cQED, qubit operations are now mature enough to be routinely executed in a lab environment with high fidelity and efficiency \cite{zhang2022high, lin202524, marxer2025above}. Thus, in this article, we assume error-free transmon operations. 

Entanglement is swapped by applying a bosonic controlled‑Z operation between the two middle \texttt{GKP} codewords. This is implemented via a non‑linear cross‑Kerr coupling of the two storage resonators using a \texttt{SNAIL} or an additional transmon device \cite{holland2015single, baskov2025exact}. We also note that in cQED systems Kerr interactions are deterministic, and the corresponding phase shifts can reach large values \cite{rebic2009giant, hoi2013giant}. Subsequently, two projective measurements in the X-basis are performed separately on each resonator to complete the protocol. Our model uses homodyne detectors for projective measurements, which contrary to microwave photon counters are available with high efficiency \cite{strandberg2024digital}. As well as entanglement generation, this process realizes the required Bell-state measurement without any joint interference measurement, and hence circumvents losses due to mode-mismatch and routing through a beamsplitter device. Recently, this method has been adapted to transfer optical noiseless linear amplifiers to the microwave domain \cite{HKMQC, khalifa2024fault}. 

In order to quantify the performance of the proposed device, we compare the remote \textit{secret key rate} generated by GBMQR to that of a microwave quantum repeater using a beamsplitter for entanglement generation and swapping~(BSMQR)~\cite{PhysRevA.111.062611, chelluri2025bosonic}. Secret key analysis is performed considering device imperfections and realistic finite-energy grid states. Additionally, we demonstrate the practical conditions under which GBMQR outperforms BSMQR.

The proposed device has great potential to become a significant addition to cQED toolbox. Precisely, it can enable secure chip-to-chip communications \cite{kurpiers2018deterministic}, and provide entanglement-resources for microwave distributed quantum computing \cite{Knörzer_2026, laracuente2025modeling}, and teleportation-assisted quantum computation \cite{babu2025gate}.

This article is organized as follows. Section ~\ref{sec:II} introduces our GBMQR by first providing a brief background of \texttt{GKP} grid states, followed by a detailed description of the ideal device performance. Section ~\ref{sec:III} considers practical limitations and imperfections of GBMQR, namely, finite-energy grid states and errors in the utilized bosonic gates. Section ~\ref{sec:IV} is dedicated to secret key analysis, as well as, practical performance comparison between GBMQR  and BSMQR. Finally, Sec.~\ref{sec:V} provides conclusions and discussions. 
\section{Gate-Based Microwave Quantum Repeater (GBMQR)}\label{sec:II}
\subsection{\texttt{GKP}-grid Mathematical Description}
This section presents the formalism for encoding a logical \texttt{GKP} qubit in a single bosonic mode~\cite{grimsmo2021quantum}.
A bosonic mode in the Fock basis \(\{\lvert n\rangle\}\) has ladder operators \(a,a^{\dagger}\), with \([a,a^{\dagger}]=1\), acting as \(a\lvert n\rangle=\sqrt{n}\,\lvert n-1\rangle\) and \(a^{\dagger}\lvert n\rangle=\sqrt{n+1}\,\lvert n+1\rangle\)~\cite{gerry2023introductory}. Equivalently, in phase space it is specified by its postition quadrature \(Q=(a+a^{\dagger})/\sqrt{2}\), and momentum quadrature \(P=(a-a^{\dagger})/\sqrt{2}\), with \([Q,P]=i\), where (\(\hbar=1\))~\cite{walls2025quantum}, and spectral resolutions \(Q=\int dz\, z\,\lvert z\rangle_{q}\,{}_{q}\langle z\vert\), \(P=\int dx\, x\,\lvert x\rangle_{p}\,{}_{p}\langle x\vert\), where \(Q\lvert z\rangle_{q}=z\,\lvert z\rangle_{q}\), \({}_{q}\langle z\vert z'\rangle_{q}=\delta(z-z')\), \(P\lvert x\rangle_{p}=x\,\lvert x\rangle_{p}\), and \({}_{p}\langle x\vert x'\rangle_{p}=\delta(x-x')\). 
\begin{figure*}   
\begin{circuitikz}[scale=1.2]   
    \draw[] (2.75,0)--(2.75,0.5)--(3,0.5);
    \draw[] (2.75,0)--(2.75,-0.65)--(3,-0.65);
    \draw[](3,0.5) to [bend right] (3,-0.65);
    \draw[] (3,0.6)--(3, 0.85)--(3, 0.85)--(4.1, 0.85)--(4.1, 0.6);
\draw[] (4.1, 0.6) to [bend right] (3,0.6);
\draw[] (3,-0.75)--(3, -0.99)--(3,-0.99)--(4.1, -0.99)--(4.1, -0.75);
\draw[] (4.1, -0.75) to [bend left] (3,-0.75);
    \draw[](4,0.5)--(4.25,0.5)--(4.25,-0.65)--(4,-0.65);
    \draw[](4,0.5) to [bend left] (4,-0.65);
    \draw[] (3.375,0.35)--(3.625,0.35)--(3.625,0.05)--(3.375,0.05)--(3.375,0.35);
    \draw [](3.475,0.05)--(3.475,-0.02);
    \draw[](3.525,0.05)--(3.525,-0.02);
    \draw[] (3.45,-0.02)--(3.55,-0.02)--(3.55,-0.12)--(3.45,-0.12)--(3.45,-0.02);
    \draw[] (3.475,-0.12)--(3.475,-0.19);
    \draw[] (3.525,-0.12)--(3.525,-0.19);
    \draw[](3.45,-0.02)--(3.55,-0.12);
    \draw[](3.55,-0.02)-- (3.45,-0.12);
   \draw[] (3.375,-0.19)--(3.625,-0.19)--(3.625,-0.49)--(3.375,-0.49)--(3.375,-0.19);
   \draw[] (3.5,-1.05)--(3.7, -1.6)--(4.2,-1.6);
   \draw[] (3.6,1)--(3.6,1) node [] {\scalebox{0.7}{\textbf{Node (A)}}};
   \draw[] (7.6,-1.2)--(7.6,-1.2) node [] {\scalebox{0.7}{\textbf{Node (B)}}};
   \draw[] (10.6,-1.2)--(10.6,-1.2) node [] {\scalebox{0.7}{\textbf{Node (C)}}};
    \draw[] (14.6,1)--(14.6,1) node [] {\scalebox{0.7}{\textbf{Node (D)}}};
   \draw[rounded corners=5pt] (4.2,-2.5) rectangle (5.8,-1.2);
   \draw[] (4.4, -1.7)--(4.9,-1.7);
  \draw[](4.9,-1.7)--(4.9,-1.55)--(5.3,-1.55)--(5.3,-1.85)--(4.9,-1.85)--(4.9,-1.7) node [xshift=6.6] {\scalebox{0.48}{$D({\frac{\sqrt{\pi}}{2}})$}};
  \draw[] (5.3,-1.7)--(5.7, -1.7);
  \draw[] (4.4, -2.2)--(5.15,-2.2);
  \filldraw[black] (5.1,-2.2) circle (1pt);
  \draw[] (5.1, -2.2)--(5.1,-1.85);
  \draw[] (5.1, -2.2)--(5.7, -2.2);
  \draw[](4.9, -2)--(4.9,-2) node [xshift = -10, yshift = -12] {\scalebox{0.45}{$\frac{1}{\sqrt{2}}(\lvert g  \rangle + \lvert e \rangle)$}};
  \draw[](4.9, -1.5)--(4.9,-1.5) node [xshift = -10, yshift = -3] {\scalebox{0.45}{$\lvert \bar{0} \rangle$}};
  \draw[](5.5, -1.35)--(5.5,-1.35) node [xshift = -4, yshift = -0] {\scalebox{0.65}{$\textbf{CD-gate} $}};
  \draw[] (4.25,0) -- (4.5,0);
  \draw[] (7.0,0.6)--(7.0, 0.85)--(7.0, 0.85)--(8.1, 0.85)--(8.1, 0.6);
\draw[] (8.1, 0.6) to [bend right] (7.0,0.6);
  \draw[fill =black!20] (4.5,0)--(4.5,0.1)--(4.5,-0.1)--(6.5,-0.1)--(6.5,0.1)--(4.5,0.1) node [xshift=35,yshift=4] {\scalebox{0.5}{\textbf{Remote CZ-gate}, $\eta$}};
  \draw[](6.5,0)--(6.75,0);
   \draw[](6.75,0)--(6.8,0);
    \draw[] (6.8,0)--(6.8,0.5)--(7.05,0.5);
    \draw[] (6.8,0)--(6.8,-0.65)--(7.05,-0.65);
    \draw[] (7,-0.75)--(7, -0.99)--(7,-0.99)--(8.1, -0.99)--(8.1, -0.75);
\draw[] (8.1, -0.75) to [bend left] (7,-0.75);
    \draw[](7.05,0.5) to [bend right] (7.05,-0.65);
    \draw[](8.05,0.5)--(8.3,0.5)--(8.3,-0.65)--(8.05,-0.65);
    \draw[](8.05,0.5) to [bend left] (8.05,-0.65);

    \draw[] (7.5,0.85)--(7.5, 1.2)--(8.75,1.2) ;
    \draw[] (8.94,1.2) node [squidshape, scale=0.5]{};
    \draw[](9.1,1.2)--(10.5,1.2)--(10.5,0.85);
    \draw[] (8.94,1.5)--(9.2, 2)--(9.6,2);
    \draw[rounded corners=5pt] (9.6,1.5) rectangle (11.2,2.8);
    \draw[](9.85,1.67)--(9.85,1.67) node [] {\scalebox{0.7}{$\lvert \overline{\texttt{GKP}}\rangle $}};
     \draw[](9.85,2.45)--(9.85,2.45) node [] {\scalebox{0.7}{$\lvert \overline{\texttt{GKP}}\rangle $}};
     \draw[] (10.85,2.65)--(10.85,2.65) node []{\scalebox{0.8}{\textbf{BSM}}};
    \draw[] (9.7, 1.85)--(10.2,1.85) ;
    \filldraw[black] (10.2,1.85) circle (1pt);
    \draw[](10.2,1.85)--(10.7,1.85) --(10.7, 2)--(11,2)--(11,1.7)--(10.7,1.7)--(10.7,1.85) node [xshift = 5.5] {\scalebox{0.65}{$\hat{M}_{x}$}};
 \draw[] (9.7, 2.3)--(10.2,2.3) ;
    \filldraw[black] (10.2,2.3) circle (1pt);
\draw[](10.2,2.3)--(10.7,2.3) --(10.7, 2.45)--(11,2.45)--(11,2.15)--(10.7,2.15)--(10.7,2.3) node [xshift = 5.25] {\scalebox{0.65}{$\hat{M}_{x}$}};
    \draw[] (10.2,1.85)--(10.2,2.3);
    \draw[] (7.38,0.35)--(7.63,0.35)--(7.63,0.05)--(7.38,0.05)--(7.38,0.35);
    \draw [](7.48,0.05)--(7.48,-0.02);
    \draw[](7.53,0.05)--(7.53,-0.02);
    
    \draw[] (7.455,-0.02)--(7.555,-0.02)--(7.555,-0.12)--(7.455,-0.12)--(7.455,-0.02);
    \draw[] (7.4755,-0.12)--(7.4755,-0.19);
    \draw[] (7.5255,-0.12)--(7.5255,-0.19);
    
    \draw[](7.455,-0.02)--(7.555,-0.12);
    \draw[](7.555,-0.02)-- (7.455,-0.12);
   \draw[] (7.3755,-0.19)--(7.6255,-0.19)--(7.6255,-0.49)--(7.3755,-0.49)--(7.3755,-0.19);

\draw[] (10,0.6)--(10, 0.85)--(10, 0.85)--(11.1, 0.85)--(11.1, 0.6);
\draw[] (11.1, 0.6) to [bend right] (10,0.6);
    \draw[] (9.8,0)--(9.8,0.5)--(10.05,0.5);
    \draw[] (9.8,0)--(9.8,-0.65)--(10.05,-0.65);
    
    \draw[](10.05,0.5) to [bend right] (10.05,-0.65);
    \draw[](11.05,0.5)--(11.3,0.5)--(11.3,-0.65)--(11.05,-0.65);
    \draw[](11.05,0.5) to [bend left] (11.05,-0.65);
    
    \draw[] (10.38,0.35)--(10.63,0.35)--(10.63,0.05)--(10.38,0.05)--(10.38,0.35);
    \draw [](10.48,0.05)--(10.48,-0.02);
    \draw[](10.53,0.05)--(10.53,-0.02);
    
    \draw[] (10.455,-0.02)--(10.555,-0.02)--(10.555,-0.12)--(10.455,-0.12)--(10.455,-0.02);
    \draw[] (10.4755,-0.12)--(10.4755,-0.19);
    \draw[] (10.5255,-0.12)--(10.5255,-0.19);
    
    \draw[](10.455,-0.02)--(10.555,-0.12);
    \draw[](10.555,-0.02)-- (10.455,-0.12);
   \draw[] (10.3755,-0.19)--(10.6255,-0.19)--(10.6255,-0.49)--(10.3755,-0.49)--(10.3755,-0.19);
\draw[] (10,-0.75)--(10, -0.99)--(10,-0.99)--(11.1, -0.99)--(11.1, -0.75);
\draw[] (11.1, -0.75) to [bend left] (10,-0.75);

      \draw[] (11.3,0)--(11.55,0);
      \draw[fill =black!20] (11.55,0)--(11.55,0.1)--(11.55,-0.1)--(13.55,-0.1)--(13.55,0.1)--(11.55,0.1)node [xshift=35,yshift=4] {\scalebox{0.5}{\textbf{Remote CZ-gate}, $\eta$}};
      \draw[] (13.55,0)--(13.80,0);

       \draw[](13.8,0)--(13.8,0);
    \draw[] (13.8,0)--(13.8,0.5)--(14.05,0.5);
    \draw[] (13.8,0)--(13.8,-0.65)--(14.05,-0.65);
    
    \draw[](14.05,0.5) to [bend right] (14.05,-0.65);
    \draw[](15.05,0.5)--(15.3,0.5)--(15.3,-0.65)--(15.05,-0.65);
    \draw[](15.05,0.5) to [bend left] (15.05,-0.65);

    \draw[] (14.38,0.35)--(14.63,0.35)--(14.63,0.05)--(14.38,0.05)--(14.38,0.35);
    \draw [](14.48,0.05)--(14.48,-0.02);
    \draw[](14.53,0.05)--(14.53,-0.02);
    
    \draw[] (14.455,-0.02)--(14.555,-0.02)--(14.555,-0.12)--(14.455,-0.12)--(14.455,-0.02);
    \draw[] (14.4755,-0.12)--(14.4755,-0.19);
    \draw[] (14.5255,-0.12)--(14.5255,-0.19);
    \draw[] (14,0.6)--(14, 0.85)--(14, 0.85)--(15.1, 0.85)--(15.1, 0.6);
\draw[] (15.1, 0.6) to [bend right] (14,0.6);
    \draw[](14.455,-0.02)--(14.555,-0.12);
    \draw[](14.555,-0.02)-- (14.455,-0.12);
   \draw[] (14.3755,-0.19)--(14.6255,-0.19)--(14.6255,-0.49)--(14.3755,-0.49)--(14.3755,-0.19);
\draw[] (14,-0.75)--(14, -0.99)--(14,-0.99)--(15.1, -0.99)--(15.1, -0.75);
\draw[] (15.1, -0.75) to [bend left] (14,-0.75);
    \end{circuitikz}
\begin{minipage}{\textwidth}
   \caption{\justifying \sloppy A schematic of the proposed GBMQR. The figure depicts two repeater segments, where nodes A and D are the end nodes of the repeater, while B and C are located in the same place to perform a swapping operation. Each node is equipped with a transmon-two-resonator system. One resonator acts as a quantum memory loaded with a \texttt{GKP} codeword, while the other is coupled to a transmission line of transmissivity $\eta$ that mediates remote sequential entanglement across each repeater segment. A controlled-displacement gate (\textbf{CD}-gate) is employed at the state preparation step as described in the main text. A swapping Bell-state measurement (\textbf{BSM}) is performed bosonically on the memory codewords via a controlled-Z operation followed by single bosonic-qubit measurements in the X-basis. This gate can be implemented either via an additional transmon or a \texttt{SNAIL} device. Ideally, the final repeater state is a bosonic Bell-state. }
    \label{fig:(1a)}
   \end{minipage}
\end{figure*}

Unitary phase‑space dynamics of a bosonic mode are generated by the displacement \(D(\lambda)=e^{\lambda a^{\dagger}-\lambda^{*}a}\) and squeezing \(S(\xi)=e^{\xi a^{\dagger 2}-\xi^{*}a^{2}}\) operators, where \(|\lambda|^{2}\) is the field intensity, and \(\xi=re^{i\theta}\), such that \(r\) is the squeezing strength, and \(\theta\) is the squeezing angle.

The logical qubit Pauli operations for \texttt{GKP} states—namely bit flip \( (X) \), phase flip \( (Z) \), and combined bit \& phase flip \( (Y) \) gates—admit a representation in terms of displacement operators in the phase-space of the  harmonic oscillator. We consider the square‑grid \texttt{GKP} code~\cite{grimsmo2021quantum}, with \(\bar{Z}=D(i\sqrt{\pi/2})\), \(\bar{X}=D(\sqrt{\pi/2})\), and \(\bar{Y}=i\bar{X}\bar{Z}\). The anticommutation \(\{\bar{X},\bar{Z}\}=0\) reflects the noncommutativity of the corresponding displacement operators \begin{align}
    [D(\lambda_{1}),D(\lambda_{2})]= D(\lambda_{2})D(\lambda_{1})(e^{\lambda_{1} \lambda_{2}^{*}-\lambda_{1}^{*}\lambda_{2}}-1),
\end{align}
where $ \lambda_{1}, \lambda_{2} \in \mathbb{C}.$

However, the two displacement operators commute when the phase‑space area they enclose is an integer multiple of \(2\pi\) (i.e., \(2\pi k\), \(k\in\mathbb{Z}_0\)). Consequently, the squared logical displacements \(X^{2}=D(\sqrt{2\pi})\) and \(Z^{2}=D(i\sqrt{2\pi})\) enclose an area of \(4\pi\) and thus form a commuting stabilizer pair whose simultaneous \(+1\) eigenstates define the \texttt{GKP} codespace.

The ideal un-normalized logical \texttt{GKP} codewords in the Z-basis can be written as 
\begin{align}
    \lvert \bar{0} \rangle &\propto \underset{k \in \mathbb{Z}}{\sum}\lvert 2k \sqrt{\pi} \rangle_{q}, \nonumber \\ 
    \lvert \bar{1} \rangle &\propto \underset{k \in \mathbb{Z}}{\sum} \lvert (2k+1) \sqrt{\pi} \rangle _{q},
\end{align}
where $\bar{Z} \lvert \bar{0} \rangle = \lvert \bar{0} \rangle$ and $\bar{Z} \lvert \bar{1} \rangle = - \lvert \bar{1} \rangle$. Analogously, the codewords in the complementary X-basis are defined as $\lvert \tilde{0} \rangle \propto \underset{k \in \mathbb{Z}}{\sum}\lvert 2k \sqrt{\pi} \rangle_{p}$ and $\lvert \tilde{1} \rangle \propto \underset{k \in \mathbb{Z}}{\sum} \lvert (2k+1) \sqrt{\pi} \rangle _{p}$, where $\bar{X} \lvert \tilde{0} \rangle = \lvert \tilde{0} \rangle$ and $\bar{X} \lvert \tilde{1} \rangle = - \lvert \tilde{1} \rangle$ (see Appendix~\ref{app:GridPauli}) for details).    

Ideal \texttt{GKP} codewords are unphysical because they have infinite energy. In practice, finite‑energy approximations are obtained by applying a Gaussian envelope \(e^{-\Delta^{2} a^{\dagger} a}\) to the ideal states, \(e^{-\Delta^{2} a^{\dagger} a}\lvert \nu\rangle\), where \(\nu\in\{\lvert\bar{0}\rangle,\lvert\bar{1}\rangle\}\) and \(\Delta\) sets the envelope width~\cite{royer2020stabilization}, which is related to the squeezing parameter $r$ as $\Delta \propto e^{-r}$ . This replaces the grid of \(\delta\)-spikes with a superposition of centered Gaussian envelopes (see Appendix~\ref{app:OVerlapExps}).
\subsection{Ideal Device Description}
 \label{sec:DeviceOperatio}
 We now describe the ideal operation of the gate-based microwave quantum repeater. In later sections, experimental imperfections are incorporated, and device performance is assessed under practical limits.

 As shown in Fig.~\ref{fig:(1a)}, two distant repeater nodes, labeled~A~and~D, aim to establish remote entanglement with the assistance of two intermediate nodes~B~and~C located at the same site. The protocol involves the three archetypal steps characteristic of a quantum repeater: entanglement generation, information storage, and entanglement swapping. However, in this proposal, information storage is achieved by leveraging quantum error correction, and swapping is performed entirely on bosonic modes, taking advantage of their extended error-corrected lifetime.  

 Initially, the transmon and storage resonator are dispersively coupled, then driven to host a joint transmon-\texttt{GKP} state. In the strong dispersive regime of superconducting circuits, the full transmon-storage resonator Hamiltonian is defined as 
\begin{align}
    H= H_{\text{transmon}} + H_{\text{storage}}+H_{\text{int}}+H_{\text{drive}},
\end{align}
where the reduced Plank's constant was set to $\hbar=1$. 
Firstly, the transmon Hamiltonian is defined by 
\begin{align}
    H_{\text{transmon}} = \omega_{b} \frac{\sigma_{z}}{2},
\end{align} 
where $\sigma_{z}$ is the qubit Pauli-Z operator of the transmon, and $\omega_{b}$ is its resonance frequency. Then, the storage resonator Hamiltonian is given by 
\begin{align}
    H_{\text{storage}}  = \omega_{c} c^{\dagger}c,
\end{align}
where $c$, $c^{\dagger}$ are the oscillator ladder operators, and $\omega_{c}$ is the resonance frequency of the storage cavity. The interaction Hamiltonian quantifies the strength of the mutual dispersive shift between the transmon and storage resonator. It is given by 
\begin{align}
    H_{\text{int}} = \chi_{bc} \sigma_{z}c^{\dagger}c,
\end{align}
where, $\chi_{bc}$ is the strength of the dispersive interaction. 

Finally, the drive Hamiltonian involves the transmon control pulses and those of the cavity responsible for displacing the stored bosonic mode.
\begin{align}
    H_{\text{drive}} &= \epsilon(t) e^{-i \omega_{c}t} c^{\dagger}+ \Omega(t)e^{-i \omega_{b}t} \sigma^{+} + \text{H.c.},
\end{align}
where, $\sigma^{+}$, $\sigma^{-}$ are the transmon ladder operators, \text{H.c.} stands for Hermitian conjugate, $D(\varsigma) = e^{\varsigma c^{\dagger} -\varsigma^{*}c}$ is the generator of displacements of the cavity mode, such that $\varsigma \propto \epsilon(t)$, and $\epsilon(t)$, $\Omega(t)$, are the temporal pulse shapes of the resonator and qubit control pulses respectively. Through a series of driven transmon rotations and storage displacements, a conditional displacement on the stored mode is realized. This displacement is dependent on the transmon state and defines the logical space of the stored \texttt{GKP} codeword \cite{campagne2020quantum,eickbusch2022fast}. 

Suppose now that at each repeater node the following transmon-\texttt{GKP} state is prepared
\begin{align}
    \lvert \tilde{\varphi} \rangle = \frac{1}{\sqrt{2}}(\lvert g \rangle + \lvert e \rangle) \otimes \lvert \bar{0} \rangle, 
    \label{eq:(4)}
\end{align}
where the transmon is in a superposition of its ground and excited states, while the \texttt{GKP} grid state is initialized in the logical-zero codeword.  After that, a controlled oscillator displacement gate is applied, where the transmon acts as a control while the \texttt{GKP} codeword is the target \cite{liu2024hybrid}: 
\begin{align}
     \textbf{CD}\lvert \tilde{\varphi} \rangle &= \Big (\lvert g \rangle {\langle g \lvert} \otimes D\Big(\sqrt{\frac{\pi}{2}}\Big) + \lvert e \rangle \langle e \lvert \otimes 
 \mathds{1}_{t} \Big) \lvert \tilde{\varphi} \rangle \nonumber \\ 
    &= \frac{1}{\sqrt{2}} \big( \lvert g \rangle _{ A} \otimes \lvert \bar{1} \rangle_{A} + \lvert e \rangle _{A}\otimes \lvert \bar{0}\rangle_{A} \big).
    \label{eq:q-gkp}
\end{align}
Here we note that the amplitude of the resonator drive pulse is set to $\epsilon(t) \propto \sqrt{\frac{\pi}{2}}$, which implements a logical \texttt{GKP} bit-flip displacement operation. In a similar manner, we assume that all nodes  have been prepared in the same state as Eq. (\ref{eq:q-gkp}).

Our next objective is to implement a remote controlled-Z gate between nodes A and B. The overall state of the two nodes is given by 
\begin{align}
    \lvert \varphi \rangle_{AB} = \frac{1}{2} \big( \lvert g g\rangle_{AB} \otimes \lvert \bar{1}\bar{1} \rangle_{AB}+  \lvert g e\rangle_{AB} \otimes \lvert \bar{1}\bar{0} \rangle_{AB} \nonumber \\ 
    + \lvert e g\rangle_{AB} \otimes \lvert \bar{0}\bar{1} \rangle_{AB} + \lvert ee\rangle_{AB} \otimes \lvert \bar{0}\bar{0} \rangle_{AB}\big).
\end{align}
At the beginning of the gate operation a drive pulse targets the transmon part of node A at the coupling frequency of the resonator-transmission line. This step populates the transmission line with a photon temporal mode propagating towards the receiving node B \cite{kurpiers2019quantum, ilves2020demand}. Accordingly, the mapping $\lvert e \rangle \lvert 0 \rangle_{\Gamma} \rightarrow \lvert e \rangle \lvert 0 \rangle_{\Gamma}$, $\lvert g \rangle \lvert 0 \rangle_{\Gamma} \rightarrow \lvert g \rangle \lvert 1 \rangle_{\Gamma}$ is implemented. After that, the state of the segment becomes
\begin{align}
&\lvert \varphi \rangle_{AB} \nonumber \\
= &\frac{1}{2}\Big(
\;\lvert g g \rangle_{AB}\otimes \lvert \bar{1}\bar{1} \rangle_{AB} \otimes \lvert 1 \rangle_{\Gamma }
+ \lvert g e \rangle_{AB}\otimes \lvert \bar{1}\bar{0} \rangle_{AB} \otimes \lvert 1 \rangle_{\Gamma } \nonumber \\
&+ \lvert e g \rangle_{AB}\otimes \lvert \bar{0}\bar{1} \rangle_{AB} \otimes \lvert 0 \rangle_{\Gamma }
+ \lvert e e \rangle_{AB}\otimes \lvert \bar{0}\bar{0} \rangle_{AB} \otimes \lvert 0 \rangle_{\Gamma }
\Big),
\end{align}
where $ \lvert 1 \rangle_{\Gamma}$ is a single photon temporal mode (see Appendix~\ref{app:PhotonTM} for details).

Upon receiving the photon wavepacket at node B, an interaction between the receiving transmon and the incoming photon is initiated. This interaction is described by the unitary $e^{-i \pi \lvert 1 \rangle_{\Gamma}{_{\Gamma}\langle 1 \lvert} \otimes \lvert g \rangle_{B} {_{B}\langle g \lvert }}$ \cite{mcintyre2025protocols}. Then, the same wavepacket is reflected back towards the sending node A, where its successful absorption completes the realization of the remote controlled-Z gate. Subsequently, the state of the two nodes becomes the following:
\begin{align}
&\lvert \varphi \rangle_{AB}\nonumber \\
&= \frac{1}{2}\Big(
 -\,\lvert g g\rangle_{AB} \otimes \lvert \bar{1}\bar{1} \rangle_{AB} \otimes \lvert 0 \rangle_{\Gamma } + \lvert g e\rangle_{AB} \otimes \lvert \bar{1}\bar{0} \rangle_{AB} \otimes \lvert 0 \rangle_{\Gamma } \nonumber \\
& \;\;+\; \lvert e g\rangle_{AB} \otimes \lvert \bar{0}\bar{1} \rangle_{AB} \otimes \lvert 0 \rangle_{\Gamma }+ \lvert e e\rangle_{AB} \otimes \lvert \bar{0}\bar{0} \rangle_{AB} \otimes \lvert 0 \rangle_{\Gamma }
\Big),
\end{align}
where the gate implements the following mapping, $\lvert g \rangle \lvert g \rangle \rightarrow - \lvert g \rangle \lvert g \rangle$, $\lvert g \rangle \lvert e \rangle \rightarrow  \lvert g \rangle \lvert e \rangle$, $\lvert e \rangle \lvert g \rangle \rightarrow  \lvert e \rangle \lvert g \rangle$, and $\lvert e \rangle \lvert e \rangle \rightarrow  \lvert e \rangle \lvert e \rangle$.

 In our proposed quantum repeater architecture the flying wavepacket mediating remote entanglement between distant nodes is unprotected. Quantum error correction is performed immediately on the stored \texttt{GKP} codewords as soon as they are loaded into their respective cavities. Following \cite{storz2023loophole}, we assume a channel attenuation factor of $\SI{2}{\deci\bel\per\kilo\meter}$. Based on this value, the duration of entanglement generation per repeater segment is approximately $\SI{3}{\mu\second}$. Ideally, the entanglement generation process is initiated simultaneously at both segments. However, practically, operational delays might occur. In order to account for potential delays between the two segments, we estimate a worst-case approximate total duration of entanglement generation as $\approx \SI{6}{\mu\second}$. Knowing that the duration of an error correction round for a stored \texttt{GKP} codeword is $\approx \SI{7}{\mu\second}$, and the achieved extended lifetime of the encoded qubit is $\approx \SI{2}{\milli\second}$ as recently reported in \cite{brock2025quantum}, it is safe to assume that the codespace is intact before the start of the swapping process.

To swap the entanglement to the bosonic modes, a Hadamard transformation is applied on each transmon followed by two projective Z-basis measurements. Based on the four possible measurement outcomes, local Pauli corrections and feedforward operations are applied on the bosonic parts, such that the two remote parties share a bosonic Bell-state
\begin{align}
\lvert \Psi \rangle_{AB} &= \frac{1}{\sqrt{2}}\left(\lvert \bar{0}  \bar{0} \rangle_{AB} - \lvert \bar{1}  \bar{1} \rangle_{AB}\right).
\label{eq: 1stEG}
\end{align}
Similarly, segment C-D executes the same sequence of operations, as a result, the state of each segment is $\lvert \Psi \rangle_{AB}$, and $\lvert \Psi \rangle_{CD}$ respectively. Then, the overall state of the four nodes reads
\begin{align}
\lvert \Psi \rangle_{ABCD} &= \frac{1}{2}\Big(
\lvert \bar{0}\bar{0}\bar{0}\bar{0}\rangle_{ABCD}
- \lvert \bar{0}\bar{0}\bar{1}\bar{1}\rangle_{ABCD} 
\nonumber \\
&\qquad \quad - \lvert \bar{1}\bar{1}\bar{0}\bar{0}\rangle_{ABCD}
+ \lvert \bar{1}\bar{1}\bar{1}\bar{1}\rangle_{ABCD}
\Big). 
\end{align}
A deterministic Bell-state measurement is now performed between nodes B and C. Using a nonlinear parametric device such as a transmon \cite{holland2015single} or a \texttt{SNAIL} \cite{baskov2025exact}, a cross-Kerr interaction between the two modes, $e^{-i Q_B Q_C}$ \cite{furusawa2011quantum} is applied. This introduces a negative sign only when both modes are in the logical-one state
\begin{align}
\lvert \Psi \rangle_{ABCD} =&\frac{1}{2} \big( 
\lvert \bar{0} \bar{0} \bar{0} \bar{0} \rangle_{ABCD}
- \lvert \bar{0} \bar{0} \bar{1} \bar{1} \rangle_{ABCD} \nonumber \\
&\qquad\quad- \lvert \bar{1} \bar{1} \bar{0} \bar{0} \rangle_{ABCD}
- \lvert \bar{1} \bar{1} \bar{1} \bar{1} \rangle_{ABCD}
\big). 
\end{align}
After that, two X-basis measurements are performed on the middle modes yielding 4 possible outcomes. Then, the final shared state can be written as 
\begin{align}
\lvert \Psi \rangle_{AD} &= \frac{1}{\sqrt{2}}\left(\lvert \bar{0}  \bar{0} \rangle_{AD} - \lvert \bar{1}  \bar{1} \rangle_{AD}\right),
\end{align}
where, similarly as in the previous step, we apply Pauli-corrections and feedforward operations to recast the output state in the above Bell basis. 
\section{Experimental Limits of GBMQR} \label{sec:III}
\subsection{Preparation Imperfections}
\begin{figure}[H]
\centering
\centering
\begin{minipage}[c]{4.7cm}
  \centering
  (a)\\[-2pt]
  \includegraphics[width=\linewidth]{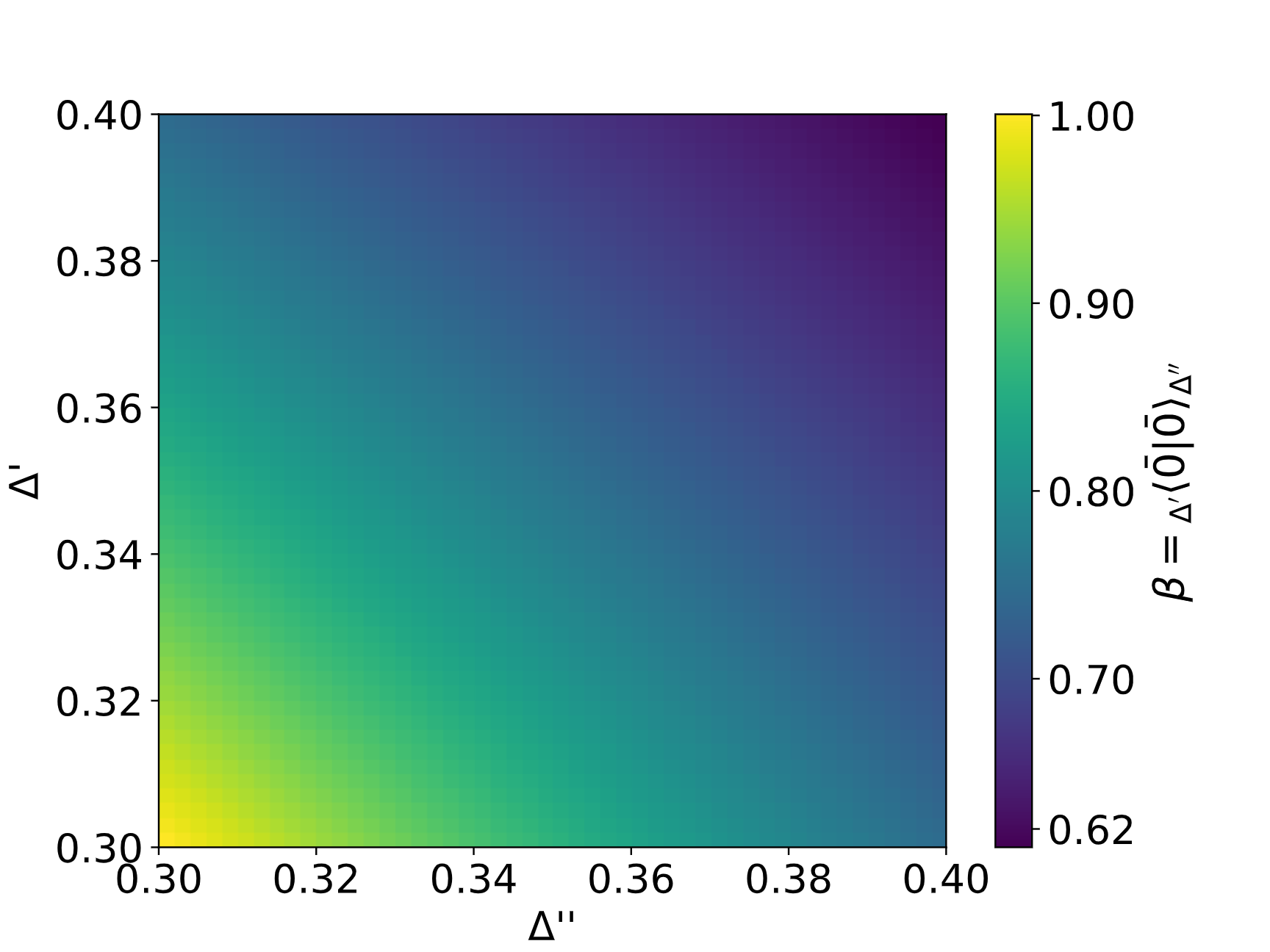}
\end{minipage}%
\begin{minipage}[c]{4.7cm}
  \centering
  (b)\\[-2pt]
  \includegraphics[width=\linewidth]{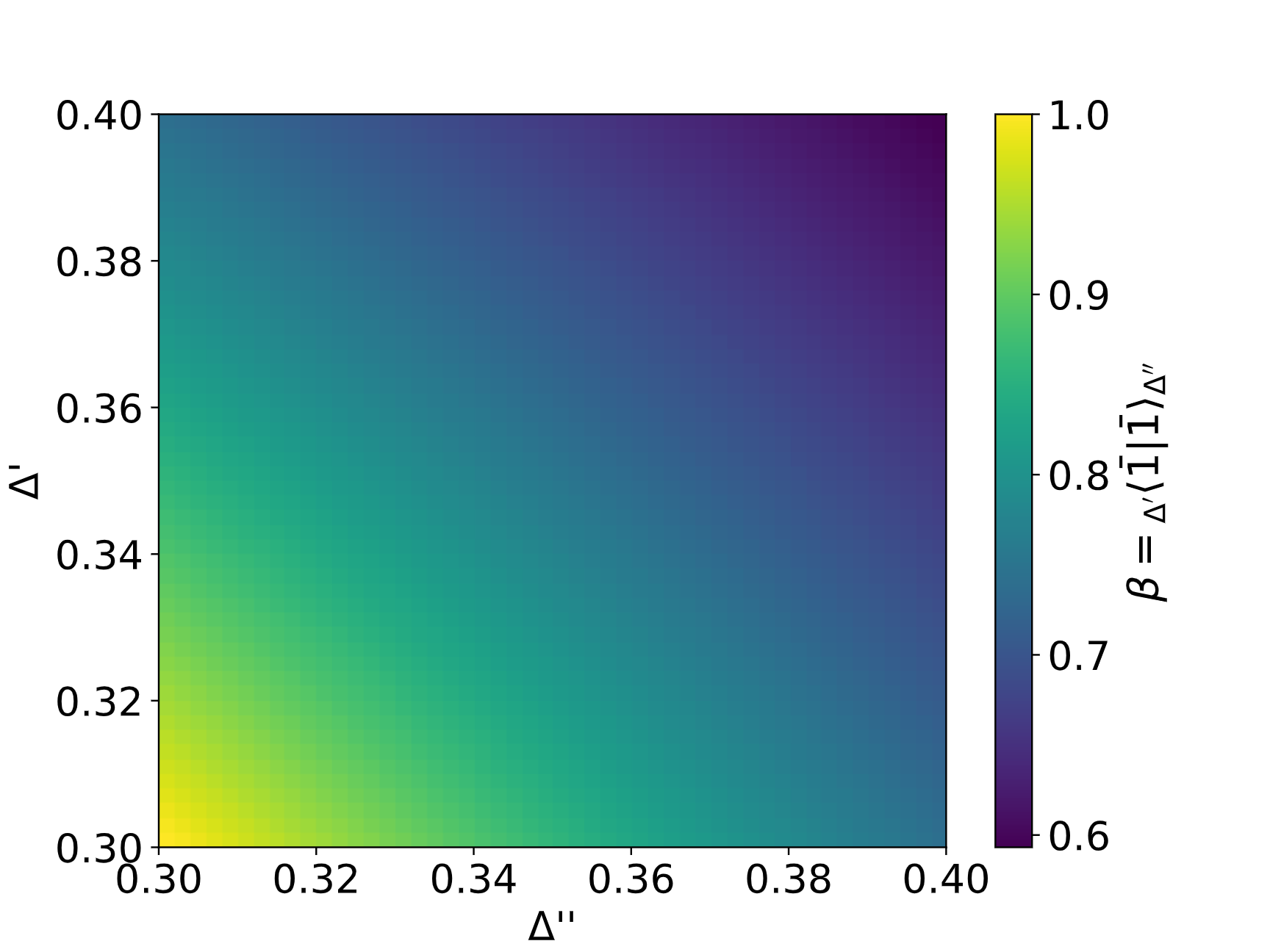}
\end{minipage}
\begin{minipage}{5.3cm}
  \centering
  (c)\\[5.2pt]
  \includegraphics[width=\linewidth]{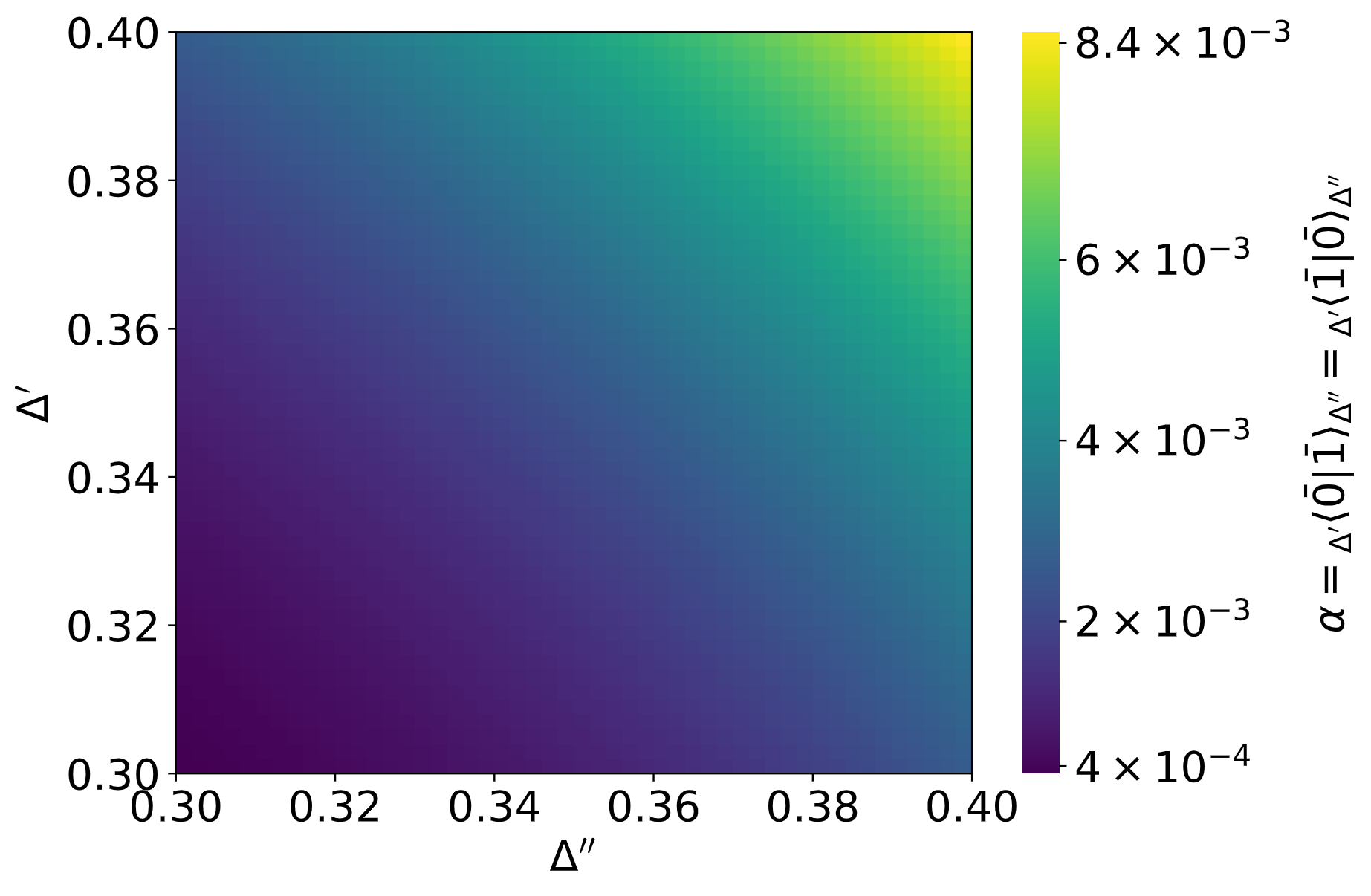}
\end{minipage}
\caption{Overlap between two logical \texttt{GKP} codewords when each is subjected to a different loss channel (Appendix \ref{app:MisError}). Panel (a) considers the logical-zero similar overlap $\beta = {_{\Delta'}\langle \bar{0} \lvert \bar{0} \rangle_{\Delta''}}$ 
as a function of $\Delta'$ and $\Delta''$, whereas panel (b) considers the logical-one similar overlap $\beta = {_{\Delta'}\langle \bar{1} \lvert \bar{1} \rangle_{\Delta''}}$. Panel (c) considers the cross-overlap $\alpha ={_{\Delta'}\langle \bar{0} \lvert \bar{1} \rangle_{\Delta''}}
={_{\Delta'}\langle \bar{1} \lvert \bar{0} \rangle_{\Delta''}}$. The initial lossless width of the codewords was assumed to be $\Delta=0.3$. Ideally, when no losses are incurred, the codewords are orthogonal, $\alpha=0$, and normalized, $\beta=1$. 
As can be seen from the plots, the similar overlaps are approximately equal. However, when stationary losses increase the effective width beyond $\Delta\simeq0.4$, the codespace becomes significantly distorted and equality is no longer maintained. This value should be understood as a reference value in the present normalization, rather than a universal threshold. 
In the limit of large losses, $\alpha \rightarrow 1$, implying that the code distance becomes zero. Consequently, the robustness of the code to losses vanishes as $\beta \rightarrow 0$.}
\label{figoverlap}
\end{figure}
Ideal \texttt{GKP} codewords cannot be realized in a laboratory setting due to finite energy constraints. Instead, a non-orthogonal, finitely squeezed pair of approximate codewords is generated. Consequently, the logical bit-flip gate becomes
\begin{align}
    D\left(\sqrt{\frac{\pi}{2}}\right) = \bar{X}_{\Delta} = \lvert \bar{0} \rangle_{\Delta} {_{\Delta}\langle \bar{1} \lvert} + \lvert \bar{1} \rangle_{\Delta} {_{\Delta}\langle \bar{0} \lvert},
\end{align}
where the subscript $\Delta$ denotes the non-ideal gate operation. This imperfection is reflected in the working of the conditional displacement gate, such that Eq.~\eqref{eq:q-gkp} becomes
\begin{align}
    \textbf{CD}_{\Delta}\lvert \varphi \rangle_{\Delta} = \frac{1}{\sqrt{2}} \left( \alpha \lvert \bar{0} \rangle_{\Delta} \lvert g \rangle + \beta \lvert \bar{1} \rangle_{\Delta} \lvert g \rangle + \lvert \bar{0} \rangle_{\Delta} \lvert e \rangle \right),
\end{align}
where $\lvert \bar{0} \rangle_{\Delta}$ and $\lvert \bar{1} \rangle_{\Delta}$ denote the approximate codewords and the coefficients are given by the overlaps $\alpha = {_{\Delta}\langle \bar{1}} \lvert \bar{0} \rangle_{\Delta}$ and $\beta = {_{\Delta}\langle \bar{0}} \lvert \bar{0} \rangle_{\Delta}$. The cross-overlap $\alpha$ is a measure of the code distance, whereas the similar overlap $\beta$ quantifies how much logical information is preserved. Additional stationary losses are effectively encoded in $\alpha$ and $\beta$ through the increased envelope width $\Delta$. The detailed expressions are given in Appendix~\ref{app:OVerlapExps}.

\subsection{Transmission Losses}
Practical transmission lines have finite transmissivity; hence, the entanglement-mediating photon wavepacket experiences channel loss. We model this as
\begin{align}
    \lvert \tilde{1} \rangle_{\Gamma} = \sqrt{\eta} \lvert 1 \rangle_{\Gamma} + \sqrt{1-\eta} \lvert 0 \rangle_{v},
\end{align}
where $\eta$ is the channel transmissivity and $\lvert 0 \rangle_{v}$ is a vacuum environment mode; see Appendix~\ref{app:Channel}.

Since the remote controlled-Z gate is conditioned on reabsorbing the transmitted wavepacket after its round trip between the two cavities, entanglement within a repeater segment is established only when the photon successfully returns to the original transmitting cavity. If the photon is lost, the applied control sequence drives the receiving transmon outside the Hilbert-space subspace defining the repeater operation. The corresponding state evolution under channel loss is given in Appendix~\ref{app:EntanglementGeneration}.

\subsection{Errors in Swapping Operation}
The following step is to perform a bosonic swapping operation on the two middle \texttt{GKP} grid states. Swapping non-idealities are manifest as recasting the controlled-Z gate in the finite-energy \texttt{GKP} basis (Appendix~\ref{app:EGSWAP}), and non-ideal projective homodyne measurements in the X-basis (Appendix \ref{app:HOM}). The main sources of error in both operations are the finite overlap between logical codewords, the accumulated stationary losses due to error correction rounds, and measurement fidelity of projective measurements. Experimentally, stationary damping rate is approximately $\kappa_{\text{damp}}^{-1}\approx$~\SIrange{25}{40}{\milli\second} with autonomous quantum error correction for \texttt{GKP} states \cite{grimsmo2021quantum}. For an initial squeezing variance of $\Delta= 0.3$, similar codeword overlap $\beta$ and cross-overlap $\alpha$ (see Fig.~\ref{figoverlap}) are calculated as $\beta\approx0.80 \text{--} 0.90$, and $\alpha\approx 4\times 10^{-4}\text{--}1 \times 10^{-3}$ respectively. As shown in Appendix~\ref{app:EGSWAP}, the controlled-Z gate succeeds with a probability of $\approx \beta^{2}$. The measurement fidelity of a homodyne projective measurement as recently reported can reach values as high as $0.95\text{--}0.99$~\cite{strandberg2024digital}, with the assistance of digital optimization methods. Thus, for the assumed accumulated damping due to multiple error correction cycles, i.e., $\kappa_{\text{damp}}^{-1} \approx$~\SIrange{25}{40}{\milli\second}, the success probability is $\mathcal{P}^{\text{succ}}_{\bar{0}/\bar{1} }\approx 0.95$ for both logical codewords. This comes from the fact that the duration of one measurement cycle is \SI{1}{\micro \second}, and the total number of measurements required to achieve this success probability is $\approx10^{3}$ \cite{strandberg2024digital}.
\section{Secret key analysis}\label{sec:IV}
One way to assess the operation of a quantum repeater is via its secret key rate. In the event of large number of transmission instances, also known as the  asymptotic limit, a secret key rate can be defined as \cite{RevModPhys.74.145, RevModPhys.81.1301, Pirandola:20}
\begin{align}
    \mathcal{R} = \mathcal{P}^{\text{succ}}_{\text{EG}} \mathcal \times \mathcal{P}^{\text{succ}}_{\text{SW}} \times \mathcal{R}_\text{raw} ,
    \label{eq:SKR}
\end{align}
where $\mathcal{P}^{\text{succ}}_{\text{EG}}$ is the success probability of entanglement generation, $\mathcal{P}^{\text{succ}}_{\text{SW}}$ is the success probability of entanglement swapping, and $\mathcal{R}_{\text{raw}}$ is the raw key rate. 

The raw key rate measures the fidelity of the final shared entangled state. It is defined via the quantum bit-error probability as
\begin{align}
    \mathcal{R}_{\text{raw}} = 1- h(E_{x})-h(E_{z}),
\end{align}
where $h(p) = -p\text{log}(p)-(1-p)\text{log}(1-p)$ is the binary entropy function, $E_{x}$ is the bit-error rate in the X-basis, whereas $E_{z}$ is the bit-error rate in the Z-basis. Bit errors arise when the detection events at the remote communicating parties, A and D (Fig.~\ref{fig:(1a)}), correspond to a different Bell-state other than the desired one. Assuming symmetric error rates in both X and Z basis, and equal probability of projecting onto any of the other three Bell-states, the raw key rate can be recast as   
\begin{align}
    \mathcal{R}_{\text{raw}} = 1- 2h(E),
\end{align}
where $E= 2/3(1-\mathcal{F})$, and $\mathcal{F}$  is the fidelity between the final shared state and the ideal Bell-basis state.

\vspace*{-1em}
\subsection{Key Rates of GBMQR vs BSMQR}
\vspace*{-0.5em}
From Eq.~(\ref{eq:SKR}), the practical performance analysis of GBMQR is evaluated. We mainly focus on the success probability of entanglement generation and swapping. A comparison is made between GBMQR and BSMQR, where the hardware efficiency of our device is demonstrated by generating higher key rate.

In a BSMQR, both entanglement generation and swapping rely on interfering two identical logical \texttt{GKP} wavefunctions on a balanced beamsplitter. Due to  \textit{Hong-Ou-Mandel} (HOM) interference of the codewords \cite{hong1987measurement}, the joint output state of the interfered modes is projected onto the \textit{singlet} Bell-state. 

Ideally, a successful Bell-state measurement happens when both of the two detectors, each placed at one of the two output ports of the beamsplitter, register no coincidence events. An important requirement for observing this phenomenon is that the interfering modes are \textit{indistinguishable} in every degree of freedom, and that the detectors are ideal photon number resolving detectors. This restrictive requirement is crucial for the success of this measurement setup, since it is unable to decode the Bell-basis subspace entirely, and can only recognize the singlet Bell-state \cite{weihs2001photon}. 

Practically, mode-mismatch due to transmission delays during routing towards a beamsplitter, and different error correction rounds when one repeater segment is created before the other, decreases the indistinguishability between the interfering codewords ~\cite{bouchard2020two}. In a BSMQR this will affect the success probability of both entanglement generation and swapping. 
Furthermore, recent experimental implementation of a microwave beamsplitter-based Bell-state measurement was only successful with high efficiency at the single-photon level \cite{gao2018programmable}. 
However, when multiphoton states were considered the HOM visibility dropped significantly, suggesting a similar behavior for our \texttt{GKP} codewords, where the average photon number per codeword is approx.~$3\text{--}4$ photons. 

To model this effect we assumed two different loss channels of transmissivites~$\eta_{1}$~and~$\eta_{2}$ acting on each codeword separately and calculated their cross and similar overlaps $\alpha$, and $\beta$, respectively. For details see Appendix~\ref{app:MisError}. When $\eta_{1}=\eta_{2}=0.6$, the similar overlap  was calculated as $\approx 0.34$. 
This value of the transmissivites corresponds to cryogenic transmission losses of \SI{2}{\deci\bel\per\kilo\meter} as recently reported in \cite{storz2023loophole}.
Since a beamsplitter-based Bell-state measurement can only identify the singlet subspace $\{ \lvert \Psi^{\pm} \rangle\}$ and cannot recognize which one of the two states was detected, the success probability is upper-bounded to $1/2$ when no coincidences are registered. 
As a result, the final success probability of  beamsplitter-based entanglement generation per repeater segment is $\mathcal{P}_{\rm EG}^{\rm succ}\approx 0.17$. A similar success probability is expected for entanglement swapping.

We have also considered an ideal BSMQR with higher theoretical transmissivities, $\eta_{1}=\eta_{2}=0.95$. Correspondingly, the success probability of  entanglement generation increases to $\mathcal{P}^{\text{succ}}_{\text{EG}}=0.32$. This choice of the transmissivities is motivated by the fact that for a multiphoton state the HOM visibility of a beamsplitter-based setup was reported to be approx.~$0.82$ \cite{gao2018programmable}. As before, entanglement swapping succeds with a similar probability.

On the other hand, for the operation of GBMQR we have considered errors due to stationary damping of the codewords, and gate errors in both sequential entanglement generation  and entanglement swapping. We assert again here that our device mitigates mismatch errors by circumventing losses due to two-mode interference Bell-state measurements, and hence only stationary losses are dominant. This is a powerful advantage of our GBMQR over BSMQR.

In entanglement generation, by assuming ideal transmon measurements, the derived expression of the success probability shows a dependence on the transmissivity of the channel $\eta$ ~\cite{takeoka2014fundamental, pirandola2017fundamental}, and the similar overlap between logical codewords, $\beta$, due to energy truncation,  and stationary damping. For a damping rate of $\kappa_{\text{damp}}^{-1}\approx$ ~\SI{25}{\milli\second}, and transmission losses of approx.~\SI{2}{\deci\bel\per\kilo\meter}, i.e., transmissivity of $\eta=0.6$ \cite{storz2023loophole}, the success probability of sequential entanglement is $\mathcal{P}^{\text{succ}}_{\text{EG}}\approx 0.60$. However, when stationary damping decreases to $\kappa_{\text{damp}}^{-1}=$~\SI{40}{\milli\second}, sequential entanglement succeeds with higher probability $\mathcal{P}^{\text{succ}}_{\text{EG}}\approx0.75$ (detailed expressions in Appendix~\ref{app:EntanglementGeneration}).
\begin{table} [H]
\centering
\scalebox{0.5}{\begin{tabular}{|l|c|c|c|c|c|c|c|}
\hline
 \shortstack{\vspace*{0em}\text{Performance } \\{ \vspace*{1.6em}\text{Parameter}} } &\shortstack{\\[-2pt]GBMQR \\$\kappa_{\text{damp}}^{-1}=$~\SI{40}{\milli\second} \\ $\eta=0.60$} &\shortstack{\\[-2pt]GBMQR \\$\kappa_{\text{damp}}^{-1}=$~\SI{25}{\milli\second} \\ $\eta=0.60$} &\shortstack{\\[-2pt]GBMQR \\$\kappa_{\text{damp}}^{-1}=$~\SI{15}{\milli\second} \\ $\eta=0.60$} & \shortstack{\\[-2pt]GBMQR \\$\kappa_{\text{damp}}^{-1}=$~\SI{10}{\milli\second} \\ $\eta=0.60$} & \shortstack{\\[-2pt]GBMQR \\$\kappa_{\text{damp}}^{-1}=$~\SI{8}{\milli\second} \\ $\eta=0.60$}& \shortstack{\\ \vspace*{1.6em}BSMQR \\ $\eta_{1}=\eta_{2}=0.95$ } & \shortstack{\\ \vspace*{1.6em}BSMQR \\ $\eta_{1}=\eta_{2}=0.60$ } \\ \hline
$\alpha$ & $4 \times 10^{-4}$ & $10\times 10^{-4}$ &$1\times 10^{-3}$& $4\times 10^{-3}$ & $6 \times 10^{-3}$&$1 \times 10^{-2}$ & $5 \times 10^{-2}$  \\ \hline
$\beta$ & 0.90 & 0.80 &0.73&0.63&0.57&0.64 & 0.34 \\ \hline
$\mathcal{P}^{\text{succ}}_{\text{EG}}$& 0.75 & 0.60 & 0.47& 0.37&{0.31}&0.32 & 0.17 \\ \hline
$\mathcal{P}^{\text{succ}}_{\text{SW}}$& 0.58 & 0.36&{0.24}&{0.14} & {0.10}&0.32 &0.17\\ \hline
\end{tabular}}
\caption{Performance parameters of GBMQR and BSMQR. The codewords cross-overlap and similar overlap are denoted $\alpha$, and $\beta$ respectively. The transmissivity of the channel of the  remote controlled-Z gate is $\eta$, where we assumed a corresponding losses equal to \SI{2}{\deci\bel\per\kilo\meter}. Mismatch losses for each codeword are denoted $\eta_{1}$, and $\eta_{2}$. Stationary losses  $\kappa^{-1}_{\rm damp}$ for GBMQR are in the range of~\SIrange{8}{40}{\milli\second}. In ideal BSMQR $\eta_{1}=\eta_{2}= 0.95 $, whereas in realistic BSMQR $\eta_{1}=\eta_{2}=0.6$ . }
\label{tab:comparison}
\end{table}
\begin{figure}[H]
    \centering
    \includegraphics[width=1\linewidth]{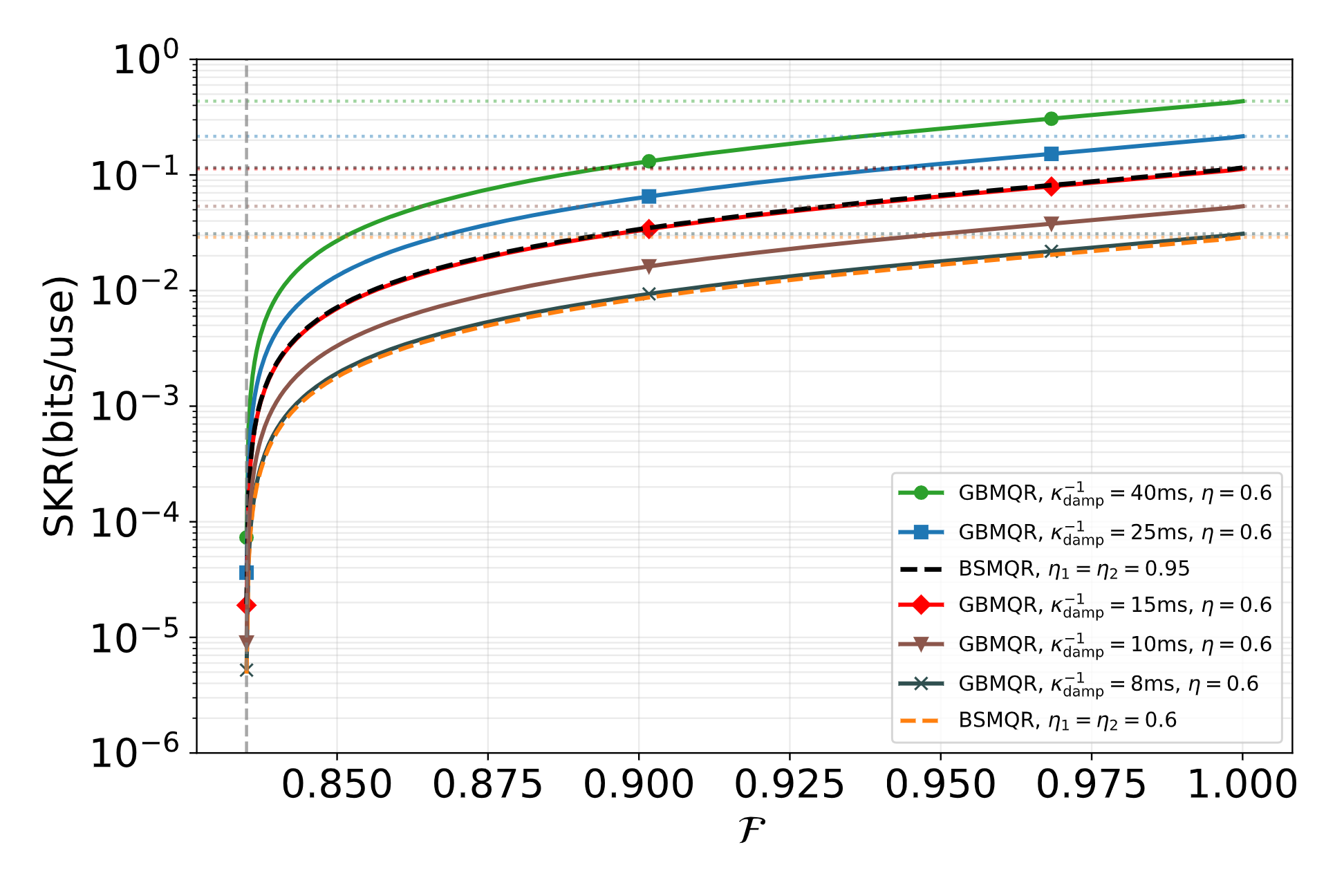}
    \caption{A comparison between GBMQR and BSMQR secret key rates as a function of the fidelity $\mathcal{F}$ of the final entangled state. The plot considers the fidelity threshold where a quantum repeater is capable of generating a positive key rate \cite{RevModPhys.81.1301}.  The best performance of GBMQR is achieved when stationary damping is $\kappa^{-1}_{\rm damp}=$~\SI{40}{\milli\second} (solid green). For stationary damping rate of $\kappa^{-1}_{\rm damp}=$~\SI{25}{\milli\second} (solid blue) GBMQR still outperforms an ideal BSMQR at overall losses of $\eta_{1}=\eta_{2}=0.95$ (dashed purple). At $\kappa^{-1}_{\rm damp}=$~\SI{15}{\milli\second} (solid orange), the performance of GBMQR coincides with that of an ideal BSMQR. When $\kappa^{-1}_{\rm damp}=$~\SI{10}{\milli\second} (solid brown), it can be seen that GBMQR is outperformed by an ideal BSMQR, while still outperforming a practical BSMQR. Decreasing the cavity lifetime further, i.e., $\kappa^{-1}_{\rm damp}=$~\SI{8}{\milli\second} (solid dark grey), GBMQR and realistic BSMQR (dotted orange) generates the smallest secret key rate. It is important to note here that damping values in the range of ${25\text{-}}\SI{40}{\milli\second}$ has been recently experimentally reported in \cite{romanenko2020three,kim2025ultracoherent}. }
  \label{fig:SKR}
\end{figure} 
\vspace*{-1em}
As for entanglement swapping in GBMQR, a Bell-state measurement is achieved by a deterministic controlled-Z gate followed by two single bosonic-qubit measurements in the X-basis. Based on the outcomes of these measurements the system follows a deterministic trajectory towards a Bell-state. Contrary to a beamsplitter Bell-state measurement where the system follows a heralded probabilistic path that depends on detection statistics.

The success probability of the gate-based approach depends on the success probability of the entangling controlled-Z gate and the two projective measurements on the intermediate bosonic modes in the X-basis. The success probability of the projective measurements involves the measurement fidelity of a homodyne POVM detector. The controlled-Z  gate has a success probability that approximately scales as the squared overlap $\beta^{2}$ (see Appendix \ref{app:EGSWAP}). When $\kappa_{\text{damp}}^{-1}=$ ~\SI{25}{\milli\second}, the overlap is $\beta \approx 0.80$, whereas when $\kappa_{\text{damp}}^{-1}=$~\SI{40}{\milli\second}, $\beta\approx0.90$ (see Appendix~\ref{app:OVerlapExps} and Fig.~\ref{figoverlap}). In both cases, the measurement fidelity of a homodyne detection is taken to be $\approx 0.95$. Accordingly, entanglement swapping success probability is $\mathcal{P}^{\text{succ}}_{\text{SW}}$  $\approx 0.36$, and $\approx 0.58$ respectively. However, as storage losses increase, the similar overlap and the success probabilities of entanglement generation and swapping both decrease (see Table \ref{tab:comparison} for details). 
In Fig.~\ref{fig:SKR} we compare the secret key rate of both GBMQR and a beamsplitter-based microwave quantum repeater. As can be seen from the figure, the best performance was achieved by GBMQR operating at damping rate 
$\kappa_{\text{damp}}^{-1}=$~\SI{40}{\milli\second}. 
When $\kappa_{\text{damp}}^{-1}=$ ~\SI{25}{\milli\second}, GBMQR is able to outperform an ideal beamsplitter-based microwave repeater and generate higher key rates. A further decrease in the storage lifetime degrades the achievable key rates. Particularly, when $\kappa_{\text{damp}}^{-1}=$ ~\SI{15}{\milli\second}, GBMQR generates secret keys at a rate coinciding with that of an ideal BSMQR. It can be seen from Fig. (\ref{fig:SKR}) that the trend continues at a damping value of $\kappa_{\text{damp}}^{-1}=$ ~\SI{10}{\milli\second}. The performance of GBMQR reaches its lowest value at $\kappa_{\text{damp}}^{-1}=$ ~\SI{8}{\milli\second}. It can be noted as well that the rate of generated secret keys coincides with that of a realistic BSMQR operating at transmissivities of $\eta_{1}=\eta_{2}=0.6$.

It is important to note here, that the considered damping rates correspond to  $150\text{--}300$ rounds of error correction, each of which lasts for $\approx$~\SI{7}{\micro\second}. Furthermore, damping values in the range of $~\SIrange{25}{40}{\milli\second}$ are within the current capabilities of superconducting technology,  as recently reported in \cite{romanenko2020three,kim2025ultracoherent}. Thus, by noting that the overall duration of a typical quantum repeater protocol is in the range of~\SIrange{1}{3}{\milli\second}~\cite{RevModPhys.83.33}, these values reasonably estimate the overall losses.

\section{Conclusion}\label{sec:V}
In this article, we proposed a novel gate‑based microwave quantum repeater (GBMQR) for secure chip‑to‑chip communication and distributed quantum computing. The device is designed to address probabilistic entanglement generation and swapping using a balanced beamsplitter, while mitigating loss during information storage prior to swapping. By leveraging autonomous error correction, GBMQR exploits the extended lifetime of error‑corrected grid states to store bosonic modes before swapping.
Operationally, GBMQR employs deterministic gates from the cQED toolbox. Entanglement generation utilizes a remote transmon–transmon controlled‑Z gate, while swapping is performed using an all‑bosonic controlled‑Z gate. Projective measurements for entanglement generation are carried out on transmons with high efficiency, and bosonic measurements for swapping are implemented via homodyne detection, eliminating the need for ideal microwave single‑photon counters. We evaluated the performance of GBMQR against a beamsplitter‑based microwave quantum repeater (BSMQR) by comparing secret key rates. Owing to the separation of stationary memory modes and propagating lossy modes, GBMQR outperformed BSMQR, which suffers from mode mismatch and routing errors.
Several challenges remain in improving GBMQR performance. We identify two key factors affecting the fidelity of the final entangled state and the achievable key rate. First, increased channel transmissivity, $\eta$, enhances the success probability of entanglement generation. Second, improved codeword orthogonality and better overlap matching $\beta$ increase gate success probabilities. These improvements can be achieved by operating at higher squeezing levels. While squeezing of \SIrange{10}{15}{\deci\bel} \cite{qiu2023broadband} introduces cross‑overlap errors of $\approx 10^{-3}$, increasing squeezing to \SI{20}{\deci\bel} can suppress errors to $\approx 10^{-5}$.
We also assumed ideal transmon operations. In practice, transmon gate errors would reduce the success probability of entanglement generation. Moreover, when multiple entanglement segments on the same chip are operated in parallel, transmon crosstalk can degrade GBMQR performance and lower the final secret key rate.

Looking forward, GBMQR could function as a bosonic remote entangling gate acting directly on the logical subspace of autonomously error-corrected grid states. This is particularly relevant for assembling bosonic square-plaquette cluster primitives into larger nearest-neighbor lattices for measurement-based quantum computing, thereby mitigating long-range wiring challenges in large cluster-state architectures. Longer microwave networks may also be constructed by concatenating elementary GBMQR units, with end-to-end rates depending on link generation, swapping, memory lifetime, and scheduling. Finally, despite physical limits on squeezing that constrain overlaps $\alpha$ and $\beta$, additional encoding layers on top of \texttt{GKP} grid states can significantly reduce error rates through code concatenation.

\section{Acknowledgment}
We acknowledge financial support from the Research Council of Finland Grant No. $359177$ and the H2Future project through the Research Council of Finland (Grant. No. $352788$) and the University of Oulu. 

\section{Author contributions }
H.K. conceived and led the research, including development of the theoretical framework, all analytical and numerical calculations, and preparation of the manuscript. M.S. provided project supervision, funding support, and manuscript review.
\pagebreak

\begin{thebibliography}{63}%
\makeatletter
\providecommand \@ifxundefined [1]{%
 \@ifx{#1\undefined}
}%
\providecommand \@ifnum [1]{%
 \ifnum #1\expandafter \@firstoftwo
 \else \expandafter \@secondoftwo
 \fi
}%
\providecommand \@ifx [1]{%
 \ifx #1\expandafter \@firstoftwo
 \else \expandafter \@secondoftwo
 \fi
}%
\providecommand \natexlab [1]{#1}%
\providecommand \enquote  [1]{``#1''}%
\providecommand \bibnamefont  [1]{#1}%
\providecommand \bibfnamefont [1]{#1}%
\providecommand \citenamefont [1]{#1}%
\providecommand \href@noop [0]{\@secondoftwo}%
\providecommand \href [0]{\begingroup \@sanitize@url \@href}%
\providecommand \@href[1]{\@@startlink{#1}\@@href}%
\providecommand \@@href[1]{\endgroup#1\@@endlink}%
\providecommand \@sanitize@url [0]{\catcode `\\12\catcode `\$12\catcode `\&12\catcode `\#12\catcode `\^12\catcode `\_12\catcode `\%12\relax}%
\providecommand \@@startlink[1]{}%
\providecommand \@@endlink[0]{}%
\providecommand \url  [0]{\begingroup\@sanitize@url \@url }%
\providecommand \@url [1]{\endgroup\@href {#1}{\urlprefix }}%
\providecommand \urlprefix  [0]{URL }%
\providecommand \Eprint [0]{\href }%
\providecommand \doibase [0]{https://doi.org/}%
\providecommand \selectlanguage [0]{\@gobble}%
\providecommand \bibinfo  [0]{\@secondoftwo}%
\providecommand \bibfield  [0]{\@secondoftwo}%
\providecommand \translation [1]{[#1]}%
\providecommand \BibitemOpen [0]{}%
\providecommand \bibitemStop [0]{}%
\providecommand \bibitemNoStop [0]{.\EOS\space}%
\providecommand \EOS [0]{\spacefactor3000\relax}%
\providecommand \BibitemShut  [1]{\csname bibitem#1\endcsname}%
\let\auto@bib@innerbib\@empty
\bibitem [{\citenamefont {Ekert}(1991)}]{ekert1991quantum}%
  \BibitemOpen
  \bibfield  {author} {\bibinfo {author} {\bibfnamefont {A.~K.}\ \bibnamefont {Ekert}},\ }\bibfield  {title} {\bibinfo {title} {Quantum cryptography based on {Bell}’s theorem},\ }\href {https://journals.aps.org/prl/abstract/10.1103/PhysRevLett.67.661} {\bibfield  {journal} {\bibinfo  {journal} {Phys. Rev. Lett.}\ }\textbf {\bibinfo {volume} {67}},\ \bibinfo {pages} {661} (\bibinfo {year} {1991})}\BibitemShut {NoStop}%
\bibitem [{\citenamefont {Bennett}\ \emph {et~al.}(1993)\citenamefont {Bennett}, \citenamefont {Brassard}, \citenamefont {Cr{\'e}peau}, \citenamefont {Jozsa}, \citenamefont {Peres},\ and\ \citenamefont {Wootters}}]{bennett1993teleporting}%
  \BibitemOpen
  \bibfield  {author} {\bibinfo {author} {\bibfnamefont {C.~H.}\ \bibnamefont {Bennett}}, \bibinfo {author} {\bibfnamefont {G.}~\bibnamefont {Brassard}}, \bibinfo {author} {\bibfnamefont {C.}~\bibnamefont {Cr{\'e}peau}}, \bibinfo {author} {\bibfnamefont {R.}~\bibnamefont {Jozsa}}, \bibinfo {author} {\bibfnamefont {A.}~\bibnamefont {Peres}},\ and\ \bibinfo {author} {\bibfnamefont {W.~K.}\ \bibnamefont {Wootters}},\ }\bibfield  {title} {\bibinfo {title} {Teleporting an unknown quantum state via dual classical and {Einstein-Podolsky-Rosen} channels},\ }\href {https://journals.aps.org/prl/abstract/10.1103/PhysRevLett.70.1895} {\bibfield  {journal} {\bibinfo  {journal} {Phys. Rev. Lett.}\ }\textbf {\bibinfo {volume} {70}},\ \bibinfo {pages} {1895} (\bibinfo {year} {1993})}\BibitemShut {NoStop}%
\bibitem [{\citenamefont {Degen}\ \emph {et~al.}(2017)\citenamefont {Degen}, \citenamefont {Reinhard},\ and\ \citenamefont {Cappellaro}}]{degen2017quantum}%
  \BibitemOpen
  \bibfield  {author} {\bibinfo {author} {\bibfnamefont {C.~L.}\ \bibnamefont {Degen}}, \bibinfo {author} {\bibfnamefont {F.}~\bibnamefont {Reinhard}},\ and\ \bibinfo {author} {\bibfnamefont {P.}~\bibnamefont {Cappellaro}},\ }\bibfield  {title} {\bibinfo {title} {Quantum sensing},\ }\href {https://journals.aps.org/rmp/abstract/10.1103/RevModPhys.89.035002} {\bibfield  {journal} {\bibinfo  {journal} {Rev. Mod. Phys.}\ }\textbf {\bibinfo {volume} {89}},\ \bibinfo {pages} {035002} (\bibinfo {year} {2017})}\BibitemShut {NoStop}%
\bibitem [{\citenamefont {Kimble}(2008)}]{kimble2008quantum}%
  \BibitemOpen
  \bibfield  {author} {\bibinfo {author} {\bibfnamefont {H.~J.}\ \bibnamefont {Kimble}},\ }\bibfield  {title} {\bibinfo {title} {The quantum internet},\ }\href {https://www.nature.com/articles/nature07127} {\bibfield  {journal} {\bibinfo  {journal} {Nature}\ }\textbf {\bibinfo {volume} {453}},\ \bibinfo {pages} {1023} (\bibinfo {year} {2008})}\BibitemShut {NoStop}%
\bibitem [{\citenamefont {Van~Meter}\ and\ \citenamefont {Devitt}(2016)}]{van2016path}%
  \BibitemOpen
  \bibfield  {author} {\bibinfo {author} {\bibfnamefont {R.}~\bibnamefont {Van~Meter}}\ and\ \bibinfo {author} {\bibfnamefont {S.~J.}\ \bibnamefont {Devitt}},\ }\bibfield  {title} {\bibinfo {title} {The path to scalable distributed quantum computing},\ }\href {https://ieeexplore.ieee.org/document/7562346} {\bibfield  {journal} {\bibinfo  {journal} {Computer}\ }\textbf {\bibinfo {volume} {49}},\ \bibinfo {pages} {31} (\bibinfo {year} {2016})}\BibitemShut {NoStop}%
\bibitem [{\citenamefont {Sangouard}\ \emph {et~al.}(2011)\citenamefont {Sangouard}, \citenamefont {Simon}, \citenamefont {de~Riedmatten},\ and\ \citenamefont {Gisin}}]{RevModPhys.83.33}%
  \BibitemOpen
  \bibfield  {author} {\bibinfo {author} {\bibfnamefont {N.}~\bibnamefont {Sangouard}}, \bibinfo {author} {\bibfnamefont {C.}~\bibnamefont {Simon}}, \bibinfo {author} {\bibfnamefont {H.}~\bibnamefont {de~Riedmatten}},\ and\ \bibinfo {author} {\bibfnamefont {N.}~\bibnamefont {Gisin}},\ }\bibfield  {title} {\bibinfo {title} {Quantum repeaters based on atomic ensembles and linear optics},\ }\href {https://doi.org/10.1103/RevModPhys.83.33} {\bibfield  {journal} {\bibinfo  {journal} {Rev. Mod. Phys.}\ }\textbf {\bibinfo {volume} {83}},\ \bibinfo {pages} {33} (\bibinfo {year} {2011})}\BibitemShut {NoStop}%
\bibitem [{\citenamefont {Pan}\ \emph {et~al.}(1998)\citenamefont {Pan}, \citenamefont {Bouwmeester}, \citenamefont {Weinfurter},\ and\ \citenamefont {Zeilinger}}]{pan1998experimental}%
  \BibitemOpen
  \bibfield  {author} {\bibinfo {author} {\bibfnamefont {J.-W.}\ \bibnamefont {Pan}}, \bibinfo {author} {\bibfnamefont {D.}~\bibnamefont {Bouwmeester}}, \bibinfo {author} {\bibfnamefont {H.}~\bibnamefont {Weinfurter}},\ and\ \bibinfo {author} {\bibfnamefont {A.}~\bibnamefont {Zeilinger}},\ }\bibfield  {title} {\bibinfo {title} {Experimental entanglement swapping: entangling photons that never interacted},\ }\href {https://journals.aps.org/prl/abstract/10.1103/PhysRevLett.80.3891} {\bibfield  {journal} {\bibinfo  {journal} {Phys. Rev. Lett.}\ }\textbf {\bibinfo {volume} {80}},\ \bibinfo {pages} {3891} (\bibinfo {year} {1998})}\BibitemShut {NoStop}%
\bibitem [{\citenamefont {Duan}\ \emph {et~al.}(2001)\citenamefont {Duan}, \citenamefont {Lukin}, \citenamefont {Cirac},\ and\ \citenamefont {Zoller}}]{duan2001long}%
  \BibitemOpen
  \bibfield  {author} {\bibinfo {author} {\bibfnamefont {L.-M.}\ \bibnamefont {Duan}}, \bibinfo {author} {\bibfnamefont {M.~D.}\ \bibnamefont {Lukin}}, \bibinfo {author} {\bibfnamefont {J.~I.}\ \bibnamefont {Cirac}},\ and\ \bibinfo {author} {\bibfnamefont {P.}~\bibnamefont {Zoller}},\ }\bibfield  {title} {\bibinfo {title} {Long-distance quantum communication with atomic ensembles and linear optics},\ }\href {https://www.nature.com/articles/35106500} {\bibfield  {journal} {\bibinfo  {journal} {Nature}\ }\textbf {\bibinfo {volume} {414}},\ \bibinfo {pages} {413} (\bibinfo {year} {2001})}\BibitemShut {NoStop}%
\bibitem [{\citenamefont {Azuma}\ \emph {et~al.}(2015)\citenamefont {Azuma}, \citenamefont {Tamaki},\ and\ \citenamefont {Lo}}]{azuma2015all}%
  \BibitemOpen
  \bibfield  {author} {\bibinfo {author} {\bibfnamefont {K.}~\bibnamefont {Azuma}}, \bibinfo {author} {\bibfnamefont {K.}~\bibnamefont {Tamaki}},\ and\ \bibinfo {author} {\bibfnamefont {H.-K.}\ \bibnamefont {Lo}},\ }\bibfield  {title} {\bibinfo {title} {All-photonic quantum repeaters},\ }\href {https://www.nature.com/articles/ncomms7787} {\bibfield  {journal} {\bibinfo  {journal} {Nat. Commun.}\ }\textbf {\bibinfo {volume} {6}},\ \bibinfo {pages} {6787} (\bibinfo {year} {2015})}\BibitemShut {NoStop}%
\bibitem [{\citenamefont {Jiang}\ \emph {et~al.}(2009)\citenamefont {Jiang}, \citenamefont {Taylor}, \citenamefont {Nemoto}, \citenamefont {Munro}, \citenamefont {Van~Meter},\ and\ \citenamefont {Lukin}}]{jiang2009quantum}%
  \BibitemOpen
  \bibfield  {author} {\bibinfo {author} {\bibfnamefont {L.}~\bibnamefont {Jiang}}, \bibinfo {author} {\bibfnamefont {J.~M.}\ \bibnamefont {Taylor}}, \bibinfo {author} {\bibfnamefont {K.}~\bibnamefont {Nemoto}}, \bibinfo {author} {\bibfnamefont {W.~J.}\ \bibnamefont {Munro}}, \bibinfo {author} {\bibfnamefont {R.}~\bibnamefont {Van~Meter}},\ and\ \bibinfo {author} {\bibfnamefont {M.~D.}\ \bibnamefont {Lukin}},\ }\bibfield  {title} {\bibinfo {title} {Quantum repeater with encoding},\ }\href {https://journals.aps.org/pra/abstract/10.1103/PhysRevA.79.032325} {\bibfield  {journal} {\bibinfo  {journal} {Phys. Rev. A}\ }\textbf {\bibinfo {volume} {79}},\ \bibinfo {pages} {032325} (\bibinfo {year} {2009})}\BibitemShut {NoStop}%
\bibitem [{\citenamefont {Chelluri}\ \emph {et~al.}(2025)\citenamefont {Chelluri}, \citenamefont {Sharma}, \citenamefont {Schmidt}, \citenamefont {Kusminskiy},\ and\ \citenamefont {van Loock}}]{chelluri2025bosonic}%
  \BibitemOpen
  \bibfield  {author} {\bibinfo {author} {\bibfnamefont {S.~S.}\ \bibnamefont {Chelluri}}, \bibinfo {author} {\bibfnamefont {S.}~\bibnamefont {Sharma}}, \bibinfo {author} {\bibfnamefont {F.}~\bibnamefont {Schmidt}}, \bibinfo {author} {\bibfnamefont {S.~V.}\ \bibnamefont {Kusminskiy}},\ and\ \bibinfo {author} {\bibfnamefont {P.}~\bibnamefont {van Loock}},\ }\bibfield  {title} {\bibinfo {title} {Bosonic quantum error correction with microwave cavities for quantum repeaters},\ }\href {https://arxiv.org/abs/2503.21569} {\bibfield  {journal} {\bibinfo  {journal} {arXiv:2503.21569}\ } (\bibinfo {year} {2025})}\BibitemShut {NoStop}%
\bibitem [{\citenamefont {Noh}\ \emph {et~al.}(2020)\citenamefont {Noh}, \citenamefont {Girvin},\ and\ \citenamefont {Jiang}}]{noh2020encoding}%
  \BibitemOpen
  \bibfield  {author} {\bibinfo {author} {\bibfnamefont {K.}~\bibnamefont {Noh}}, \bibinfo {author} {\bibfnamefont {S.}~\bibnamefont {Girvin}},\ and\ \bibinfo {author} {\bibfnamefont {L.}~\bibnamefont {Jiang}},\ }\bibfield  {title} {\bibinfo {title} {Encoding an oscillator into many oscillators},\ }\href {https://journals.aps.org/prl/abstract/10.1103/PhysRevLett.125.080503} {\bibfield  {journal} {\bibinfo  {journal} {Phys. Rev. Lett.}\ }\textbf {\bibinfo {volume} {125}},\ \bibinfo {pages} {080503} (\bibinfo {year} {2020})}\BibitemShut {NoStop}%
\bibitem [{\citenamefont {Sivak}\ \emph {et~al.}(2023)\citenamefont {Sivak}, \citenamefont {Eickbusch}, \citenamefont {Royer}, \citenamefont {Singh}, \citenamefont {Tsioutsios}, \citenamefont {Ganjam}, \citenamefont {Miano}, \citenamefont {Brock}, \citenamefont {Ding}, \citenamefont {Frunzio} \emph {et~al.}}]{sivak2023real}%
  \BibitemOpen
  \bibfield  {author} {\bibinfo {author} {\bibfnamefont {V.~V.}\ \bibnamefont {Sivak}}, \bibinfo {author} {\bibfnamefont {A.}~\bibnamefont {Eickbusch}}, \bibinfo {author} {\bibfnamefont {B.}~\bibnamefont {Royer}}, \bibinfo {author} {\bibfnamefont {S.}~\bibnamefont {Singh}}, \bibinfo {author} {\bibfnamefont {I.}~\bibnamefont {Tsioutsios}}, \bibinfo {author} {\bibfnamefont {S.}~\bibnamefont {Ganjam}}, \bibinfo {author} {\bibfnamefont {A.}~\bibnamefont {Miano}}, \bibinfo {author} {\bibfnamefont {B.}~\bibnamefont {Brock}}, \bibinfo {author} {\bibfnamefont {A.}~\bibnamefont {Ding}}, \bibinfo {author} {\bibfnamefont {L.}~\bibnamefont {Frunzio}}, \emph {et~al.},\ }\bibfield  {title} {\bibinfo {title} {Real-time quantum error correction beyond break-even},\ }\href {https://www.nature.com/articles/s41586-023-05782-6} {\bibfield  {journal} {\bibinfo  {journal} {Nature}\ }\textbf {\bibinfo {volume} {616}},\ \bibinfo {pages} {50} (\bibinfo {year} {2023})}\BibitemShut {NoStop}%
\bibitem [{\citenamefont {Brock}\ \emph {et~al.}(2025)\citenamefont {Brock}, \citenamefont {Singh}, \citenamefont {Eickbusch}, \citenamefont {Sivak}, \citenamefont {Ding}, \citenamefont {Frunzio}, \citenamefont {Girvin},\ and\ \citenamefont {Devoret}}]{brock2025quantum}%
  \BibitemOpen
  \bibfield  {author} {\bibinfo {author} {\bibfnamefont {B.~L.}\ \bibnamefont {Brock}}, \bibinfo {author} {\bibfnamefont {S.}~\bibnamefont {Singh}}, \bibinfo {author} {\bibfnamefont {A.}~\bibnamefont {Eickbusch}}, \bibinfo {author} {\bibfnamefont {V.~V.}\ \bibnamefont {Sivak}}, \bibinfo {author} {\bibfnamefont {A.~Z.}\ \bibnamefont {Ding}}, \bibinfo {author} {\bibfnamefont {L.}~\bibnamefont {Frunzio}}, \bibinfo {author} {\bibfnamefont {S.~M.}\ \bibnamefont {Girvin}},\ and\ \bibinfo {author} {\bibfnamefont {M.~H.}\ \bibnamefont {Devoret}},\ }\bibfield  {title} {\bibinfo {title} {Quantum error correction of qudits beyond break-even},\ }\href {https://www.nature.com/articles/s41586-025-08899-y} {\bibfield  {journal} {\bibinfo  {journal} {Nature}\ }\textbf {\bibinfo {volume} {641}},\ \bibinfo {pages} {612} (\bibinfo {year} {2025})}\BibitemShut {NoStop}%
\bibitem [{\citenamefont {Krantz}\ \emph {et~al.}(2019)\citenamefont {Krantz}, \citenamefont {Kjaergaard}, \citenamefont {Yan}, \citenamefont {Orlando}, \citenamefont {Gustavsson},\ and\ \citenamefont {Oliver}}]{krantz2019quantum}%
  \BibitemOpen
  \bibfield  {author} {\bibinfo {author} {\bibfnamefont {P.}~\bibnamefont {Krantz}}, \bibinfo {author} {\bibfnamefont {M.}~\bibnamefont {Kjaergaard}}, \bibinfo {author} {\bibfnamefont {F.}~\bibnamefont {Yan}}, \bibinfo {author} {\bibfnamefont {T.~P.}\ \bibnamefont {Orlando}}, \bibinfo {author} {\bibfnamefont {S.}~\bibnamefont {Gustavsson}},\ and\ \bibinfo {author} {\bibfnamefont {W.~D.}\ \bibnamefont {Oliver}},\ }\bibfield  {title} {\bibinfo {title} {A quantum engineer's guide to superconducting qubits},\ }\href {https://pubs.aip.org/aip/apr/article/6/2/021318/570326} {\bibfield  {journal} {\bibinfo  {journal} {App. Phys. Rev.}\ }\textbf {\bibinfo {volume} {6}} (\bibinfo {year} {2019})}\BibitemShut {NoStop}%
\bibitem [{\citenamefont {Royer}\ \emph {et~al.}(2020)\citenamefont {Royer}, \citenamefont {Singh},\ and\ \citenamefont {Girvin}}]{royer2020stabilization}%
  \BibitemOpen
  \bibfield  {author} {\bibinfo {author} {\bibfnamefont {B.}~\bibnamefont {Royer}}, \bibinfo {author} {\bibfnamefont {S.}~\bibnamefont {Singh}},\ and\ \bibinfo {author} {\bibfnamefont {S.}~\bibnamefont {Girvin}},\ }\bibfield  {title} {\bibinfo {title} {Stabilization of finite-energy {Gottesman-Kitaev-Preskill} states},\ }\href {https://journals.aps.org/prl/abstract/10.1103/PhysRevLett.125.260509} {\bibfield  {journal} {\bibinfo  {journal} {Phys. Rev. Lett.}\ }\textbf {\bibinfo {volume} {125}},\ \bibinfo {pages} {260509} (\bibinfo {year} {2020})}\BibitemShut {NoStop}%
\bibitem [{\citenamefont {Campagne-Ibarcq}\ \emph {et~al.}(2018)\citenamefont {Campagne-Ibarcq}, \citenamefont {Zalys-Geller}, \citenamefont {Narla}, \citenamefont {Shankar}, \citenamefont {Reinhold}, \citenamefont {Burkhart}, \citenamefont {Axline}, \citenamefont {Pfaff}, \citenamefont {Frunzio}, \citenamefont {Schoelkopf} \emph {et~al.}}]{campagne2018deterministic}%
  \BibitemOpen
  \bibfield  {author} {\bibinfo {author} {\bibfnamefont {P.}~\bibnamefont {Campagne-Ibarcq}}, \bibinfo {author} {\bibfnamefont {E.}~\bibnamefont {Zalys-Geller}}, \bibinfo {author} {\bibfnamefont {A.}~\bibnamefont {Narla}}, \bibinfo {author} {\bibfnamefont {S.}~\bibnamefont {Shankar}}, \bibinfo {author} {\bibfnamefont {P.}~\bibnamefont {Reinhold}}, \bibinfo {author} {\bibfnamefont {L.}~\bibnamefont {Burkhart}}, \bibinfo {author} {\bibfnamefont {C.}~\bibnamefont {Axline}}, \bibinfo {author} {\bibfnamefont {W.}~\bibnamefont {Pfaff}}, \bibinfo {author} {\bibfnamefont {L.}~\bibnamefont {Frunzio}}, \bibinfo {author} {\bibfnamefont {R.}~\bibnamefont {Schoelkopf}}, \emph {et~al.},\ }\bibfield  {title} {\bibinfo {title} {Deterministic remote entanglement of superconducting circuits through microwave two-photon transitions},\ }\href {https://journals.aps.org/prl/abstract/10.1103/PhysRevLett.120.200501} {\bibfield  {journal} {\bibinfo  {journal} {Phys. Rev. Lett.}\ }\textbf {\bibinfo {volume} {120}},\ \bibinfo {pages}
  {200501} (\bibinfo {year} {2018})}\BibitemShut {NoStop}%
\bibitem [{\citenamefont {Narla}\ \emph {et~al.}(2016)\citenamefont {Narla}, \citenamefont {Shankar}, \citenamefont {Hatridge}, \citenamefont {Leghtas}, \citenamefont {Sliwa}, \citenamefont {Zalys-Geller}, \citenamefont {Mundhada}, \citenamefont {Pfaff}, \citenamefont {Frunzio}, \citenamefont {Schoelkopf} \emph {et~al.}}]{narla2016robust}%
  \BibitemOpen
  \bibfield  {author} {\bibinfo {author} {\bibfnamefont {A.}~\bibnamefont {Narla}}, \bibinfo {author} {\bibfnamefont {S.}~\bibnamefont {Shankar}}, \bibinfo {author} {\bibfnamefont {M.}~\bibnamefont {Hatridge}}, \bibinfo {author} {\bibfnamefont {Z.}~\bibnamefont {Leghtas}}, \bibinfo {author} {\bibfnamefont {K.~M.}\ \bibnamefont {Sliwa}}, \bibinfo {author} {\bibfnamefont {E.}~\bibnamefont {Zalys-Geller}}, \bibinfo {author} {\bibfnamefont {S.~O.}\ \bibnamefont {Mundhada}}, \bibinfo {author} {\bibfnamefont {W.}~\bibnamefont {Pfaff}}, \bibinfo {author} {\bibfnamefont {L.}~\bibnamefont {Frunzio}}, \bibinfo {author} {\bibfnamefont {R.~J.}\ \bibnamefont {Schoelkopf}}, \emph {et~al.},\ }\bibfield  {title} {\bibinfo {title} {Robust concurrent remote entanglement between two superconducting qubits},\ }\href {https://journals.aps.org/prx/abstract/10.1103/PhysRevX.6.031036} {\bibfield  {journal} {\bibinfo  {journal} {Phys. Rev. X}\ }\textbf {\bibinfo {volume} {6}},\ \bibinfo {pages} {031036} (\bibinfo {year}
  {2016})}\BibitemShut {NoStop}%
\bibitem [{\citenamefont {Casariego}\ \emph {et~al.}(2023)\citenamefont {Casariego}, \citenamefont {Cruzeiro}, \citenamefont {Gherardini}, \citenamefont {Gonzalez-Raya}, \citenamefont {Andr{\'e}}, \citenamefont {Fraz{\~a}o}, \citenamefont {Catto}, \citenamefont {M{\"o}tt{\"o}nen}, \citenamefont {Datta}, \citenamefont {Viisanen} \emph {et~al.}}]{casariego2023propagating}%
  \BibitemOpen
  \bibfield  {author} {\bibinfo {author} {\bibfnamefont {M.}~\bibnamefont {Casariego}}, \bibinfo {author} {\bibfnamefont {E.~Z.}\ \bibnamefont {Cruzeiro}}, \bibinfo {author} {\bibfnamefont {S.}~\bibnamefont {Gherardini}}, \bibinfo {author} {\bibfnamefont {T.}~\bibnamefont {Gonzalez-Raya}}, \bibinfo {author} {\bibfnamefont {R.}~\bibnamefont {Andr{\'e}}}, \bibinfo {author} {\bibfnamefont {G.}~\bibnamefont {Fraz{\~a}o}}, \bibinfo {author} {\bibfnamefont {G.}~\bibnamefont {Catto}}, \bibinfo {author} {\bibfnamefont {M.}~\bibnamefont {M{\"o}tt{\"o}nen}}, \bibinfo {author} {\bibfnamefont {D.}~\bibnamefont {Datta}}, \bibinfo {author} {\bibfnamefont {K.}~\bibnamefont {Viisanen}}, \emph {et~al.},\ }\bibfield  {title} {\bibinfo {title} {Propagating quantum microwaves: {Towards} applications in communication and sensing},\ }\href {https://iopscience.iop.org/article/10.1088/2058-9565/acc4af} {\bibfield  {journal} {\bibinfo  {journal} {Quantum Sci. Technol.}\ }\textbf {\bibinfo {volume} {8}},\ \bibinfo {pages} {023001}
  (\bibinfo {year} {2023})}\BibitemShut {NoStop}%
\bibitem [{\citenamefont {Zhang}\ \emph {et~al.}(2022)\citenamefont {Zhang}, \citenamefont {Srinivasan}, \citenamefont {Sundaresan}, \citenamefont {Bogorin}, \citenamefont {Martin}, \citenamefont {Hertzberg}, \citenamefont {Timmerwilke}, \citenamefont {Pritchett}, \citenamefont {Yau}, \citenamefont {Wang} \emph {et~al.}}]{zhang2022high}%
  \BibitemOpen
  \bibfield  {author} {\bibinfo {author} {\bibfnamefont {E.~J.}\ \bibnamefont {Zhang}}, \bibinfo {author} {\bibfnamefont {S.}~\bibnamefont {Srinivasan}}, \bibinfo {author} {\bibfnamefont {N.}~\bibnamefont {Sundaresan}}, \bibinfo {author} {\bibfnamefont {D.~F.}\ \bibnamefont {Bogorin}}, \bibinfo {author} {\bibfnamefont {Y.}~\bibnamefont {Martin}}, \bibinfo {author} {\bibfnamefont {J.~B.}\ \bibnamefont {Hertzberg}}, \bibinfo {author} {\bibfnamefont {J.}~\bibnamefont {Timmerwilke}}, \bibinfo {author} {\bibfnamefont {E.~J.}\ \bibnamefont {Pritchett}}, \bibinfo {author} {\bibfnamefont {J.-B.}\ \bibnamefont {Yau}}, \bibinfo {author} {\bibfnamefont {C.}~\bibnamefont {Wang}}, \emph {et~al.},\ }\bibfield  {title} {\bibinfo {title} {High-performance superconducting quantum processors via laser annealing of transmon qubits},\ }\href {https://www.science.org/doi/pdf/10.1126/sciadv.abi6690} {\bibfield  {journal} {\bibinfo  {journal} {Sci. Adv.}\ }\textbf {\bibinfo {volume} {8}},\ \bibinfo {pages} {eabi6690} (\bibinfo
  {year} {2022})}\BibitemShut {NoStop}%
\bibitem [{\citenamefont {Lin}\ \emph {et~al.}(2025)\citenamefont {Lin}, \citenamefont {Cho}, \citenamefont {Chen}, \citenamefont {Vavilov}, \citenamefont {Wang},\ and\ \citenamefont {Manucharyan}}]{lin202524}%
  \BibitemOpen
  \bibfield  {author} {\bibinfo {author} {\bibfnamefont {W.-J.}\ \bibnamefont {Lin}}, \bibinfo {author} {\bibfnamefont {H.}~\bibnamefont {Cho}}, \bibinfo {author} {\bibfnamefont {Y.}~\bibnamefont {Chen}}, \bibinfo {author} {\bibfnamefont {M.~G.}\ \bibnamefont {Vavilov}}, \bibinfo {author} {\bibfnamefont {C.}~\bibnamefont {Wang}},\ and\ \bibinfo {author} {\bibfnamefont {V.~E.}\ \bibnamefont {Manucharyan}},\ }\bibfield  {title} {\bibinfo {title} {24 days-stable {CNOT} gate on fluxonium qubits with over 99.9\% fidelity},\ }\href {https://journals.aps.org/prxquantum/abstract/10.1103/PRXQuantum.6.010349} {\bibfield  {journal} {\bibinfo  {journal} {PRX Quantum}\ }\textbf {\bibinfo {volume} {6}},\ \bibinfo {pages} {010349} (\bibinfo {year} {2025})}\BibitemShut {NoStop}%
\bibitem [{\citenamefont {Marxer}\ \emph {et~al.}(2025)\citenamefont {Marxer}, \citenamefont {Mro{\.z}ek}, \citenamefont {Andersson}, \citenamefont {Abdurakhimov}, \citenamefont {Adam}, \citenamefont {Bergholm}, \citenamefont {Beriwal}, \citenamefont {Chan}, \citenamefont {Dahl}, \citenamefont {Das} \emph {et~al.}}]{marxer2025above}%
  \BibitemOpen
  \bibfield  {author} {\bibinfo {author} {\bibfnamefont {F.}~\bibnamefont {Marxer}}, \bibinfo {author} {\bibfnamefont {J.}~\bibnamefont {Mro{\.z}ek}}, \bibinfo {author} {\bibfnamefont {J.}~\bibnamefont {Andersson}}, \bibinfo {author} {\bibfnamefont {L.}~\bibnamefont {Abdurakhimov}}, \bibinfo {author} {\bibfnamefont {J.}~\bibnamefont {Adam}}, \bibinfo {author} {\bibfnamefont {V.}~\bibnamefont {Bergholm}}, \bibinfo {author} {\bibfnamefont {R.}~\bibnamefont {Beriwal}}, \bibinfo {author} {\bibfnamefont {C.~F.}\ \bibnamefont {Chan}}, \bibinfo {author} {\bibfnamefont {S.}~\bibnamefont {Dahl}}, \bibinfo {author} {\bibfnamefont {S.~R.}\ \bibnamefont {Das}}, \emph {et~al.},\ }\bibfield  {title} {\bibinfo {title} {Above 99.9\% fidelity single-qubit gates, two-qubit gates, and readout in a single superconducting quantum device},\ }\href {https://arxiv.org/abs/2508.16437} {\bibfield  {journal} {\bibinfo  {journal} {arXiv:2508.16437}\ } (\bibinfo {year} {2025})}\BibitemShut {NoStop}%
\bibitem [{\citenamefont {Holland}\ \emph {et~al.}(2015)\citenamefont {Holland}, \citenamefont {Vlastakis}, \citenamefont {Heeres}, \citenamefont {Reagor}, \citenamefont {Vool}, \citenamefont {Leghtas}, \citenamefont {Frunzio}, \citenamefont {Kirchmair}, \citenamefont {Devoret}, \citenamefont {Mirrahimi} \emph {et~al.}}]{holland2015single}%
  \BibitemOpen
  \bibfield  {author} {\bibinfo {author} {\bibfnamefont {E.~T.}\ \bibnamefont {Holland}}, \bibinfo {author} {\bibfnamefont {B.}~\bibnamefont {Vlastakis}}, \bibinfo {author} {\bibfnamefont {R.~W.}\ \bibnamefont {Heeres}}, \bibinfo {author} {\bibfnamefont {M.~J.}\ \bibnamefont {Reagor}}, \bibinfo {author} {\bibfnamefont {U.}~\bibnamefont {Vool}}, \bibinfo {author} {\bibfnamefont {Z.}~\bibnamefont {Leghtas}}, \bibinfo {author} {\bibfnamefont {L.}~\bibnamefont {Frunzio}}, \bibinfo {author} {\bibfnamefont {G.}~\bibnamefont {Kirchmair}}, \bibinfo {author} {\bibfnamefont {M.~H.}\ \bibnamefont {Devoret}}, \bibinfo {author} {\bibfnamefont {M.}~\bibnamefont {Mirrahimi}}, \emph {et~al.},\ }\bibfield  {title} {\bibinfo {title} {Single-photon-resolved cross-kerr interaction for autonomous stabilization of photon-number states},\ }\href {https://journals.aps.org/prl/abstract/10.1103/PhysRevLett.115.180501} {\bibfield  {journal} {\bibinfo  {journal} {Phys. Rev. Lett.}\ }\textbf {\bibinfo {volume} {115}},\ \bibinfo {pages}
  {180501} (\bibinfo {year} {2015})}\BibitemShut {NoStop}%
\bibitem [{\citenamefont {Baskov}\ \emph {et~al.}(2025)\citenamefont {Baskov}, \citenamefont {Weiss},\ and\ \citenamefont {Girvin}}]{baskov2025exact}%
  \BibitemOpen
  \bibfield  {author} {\bibinfo {author} {\bibfnamefont {R.}~\bibnamefont {Baskov}}, \bibinfo {author} {\bibfnamefont {D.~K.}\ \bibnamefont {Weiss}},\ and\ \bibinfo {author} {\bibfnamefont {S.~M.}\ \bibnamefont {Girvin}},\ }\bibfield  {title} {\bibinfo {title} {Exact amplitudes of parametric processes in driven {Josephson} circuits},\ }\href {https://doi.org/10.1103/zpl5-lztx} {\bibfield  {journal} {\bibinfo  {journal} {Phys. Rev. Applied}\ }\textbf {\bibinfo {volume} {24}},\ \bibinfo {pages} {054038} (\bibinfo {year} {2025})}\BibitemShut {NoStop}%
\bibitem [{\citenamefont {Rebi{\'c}}\ \emph {et~al.}(2009)\citenamefont {Rebi{\'c}}, \citenamefont {Twamley},\ and\ \citenamefont {Milburn}}]{rebic2009giant}%
  \BibitemOpen
  \bibfield  {author} {\bibinfo {author} {\bibfnamefont {S.}~\bibnamefont {Rebi{\'c}}}, \bibinfo {author} {\bibfnamefont {J.}~\bibnamefont {Twamley}},\ and\ \bibinfo {author} {\bibfnamefont {G.~J.}\ \bibnamefont {Milburn}},\ }\bibfield  {title} {\bibinfo {title} {Giant {Kerr} nonlinearities in circuit quantum electrodynamics},\ }\href {https://journals.aps.org/prl/abstract/10.1103/PhysRevLett.103.150503} {\bibfield  {journal} {\bibinfo  {journal} {Phys. Rev. Lett.}\ }\textbf {\bibinfo {volume} {103}},\ \bibinfo {pages} {150503} (\bibinfo {year} {2009})}\BibitemShut {NoStop}%
\bibitem [{\citenamefont {Hoi}\ \emph {et~al.}(2013)\citenamefont {Hoi}, \citenamefont {Kockum}, \citenamefont {Palomaki}, \citenamefont {Stace}, \citenamefont {Fan}, \citenamefont {Tornberg}, \citenamefont {Sathyamoorthy}, \citenamefont {Johansson}, \citenamefont {Delsing},\ and\ \citenamefont {Wilson}}]{hoi2013giant}%
  \BibitemOpen
  \bibfield  {author} {\bibinfo {author} {\bibfnamefont {I.-C.}\ \bibnamefont {Hoi}}, \bibinfo {author} {\bibfnamefont {A.~F.}\ \bibnamefont {Kockum}}, \bibinfo {author} {\bibfnamefont {T.}~\bibnamefont {Palomaki}}, \bibinfo {author} {\bibfnamefont {T.~M.}\ \bibnamefont {Stace}}, \bibinfo {author} {\bibfnamefont {B.}~\bibnamefont {Fan}}, \bibinfo {author} {\bibfnamefont {L.}~\bibnamefont {Tornberg}}, \bibinfo {author} {\bibfnamefont {S.~R.}\ \bibnamefont {Sathyamoorthy}}, \bibinfo {author} {\bibfnamefont {G.}~\bibnamefont {Johansson}}, \bibinfo {author} {\bibfnamefont {P.}~\bibnamefont {Delsing}},\ and\ \bibinfo {author} {\bibfnamefont {C.}~\bibnamefont {Wilson}},\ }\bibfield  {title} {\bibinfo {title} {Giant cross--{Kerr} effect for propagating microwaves induced by an artificial atom},\ }\href {https://journals.aps.org/prl/abstract/10.1103/PhysRevLett.111.053601} {\bibfield  {journal} {\bibinfo  {journal} {Phys. Rev. Lett.}\ }\textbf {\bibinfo {volume} {111}},\ \bibinfo {pages} {053601} (\bibinfo {year}
  {2013})}\BibitemShut {NoStop}%
\bibitem [{\citenamefont {Strandberg}\ \emph {et~al.}(2024)\citenamefont {Strandberg}, \citenamefont {Eriksson}, \citenamefont {Royer}, \citenamefont {Kervinen},\ and\ \citenamefont {Gasparinetti}}]{strandberg2024digital}%
  \BibitemOpen
  \bibfield  {author} {\bibinfo {author} {\bibfnamefont {I.}~\bibnamefont {Strandberg}}, \bibinfo {author} {\bibfnamefont {A.~M.}\ \bibnamefont {Eriksson}}, \bibinfo {author} {\bibfnamefont {B.}~\bibnamefont {Royer}}, \bibinfo {author} {\bibfnamefont {M.}~\bibnamefont {Kervinen}},\ and\ \bibinfo {author} {\bibfnamefont {S.}~\bibnamefont {Gasparinetti}},\ }\bibfield  {title} {\bibinfo {title} {Digital homodyne and heterodyne detection for stationary bosonic modes},\ }\href {https://journals.aps.org/prl/abstract/10.1103/PhysRevLett.133.063601} {\bibfield  {journal} {\bibinfo  {journal} {Phys. Rev. Lett.}\ }\textbf {\bibinfo {volume} {133}},\ \bibinfo {pages} {063601} (\bibinfo {year} {2024})}\BibitemShut {NoStop}%
\bibitem [{\citenamefont {Khalifa}(2024)}]{HKMQC}%
  \BibitemOpen
  \bibfield  {author} {\bibinfo {author} {\bibfnamefont {H.}~\bibnamefont {Khalifa}},\ }\emph {\bibinfo {title} {Microwave quantum communications: new approaches to sensing and mitigation of the bosonic pure-loss channel}},\ \href {https://aaltodoc.aalto.fi/items/ceec4029-8bf2-4a56-99a6-8e5dac66d102} {Ph.D. thesis},\ \bibinfo  {school} {Aalto University}, \bibinfo {address} {Espoo, Finland} (\bibinfo {year} {2024})\BibitemShut {NoStop}%
\bibitem [{\citenamefont {Khalifa}\ \emph {et~al.}(2024)\citenamefont {Khalifa}, \citenamefont {J{\"a}ntti},\ and\ \citenamefont {Paraoanu}}]{khalifa2024fault}%
  \BibitemOpen
  \bibfield  {author} {\bibinfo {author} {\bibfnamefont {H.}~\bibnamefont {Khalifa}}, \bibinfo {author} {\bibfnamefont {R.}~\bibnamefont {J{\"a}ntti}},\ and\ \bibinfo {author} {\bibfnamefont {G.~S.}\ \bibnamefont {Paraoanu}},\ }\bibfield  {title} {\bibinfo {title} {Fault-tolerant one-way noiseless amplification for microwave bosonic quantum information processing},\ }\href {https://ieeexplore.ieee.org/document/10629178} {\bibfield  {journal} {\bibinfo  {journal} {IEEE Trans. Quantum Eng.}\ }\textbf {\bibinfo {volume} {5}},\ \bibinfo {pages} {4100917} (\bibinfo {year} {2024})}\BibitemShut {NoStop}%
\bibitem [{\citenamefont {H\"aussler}\ and\ \citenamefont {van Loock}(2025)}]{PhysRevA.111.062611}%
  \BibitemOpen
  \bibfield  {author} {\bibinfo {author} {\bibfnamefont {S.}~\bibnamefont {H\"aussler}}\ and\ \bibinfo {author} {\bibfnamefont {P.}~\bibnamefont {van Loock}},\ }\bibfield  {title} {\bibinfo {title} {Quantum repeaters based on stationary {Gottesman-Kitaev-Preskill} qubits},\ }\href {https://doi.org/10.1103/PhysRevA.111.062611} {\bibfield  {journal} {\bibinfo  {journal} {Phys. Rev. A}\ }\textbf {\bibinfo {volume} {111}},\ \bibinfo {pages} {062611} (\bibinfo {year} {2025})}\BibitemShut {NoStop}%
\bibitem [{\citenamefont {Kurpiers}\ \emph {et~al.}(2018)\citenamefont {Kurpiers}, \citenamefont {Magnard}, \citenamefont {Walter}, \citenamefont {Royer}, \citenamefont {Pechal}, \citenamefont {Heinsoo}, \citenamefont {Salath{\'e}}, \citenamefont {Akin}, \citenamefont {Storz}, \citenamefont {Besse} \emph {et~al.}}]{kurpiers2018deterministic}%
  \BibitemOpen
  \bibfield  {author} {\bibinfo {author} {\bibfnamefont {P.}~\bibnamefont {Kurpiers}}, \bibinfo {author} {\bibfnamefont {P.}~\bibnamefont {Magnard}}, \bibinfo {author} {\bibfnamefont {T.}~\bibnamefont {Walter}}, \bibinfo {author} {\bibfnamefont {B.}~\bibnamefont {Royer}}, \bibinfo {author} {\bibfnamefont {M.}~\bibnamefont {Pechal}}, \bibinfo {author} {\bibfnamefont {J.}~\bibnamefont {Heinsoo}}, \bibinfo {author} {\bibfnamefont {Y.}~\bibnamefont {Salath{\'e}}}, \bibinfo {author} {\bibfnamefont {A.}~\bibnamefont {Akin}}, \bibinfo {author} {\bibfnamefont {S.}~\bibnamefont {Storz}}, \bibinfo {author} {\bibfnamefont {J.-C.}\ \bibnamefont {Besse}}, \emph {et~al.},\ }\bibfield  {title} {\bibinfo {title} {Deterministic quantum state transfer and remote entanglement using microwave photons},\ }\href {https://www.nature.com/articles/s41586-018-0195-y} {\bibfield  {journal} {\bibinfo  {journal} {Nature}\ }\textbf {\bibinfo {volume} {558}},\ \bibinfo {pages} {264} (\bibinfo {year} {2018})}\BibitemShut {NoStop}%
\bibitem [{\citenamefont {Knörzer}\ \emph {et~al.}(2026)\citenamefont {Knörzer}, \citenamefont {Liu}, \citenamefont {Schiffer},\ and\ \citenamefont {Tura}}]{Knörzer_2026}%
  \BibitemOpen
  \bibfield  {author} {\bibinfo {author} {\bibfnamefont {J.}~\bibnamefont {Knörzer}}, \bibinfo {author} {\bibfnamefont {X.}~\bibnamefont {Liu}}, \bibinfo {author} {\bibfnamefont {B.~F.}\ \bibnamefont {Schiffer}},\ and\ \bibinfo {author} {\bibfnamefont {J.}~\bibnamefont {Tura}},\ }\bibfield  {title} {\bibinfo {title} {Distributed quantum information processing: a review of recent progress},\ }\href {https://doi.org/10.1088/1361-6633/ae74e0} {\bibfield  {journal} {\bibinfo  {journal} {Reports on Progress in Physics}\ }\textbf {\bibinfo {volume} {89}},\ \bibinfo {pages} {074401} (\bibinfo {year} {2026})}\BibitemShut {NoStop}%
\bibitem [{\citenamefont {LaRacuente}\ \emph {et~al.}(2025)\citenamefont {LaRacuente}, \citenamefont {Smith}, \citenamefont {Imany}, \citenamefont {Silverman},\ and\ \citenamefont {Chong}}]{laracuente2025modeling}%
  \BibitemOpen
  \bibfield  {author} {\bibinfo {author} {\bibfnamefont {N.}~\bibnamefont {LaRacuente}}, \bibinfo {author} {\bibfnamefont {K.~N.}\ \bibnamefont {Smith}}, \bibinfo {author} {\bibfnamefont {P.}~\bibnamefont {Imany}}, \bibinfo {author} {\bibfnamefont {K.~L.}\ \bibnamefont {Silverman}},\ and\ \bibinfo {author} {\bibfnamefont {F.~T.}\ \bibnamefont {Chong}},\ }\bibfield  {title} {\bibinfo {title} {Modeling short-range microwave networks to scale superconducting quantum computation},\ }\href {https://quantum-journal.org/papers/q-2025-01-08-1581/} {\bibfield  {journal} {\bibinfo  {journal} {Quantum}\ }\textbf {\bibinfo {volume} {9}},\ \bibinfo {pages} {1581} (\bibinfo {year} {2025})}\BibitemShut {NoStop}%
\bibitem [{\citenamefont {Babu}\ \emph {et~al.}(2025)\citenamefont {Babu}, \citenamefont {Kerppo}, \citenamefont {Mu{\~n}oz-Moller}, \citenamefont {Haghparast},\ and\ \citenamefont {Silveri}}]{babu2025gate}%
  \BibitemOpen
  \bibfield  {author} {\bibinfo {author} {\bibfnamefont {A.~P.}\ \bibnamefont {Babu}}, \bibinfo {author} {\bibfnamefont {O.}~\bibnamefont {Kerppo}}, \bibinfo {author} {\bibfnamefont {A.}~\bibnamefont {Mu{\~n}oz-Moller}}, \bibinfo {author} {\bibfnamefont {M.}~\bibnamefont {Haghparast}},\ and\ \bibinfo {author} {\bibfnamefont {M.}~\bibnamefont {Silveri}},\ }\bibfield  {title} {\bibinfo {title} {Gate teleportation-assisted routing for quantum algorithms},\ }\href {https://iopscience.iop.org/article/10.1088/2058-9565/adcae4/pdf} {\bibfield  {journal} {\bibinfo  {journal} {Quantum Sci. Technol.}\ }\textbf {\bibinfo {volume} {10}},\ \bibinfo {pages} {035004} (\bibinfo {year} {2025})}\BibitemShut {NoStop}%
\bibitem [{\citenamefont {Grimsmo}\ and\ \citenamefont {Puri}(2021)}]{grimsmo2021quantum}%
  \BibitemOpen
  \bibfield  {author} {\bibinfo {author} {\bibfnamefont {A.~L.}\ \bibnamefont {Grimsmo}}\ and\ \bibinfo {author} {\bibfnamefont {S.}~\bibnamefont {Puri}},\ }\bibfield  {title} {\bibinfo {title} {Quantum error correction with the {Gottesman-Kitaev-Preskill} code},\ }\href {https://journals.aps.org/prxquantum/abstract/10.1103/PRXQuantum.2.020101} {\bibfield  {journal} {\bibinfo  {journal} {PRX Quantum}\ }\textbf {\bibinfo {volume} {2}},\ \bibinfo {pages} {020101} (\bibinfo {year} {2021})}\BibitemShut {NoStop}%
\bibitem [{\citenamefont {Gerry}\ and\ \citenamefont {Knight}(2023)}]{gerry2023introductory}%
  \BibitemOpen
  \bibfield  {author} {\bibinfo {author} {\bibfnamefont {C.~C.}\ \bibnamefont {Gerry}}\ and\ \bibinfo {author} {\bibfnamefont {P.~L.}\ \bibnamefont {Knight}},\ }\href {https://www.cambridge.org/core/books/introductory-quantum-optics/B9866F1F40C45936A81D03AF7617CF44} {\emph {\bibinfo {title} {Introductory quantum optics}}}\ (\bibinfo  {publisher} {Cambridge university press},\ \bibinfo {address} {Cambridge, UK},\ \bibinfo {year} {2023})\BibitemShut {NoStop}%
\bibitem [{\citenamefont {Walls}\ and\ \citenamefont {Milburn}(2025)}]{walls2025quantum}%
  \BibitemOpen
  \bibfield  {author} {\bibinfo {author} {\bibfnamefont {D.}~\bibnamefont {Walls}}\ and\ \bibinfo {author} {\bibfnamefont {G.~J.}\ \bibnamefont {Milburn}},\ }\href {https://link.springer.com/book/10.1007/978-3-031-84177-4} {\emph {\bibinfo {title} {Quantum Optics}}}\ (\bibinfo  {publisher} {Springer},\ \bibinfo {address} {Berlin},\ \bibinfo {year} {2025})\BibitemShut {NoStop}%
\bibitem [{\citenamefont {Campagne-Ibarcq}\ \emph {et~al.}(2020)\citenamefont {Campagne-Ibarcq}, \citenamefont {Eickbusch}, \citenamefont {Touzard}, \citenamefont {Zalys-Geller}, \citenamefont {Frattini}, \citenamefont {Sivak}, \citenamefont {Reinhold}, \citenamefont {Puri}, \citenamefont {Shankar}, \citenamefont {Schoelkopf} \emph {et~al.}}]{campagne2020quantum}%
  \BibitemOpen
  \bibfield  {author} {\bibinfo {author} {\bibfnamefont {P.}~\bibnamefont {Campagne-Ibarcq}}, \bibinfo {author} {\bibfnamefont {A.}~\bibnamefont {Eickbusch}}, \bibinfo {author} {\bibfnamefont {S.}~\bibnamefont {Touzard}}, \bibinfo {author} {\bibfnamefont {E.}~\bibnamefont {Zalys-Geller}}, \bibinfo {author} {\bibfnamefont {N.~E.}\ \bibnamefont {Frattini}}, \bibinfo {author} {\bibfnamefont {V.~V.}\ \bibnamefont {Sivak}}, \bibinfo {author} {\bibfnamefont {P.}~\bibnamefont {Reinhold}}, \bibinfo {author} {\bibfnamefont {S.}~\bibnamefont {Puri}}, \bibinfo {author} {\bibfnamefont {S.}~\bibnamefont {Shankar}}, \bibinfo {author} {\bibfnamefont {R.~J.}\ \bibnamefont {Schoelkopf}}, \emph {et~al.},\ }\bibfield  {title} {\bibinfo {title} {Quantum error correction of a qubit encoded in grid states of an oscillator},\ }\href {https://www.nature.com/articles/s41586-020-2603-3} {\bibfield  {journal} {\bibinfo  {journal} {Nature}\ }\textbf {\bibinfo {volume} {584}},\ \bibinfo {pages} {368} (\bibinfo {year} {2020})}\BibitemShut
  {NoStop}%
\bibitem [{\citenamefont {Eickbusch}\ \emph {et~al.}(2022)\citenamefont {Eickbusch}, \citenamefont {Sivak}, \citenamefont {Ding}, \citenamefont {Elder}, \citenamefont {Jha}, \citenamefont {Venkatraman}, \citenamefont {Royer}, \citenamefont {Girvin}, \citenamefont {Schoelkopf},\ and\ \citenamefont {Devoret}}]{eickbusch2022fast}%
  \BibitemOpen
  \bibfield  {author} {\bibinfo {author} {\bibfnamefont {A.}~\bibnamefont {Eickbusch}}, \bibinfo {author} {\bibfnamefont {V.}~\bibnamefont {Sivak}}, \bibinfo {author} {\bibfnamefont {A.~Z.}\ \bibnamefont {Ding}}, \bibinfo {author} {\bibfnamefont {S.~S.}\ \bibnamefont {Elder}}, \bibinfo {author} {\bibfnamefont {S.~R.}\ \bibnamefont {Jha}}, \bibinfo {author} {\bibfnamefont {J.}~\bibnamefont {Venkatraman}}, \bibinfo {author} {\bibfnamefont {B.}~\bibnamefont {Royer}}, \bibinfo {author} {\bibfnamefont {S.~M.}\ \bibnamefont {Girvin}}, \bibinfo {author} {\bibfnamefont {R.~J.}\ \bibnamefont {Schoelkopf}},\ and\ \bibinfo {author} {\bibfnamefont {M.~H.}\ \bibnamefont {Devoret}},\ }\bibfield  {title} {\bibinfo {title} {Fast universal control of an oscillator with weak dispersive coupling to a qubit},\ }\href {https://www.nature.com/articles/s41567-022-01776-9} {\bibfield  {journal} {\bibinfo  {journal} {Nat. Phys.}\ }\textbf {\bibinfo {volume} {18}},\ \bibinfo {pages} {1464} (\bibinfo {year} {2022})}\BibitemShut
  {NoStop}%
\bibitem [{\citenamefont {Liu}\ \emph {et~al.}(2025)\citenamefont {Liu}, \citenamefont {Singh}, \citenamefont {Smith}, \citenamefont {Crane}, \citenamefont {Martyn}, \citenamefont {Eickbusch}, \citenamefont {Schuckert}, \citenamefont {Li}, \citenamefont {Sinanan-Singh}, \citenamefont {Soley}, \citenamefont {Tsunoda}, \citenamefont {Chuang}, \citenamefont {Wiebe},\ and\ \citenamefont {Girvin}}]{liu2024hybrid}%
  \BibitemOpen
  \bibfield  {author} {\bibinfo {author} {\bibfnamefont {Y.}~\bibnamefont {Liu}}, \bibinfo {author} {\bibfnamefont {S.}~\bibnamefont {Singh}}, \bibinfo {author} {\bibfnamefont {K.~C.}\ \bibnamefont {Smith}}, \bibinfo {author} {\bibfnamefont {E.}~\bibnamefont {Crane}}, \bibinfo {author} {\bibfnamefont {J.~M.}\ \bibnamefont {Martyn}}, \bibinfo {author} {\bibfnamefont {A.}~\bibnamefont {Eickbusch}}, \bibinfo {author} {\bibfnamefont {A.}~\bibnamefont {Schuckert}}, \bibinfo {author} {\bibfnamefont {R.~D.}\ \bibnamefont {Li}}, \bibinfo {author} {\bibfnamefont {J.}~\bibnamefont {Sinanan-Singh}}, \bibinfo {author} {\bibfnamefont {M.~B.}\ \bibnamefont {Soley}}, \bibinfo {author} {\bibfnamefont {T.}~\bibnamefont {Tsunoda}}, \bibinfo {author} {\bibfnamefont {I.~L.}\ \bibnamefont {Chuang}}, \bibinfo {author} {\bibfnamefont {N.}~\bibnamefont {Wiebe}},\ and\ \bibinfo {author} {\bibfnamefont {S.~M.}\ \bibnamefont {Girvin}},\ }\bibfield  {title} {\bibinfo {title} {Hybrid oscillator-qubit quantum processors: Instruction set
  architectures, abstract machine models, and applications},\ }\href {https://doi.org/10.1103/4rf7-9tfx} {\bibfield  {journal} {\bibinfo  {journal} {PRX Quantum}\ }\textbf {\bibinfo {volume} {7}},\ \bibinfo {pages} {010201} (\bibinfo {year} {2025})}\BibitemShut {NoStop}%
\bibitem [{\citenamefont {Kurpiers}\ \emph {et~al.}(2019)\citenamefont {Kurpiers}, \citenamefont {Pechal}, \citenamefont {Royer}, \citenamefont {Magnard}, \citenamefont {Walter}, \citenamefont {Heinsoo}, \citenamefont {Salath{\'e}}, \citenamefont {Akin}, \citenamefont {Storz}, \citenamefont {Besse} \emph {et~al.}}]{kurpiers2019quantum}%
  \BibitemOpen
  \bibfield  {author} {\bibinfo {author} {\bibfnamefont {P.}~\bibnamefont {Kurpiers}}, \bibinfo {author} {\bibfnamefont {M.}~\bibnamefont {Pechal}}, \bibinfo {author} {\bibfnamefont {B.}~\bibnamefont {Royer}}, \bibinfo {author} {\bibfnamefont {P.}~\bibnamefont {Magnard}}, \bibinfo {author} {\bibfnamefont {T.}~\bibnamefont {Walter}}, \bibinfo {author} {\bibfnamefont {J.}~\bibnamefont {Heinsoo}}, \bibinfo {author} {\bibfnamefont {Y.}~\bibnamefont {Salath{\'e}}}, \bibinfo {author} {\bibfnamefont {A.}~\bibnamefont {Akin}}, \bibinfo {author} {\bibfnamefont {S.}~\bibnamefont {Storz}}, \bibinfo {author} {\bibfnamefont {J.-C.}\ \bibnamefont {Besse}}, \emph {et~al.},\ }\bibfield  {title} {\bibinfo {title} {Quantum communication with time-bin encoded microwave photons},\ }\href {https://journals.aps.org/prapplied/abstract/10.1103/PhysRevApplied.12.044067} {\bibfield  {journal} {\bibinfo  {journal} {Phys. Rev. Appl.}\ }\textbf {\bibinfo {volume} {12}},\ \bibinfo {pages} {044067} (\bibinfo {year} {2019})}\BibitemShut
  {NoStop}%
\bibitem [{\citenamefont {Ilves}\ \emph {et~al.}(2020)\citenamefont {Ilves}, \citenamefont {Kono}, \citenamefont {Sunada}, \citenamefont {Yamazaki}, \citenamefont {Kim}, \citenamefont {Koshino},\ and\ \citenamefont {Nakamura}}]{ilves2020demand}%
  \BibitemOpen
  \bibfield  {author} {\bibinfo {author} {\bibfnamefont {J.}~\bibnamefont {Ilves}}, \bibinfo {author} {\bibfnamefont {S.}~\bibnamefont {Kono}}, \bibinfo {author} {\bibfnamefont {Y.}~\bibnamefont {Sunada}}, \bibinfo {author} {\bibfnamefont {S.}~\bibnamefont {Yamazaki}}, \bibinfo {author} {\bibfnamefont {M.}~\bibnamefont {Kim}}, \bibinfo {author} {\bibfnamefont {K.}~\bibnamefont {Koshino}},\ and\ \bibinfo {author} {\bibfnamefont {Y.}~\bibnamefont {Nakamura}},\ }\bibfield  {title} {\bibinfo {title} {On-demand generation and characterization of a microwave time-bin qubit},\ }\href {https://www.nature.com/articles/s41534-020-0266-4} {\bibfield  {journal} {\bibinfo  {journal} {npj Quantum Inf.}\ }\textbf {\bibinfo {volume} {6}},\ \bibinfo {pages} {34} (\bibinfo {year} {2020})}\BibitemShut {NoStop}%
\bibitem [{\citenamefont {McIntyre}\ and\ \citenamefont {Coish}(2025)}]{mcintyre2025protocols}%
  \BibitemOpen
  \bibfield  {author} {\bibinfo {author} {\bibfnamefont {Z.}~\bibnamefont {McIntyre}}\ and\ \bibinfo {author} {\bibfnamefont {W.}~\bibnamefont {Coish}},\ }\bibfield  {title} {\bibinfo {title} {Protocols for intermodule two-qubit gates mediated by time-bin encoded photons},\ }\href {https://journals.aps.org/prresearch/abstract/10.1103/xwyt-1ck7} {\bibfield  {journal} {\bibinfo  {journal} {Phys. Rev. Research}\ }\textbf {\bibinfo {volume} {7}},\ \bibinfo {pages} {023255} (\bibinfo {year} {2025})}\BibitemShut {NoStop}%
\bibitem [{\citenamefont {Storz}\ \emph {et~al.}(2023)\citenamefont {Storz}, \citenamefont {Sch{\"a}r}, \citenamefont {Kulikov}, \citenamefont {Magnard}, \citenamefont {Kurpiers}, \citenamefont {L{\"u}tolf}, \citenamefont {Walter}, \citenamefont {Copetudo}, \citenamefont {Reuer}, \citenamefont {Akin} \emph {et~al.}}]{storz2023loophole}%
  \BibitemOpen
  \bibfield  {author} {\bibinfo {author} {\bibfnamefont {S.}~\bibnamefont {Storz}}, \bibinfo {author} {\bibfnamefont {J.}~\bibnamefont {Sch{\"a}r}}, \bibinfo {author} {\bibfnamefont {A.}~\bibnamefont {Kulikov}}, \bibinfo {author} {\bibfnamefont {P.}~\bibnamefont {Magnard}}, \bibinfo {author} {\bibfnamefont {P.}~\bibnamefont {Kurpiers}}, \bibinfo {author} {\bibfnamefont {J.}~\bibnamefont {L{\"u}tolf}}, \bibinfo {author} {\bibfnamefont {T.}~\bibnamefont {Walter}}, \bibinfo {author} {\bibfnamefont {A.}~\bibnamefont {Copetudo}}, \bibinfo {author} {\bibfnamefont {K.}~\bibnamefont {Reuer}}, \bibinfo {author} {\bibfnamefont {A.}~\bibnamefont {Akin}}, \emph {et~al.},\ }\bibfield  {title} {\bibinfo {title} {Loophole-free {Bell} inequality violation with superconducting circuits},\ }\href {https://www.nature.com/articles/s41586-023-05885-0} {\bibfield  {journal} {\bibinfo  {journal} {Nature}\ }\textbf {\bibinfo {volume} {617}},\ \bibinfo {pages} {265} (\bibinfo {year} {2023})}\BibitemShut {NoStop}%
\bibitem [{\citenamefont {Furusawa}\ and\ \citenamefont {Van~Loock}(2011)}]{furusawa2011quantum}%
  \BibitemOpen
  \bibfield  {author} {\bibinfo {author} {\bibfnamefont {A.}~\bibnamefont {Furusawa}}\ and\ \bibinfo {author} {\bibfnamefont {P.}~\bibnamefont {Van~Loock}},\ }\href {https://onlinelibrary.wiley.com/doi/book/10.1002/9783527635283?msockid=183035cb52f66d58223e20c453796c0a} {\emph {\bibinfo {title} {Quantum teleportation and entanglement: a hybrid approach to optical quantum information processing}}}\ (\bibinfo  {publisher} {John Wiley \& Sons},\ \bibinfo {address} {Weinheim},\ \bibinfo {year} {2011})\BibitemShut {NoStop}%
\bibitem [{\citenamefont {Gisin}\ \emph {et~al.}(2002)\citenamefont {Gisin}, \citenamefont {Ribordy}, \citenamefont {Tittel},\ and\ \citenamefont {Zbinden}}]{RevModPhys.74.145}%
  \BibitemOpen
  \bibfield  {author} {\bibinfo {author} {\bibfnamefont {N.}~\bibnamefont {Gisin}}, \bibinfo {author} {\bibfnamefont {G.}~\bibnamefont {Ribordy}}, \bibinfo {author} {\bibfnamefont {W.}~\bibnamefont {Tittel}},\ and\ \bibinfo {author} {\bibfnamefont {H.}~\bibnamefont {Zbinden}},\ }\bibfield  {title} {\bibinfo {title} {Quantum cryptography},\ }\href {https://doi.org/10.1103/RevModPhys.74.145} {\bibfield  {journal} {\bibinfo  {journal} {Rev. Mod. Phys.}\ }\textbf {\bibinfo {volume} {74}},\ \bibinfo {pages} {145} (\bibinfo {year} {2002})}\BibitemShut {NoStop}%
\bibitem [{\citenamefont {Scarani}\ \emph {et~al.}(2009)\citenamefont {Scarani}, \citenamefont {Bechmann-Pasquinucci}, \citenamefont {Cerf}, \citenamefont {Du\ifmmode~\check{s}\else \v{s}\fi{}ek}, \citenamefont {L\"utkenhaus},\ and\ \citenamefont {Peev}}]{RevModPhys.81.1301}%
  \BibitemOpen
  \bibfield  {author} {\bibinfo {author} {\bibfnamefont {V.}~\bibnamefont {Scarani}}, \bibinfo {author} {\bibfnamefont {H.}~\bibnamefont {Bechmann-Pasquinucci}}, \bibinfo {author} {\bibfnamefont {N.~J.}\ \bibnamefont {Cerf}}, \bibinfo {author} {\bibfnamefont {M.}~\bibnamefont {Du\ifmmode~\check{s}\else \v{s}\fi{}ek}}, \bibinfo {author} {\bibfnamefont {N.}~\bibnamefont {L\"utkenhaus}},\ and\ \bibinfo {author} {\bibfnamefont {M.}~\bibnamefont {Peev}},\ }\bibfield  {title} {\bibinfo {title} {The security of practical quantum key distribution},\ }\href {https://doi.org/10.1103/RevModPhys.81.1301} {\bibfield  {journal} {\bibinfo  {journal} {Rev. Mod. Phys.}\ }\textbf {\bibinfo {volume} {81}},\ \bibinfo {pages} {1301} (\bibinfo {year} {2009})}\BibitemShut {NoStop}%
\bibitem [{\citenamefont {Pirandola}\ \emph {et~al.}(2020)\citenamefont {Pirandola}, \citenamefont {Andersen}, \citenamefont {Banchi}, \citenamefont {Berta}, \citenamefont {Bunandar}, \citenamefont {Colbeck}, \citenamefont {Englund}, \citenamefont {Gehring}, \citenamefont {Lupo}, \citenamefont {Ottaviani}, \citenamefont {Pereira}, \citenamefont {Razavi}, \citenamefont {Shaari}, \citenamefont {Tomamichel}, \citenamefont {Usenko}, \citenamefont {Vallone}, \citenamefont {Villoresi},\ and\ \citenamefont {Wallden}}]{Pirandola:20}%
  \BibitemOpen
  \bibfield  {author} {\bibinfo {author} {\bibfnamefont {S.}~\bibnamefont {Pirandola}}, \bibinfo {author} {\bibfnamefont {U.~L.}\ \bibnamefont {Andersen}}, \bibinfo {author} {\bibfnamefont {L.}~\bibnamefont {Banchi}}, \bibinfo {author} {\bibfnamefont {M.}~\bibnamefont {Berta}}, \bibinfo {author} {\bibfnamefont {D.}~\bibnamefont {Bunandar}}, \bibinfo {author} {\bibfnamefont {R.}~\bibnamefont {Colbeck}}, \bibinfo {author} {\bibfnamefont {D.}~\bibnamefont {Englund}}, \bibinfo {author} {\bibfnamefont {T.}~\bibnamefont {Gehring}}, \bibinfo {author} {\bibfnamefont {C.}~\bibnamefont {Lupo}}, \bibinfo {author} {\bibfnamefont {C.}~\bibnamefont {Ottaviani}}, \bibinfo {author} {\bibfnamefont {J.~L.}\ \bibnamefont {Pereira}}, \bibinfo {author} {\bibfnamefont {M.}~\bibnamefont {Razavi}}, \bibinfo {author} {\bibfnamefont {J.~S.}\ \bibnamefont {Shaari}}, \bibinfo {author} {\bibfnamefont {M.}~\bibnamefont {Tomamichel}}, \bibinfo {author} {\bibfnamefont {V.~C.}\ \bibnamefont {Usenko}}, \bibinfo {author} {\bibfnamefont
  {G.}~\bibnamefont {Vallone}}, \bibinfo {author} {\bibfnamefont {P.}~\bibnamefont {Villoresi}},\ and\ \bibinfo {author} {\bibfnamefont {P.}~\bibnamefont {Wallden}},\ }\bibfield  {title} {\bibinfo {title} {Advances in quantum cryptography},\ }\href {https://doi.org/10.1364/AOP.361502} {\bibfield  {journal} {\bibinfo  {journal} {Adv. Opt. Photon.}\ }\textbf {\bibinfo {volume} {12}},\ \bibinfo {pages} {1012} (\bibinfo {year} {2020})}\BibitemShut {NoStop}%
\bibitem [{\citenamefont {Hong}\ \emph {et~al.}(1987)\citenamefont {Hong}, \citenamefont {Ou},\ and\ \citenamefont {Mandel}}]{hong1987measurement}%
  \BibitemOpen
  \bibfield  {author} {\bibinfo {author} {\bibfnamefont {C.-K.}\ \bibnamefont {Hong}}, \bibinfo {author} {\bibfnamefont {Z.-Y.}\ \bibnamefont {Ou}},\ and\ \bibinfo {author} {\bibfnamefont {L.}~\bibnamefont {Mandel}},\ }\bibfield  {title} {\bibinfo {title} {Measurement of subpicosecond time intervals between two photons by interference},\ }\href {https://journals.aps.org/prl/abstract/10.1103/PhysRevLett.59.2044} {\bibfield  {journal} {\bibinfo  {journal} {Phys. Rev. Lett.}\ }\textbf {\bibinfo {volume} {59}},\ \bibinfo {pages} {2044} (\bibinfo {year} {1987})}\BibitemShut {NoStop}%
\bibitem [{\citenamefont {Weihs}\ and\ \citenamefont {Zeilinger}(2001)}]{weihs2001photon}%
  \BibitemOpen
  \bibfield  {author} {\bibinfo {author} {\bibfnamefont {G.}~\bibnamefont {Weihs}}\ and\ \bibinfo {author} {\bibfnamefont {A.}~\bibnamefont {Zeilinger}},\ }\bibfield  {title} {\bibinfo {title} {Photon statistics at beam splitters: an essential tool in quantum information and teleportation},\ }in\ \href {https://painterlab.caltech.edu/wp-content/uploads/2019/06/iqd_photon_stats_at_beamsplitters.pdf} {\emph {\bibinfo {booktitle} {Coherence and Statistics of Photons and Atoms}}},\ \bibinfo {editor} {edited by\ \bibinfo {editor} {\bibfnamefont {J.}~\bibnamefont {Pe{\v r}ina}}}\ (\bibinfo  {publisher} {Wiley},\ \bibinfo {address} {New Jersey},\ \bibinfo {year} {2001})\ pp.\ \bibinfo {pages} {262--288}\BibitemShut {NoStop}%
\bibitem [{\citenamefont {Bouchard}\ \emph {et~al.}(2020)\citenamefont {Bouchard}, \citenamefont {Sit}, \citenamefont {Zhang}, \citenamefont {Fickler}, \citenamefont {Miatto}, \citenamefont {Yao}, \citenamefont {Sciarrino},\ and\ \citenamefont {Karimi}}]{bouchard2020two}%
  \BibitemOpen
  \bibfield  {author} {\bibinfo {author} {\bibfnamefont {F.}~\bibnamefont {Bouchard}}, \bibinfo {author} {\bibfnamefont {A.}~\bibnamefont {Sit}}, \bibinfo {author} {\bibfnamefont {Y.}~\bibnamefont {Zhang}}, \bibinfo {author} {\bibfnamefont {R.}~\bibnamefont {Fickler}}, \bibinfo {author} {\bibfnamefont {F.~M.}\ \bibnamefont {Miatto}}, \bibinfo {author} {\bibfnamefont {Y.}~\bibnamefont {Yao}}, \bibinfo {author} {\bibfnamefont {F.}~\bibnamefont {Sciarrino}},\ and\ \bibinfo {author} {\bibfnamefont {E.}~\bibnamefont {Karimi}},\ }\bibfield  {title} {\bibinfo {title} {Two-photon interference: the {Hong}--{Ou}--{Mandel} effect},\ }\href {https://iopscience.iop.org/article/10.1088/1361-6633/abcd7a} {\bibfield  {journal} {\bibinfo  {journal} {Rep. Prog. Phys.}\ }\textbf {\bibinfo {volume} {84}},\ \bibinfo {pages} {012402} (\bibinfo {year} {2020})}\BibitemShut {NoStop}%
\bibitem [{\citenamefont {Gao}\ \emph {et~al.}(2018)\citenamefont {Gao}, \citenamefont {Lester}, \citenamefont {Zhang}, \citenamefont {Wang}, \citenamefont {Rosenblum}, \citenamefont {Frunzio}, \citenamefont {Jiang}, \citenamefont {Girvin},\ and\ \citenamefont {Schoelkopf}}]{gao2018programmable}%
  \BibitemOpen
  \bibfield  {author} {\bibinfo {author} {\bibfnamefont {Y.~Y.}\ \bibnamefont {Gao}}, \bibinfo {author} {\bibfnamefont {B.~J.}\ \bibnamefont {Lester}}, \bibinfo {author} {\bibfnamefont {Y.}~\bibnamefont {Zhang}}, \bibinfo {author} {\bibfnamefont {C.}~\bibnamefont {Wang}}, \bibinfo {author} {\bibfnamefont {S.}~\bibnamefont {Rosenblum}}, \bibinfo {author} {\bibfnamefont {L.}~\bibnamefont {Frunzio}}, \bibinfo {author} {\bibfnamefont {L.}~\bibnamefont {Jiang}}, \bibinfo {author} {\bibfnamefont {S.}~\bibnamefont {Girvin}},\ and\ \bibinfo {author} {\bibfnamefont {R.~J.}\ \bibnamefont {Schoelkopf}},\ }\bibfield  {title} {\bibinfo {title} {Programmable interference between two microwave quantum memories},\ }\href {https://journals.aps.org/prx/abstract/10.1103/PhysRevX.8.021073} {\bibfield  {journal} {\bibinfo  {journal} {Phys. Rev. X}\ }\textbf {\bibinfo {volume} {8}},\ \bibinfo {pages} {021073} (\bibinfo {year} {2018})}\BibitemShut {NoStop}%
\bibitem [{\citenamefont {Takeoka}\ \emph {et~al.}(2014)\citenamefont {Takeoka}, \citenamefont {Guha},\ and\ \citenamefont {Wilde}}]{takeoka2014fundamental}%
  \BibitemOpen
  \bibfield  {author} {\bibinfo {author} {\bibfnamefont {M.}~\bibnamefont {Takeoka}}, \bibinfo {author} {\bibfnamefont {S.}~\bibnamefont {Guha}},\ and\ \bibinfo {author} {\bibfnamefont {M.~M.}\ \bibnamefont {Wilde}},\ }\bibfield  {title} {\bibinfo {title} {Fundamental rate-loss tradeoff for optical quantum key distribution},\ }\href {https://www.nature.com/articles/ncomms6235} {\bibfield  {journal} {\bibinfo  {journal} {Nat. Commun.}\ }\textbf {\bibinfo {volume} {5}},\ \bibinfo {pages} {5235} (\bibinfo {year} {2014})}\BibitemShut {NoStop}%
\bibitem [{\citenamefont {Pirandola}\ \emph {et~al.}(2017)\citenamefont {Pirandola}, \citenamefont {Laurenza}, \citenamefont {Ottaviani},\ and\ \citenamefont {Banchi}}]{pirandola2017fundamental}%
  \BibitemOpen
  \bibfield  {author} {\bibinfo {author} {\bibfnamefont {S.}~\bibnamefont {Pirandola}}, \bibinfo {author} {\bibfnamefont {R.}~\bibnamefont {Laurenza}}, \bibinfo {author} {\bibfnamefont {C.}~\bibnamefont {Ottaviani}},\ and\ \bibinfo {author} {\bibfnamefont {L.}~\bibnamefont {Banchi}},\ }\bibfield  {title} {\bibinfo {title} {Fundamental limits of repeaterless quantum communications},\ }\href {https://www.nature.com/articles/ncomms15043} {\bibfield  {journal} {\bibinfo  {journal} {Nat. Commun.}\ }\textbf {\bibinfo {volume} {8}},\ \bibinfo {pages} {15043} (\bibinfo {year} {2017})}\BibitemShut {NoStop}%
\bibitem [{\citenamefont {Romanenko}\ \emph {et~al.}(2020)\citenamefont {Romanenko}, \citenamefont {Pilipenko}, \citenamefont {Zorzetti}, \citenamefont {Frolov}, \citenamefont {Awida}, \citenamefont {Belomestnykh}, \citenamefont {Posen},\ and\ \citenamefont {Grassellino}}]{romanenko2020three}%
  \BibitemOpen
  \bibfield  {author} {\bibinfo {author} {\bibfnamefont {A.}~\bibnamefont {Romanenko}}, \bibinfo {author} {\bibfnamefont {R.}~\bibnamefont {Pilipenko}}, \bibinfo {author} {\bibfnamefont {S.}~\bibnamefont {Zorzetti}}, \bibinfo {author} {\bibfnamefont {D.}~\bibnamefont {Frolov}}, \bibinfo {author} {\bibfnamefont {M.}~\bibnamefont {Awida}}, \bibinfo {author} {\bibfnamefont {S.}~\bibnamefont {Belomestnykh}}, \bibinfo {author} {\bibfnamefont {S.}~\bibnamefont {Posen}},\ and\ \bibinfo {author} {\bibfnamefont {A.}~\bibnamefont {Grassellino}},\ }\bibfield  {title} {\bibinfo {title} {Three-dimensional superconducting resonators at $t<$~20~mk with photon lifetimes up to $\tau$=~2~s},\ }\href {https://link.aps.org/doi/10.1103/PhysRevApplied.13.034032} {\bibfield  {journal} {\bibinfo  {journal} {Phys. Rev. App.}\ }\textbf {\bibinfo {volume} {13}},\ \bibinfo {pages} {034032} (\bibinfo {year} {2020})}\BibitemShut {NoStop}%
\bibitem [{\citenamefont {Kim}\ \emph {et~al.}(2025)\citenamefont {Kim}, \citenamefont {Roy}, \citenamefont {You}, \citenamefont {Li}, \citenamefont {Lamm}, \citenamefont {Pronitchev}, \citenamefont {Bal}, \citenamefont {Garattoni}, \citenamefont {Crisa}, \citenamefont {Bafia} \emph {et~al.}}]{kim2025ultracoherent}%
  \BibitemOpen
  \bibfield  {author} {\bibinfo {author} {\bibfnamefont {T.}~\bibnamefont {Kim}}, \bibinfo {author} {\bibfnamefont {T.}~\bibnamefont {Roy}}, \bibinfo {author} {\bibfnamefont {X.}~\bibnamefont {You}}, \bibinfo {author} {\bibfnamefont {A.~C.}\ \bibnamefont {Li}}, \bibinfo {author} {\bibfnamefont {H.}~\bibnamefont {Lamm}}, \bibinfo {author} {\bibfnamefont {O.}~\bibnamefont {Pronitchev}}, \bibinfo {author} {\bibfnamefont {M.}~\bibnamefont {Bal}}, \bibinfo {author} {\bibfnamefont {S.}~\bibnamefont {Garattoni}}, \bibinfo {author} {\bibfnamefont {F.}~\bibnamefont {Crisa}}, \bibinfo {author} {\bibfnamefont {D.}~\bibnamefont {Bafia}}, \emph {et~al.},\ }\bibfield  {title} {\bibinfo {title} {Ultracoherent superconducting cavity-based multiqudit platform with error-resilient control},\ }\href {https://arxiv.org/pdf/2506.03286v1} {\bibfield  {journal} {\bibinfo  {journal} {arXiv:2506.03286}\ } (\bibinfo {year} {2025})}\BibitemShut {NoStop}%
\bibitem [{\citenamefont {Qiu}\ \emph {et~al.}(2023)\citenamefont {Qiu}, \citenamefont {Grimsmo}, \citenamefont {Peng}, \citenamefont {Kannan}, \citenamefont {Lienhard}, \citenamefont {Sung}, \citenamefont {Krantz}, \citenamefont {Bolkhovsky}, \citenamefont {Calusine}, \citenamefont {Kim} \emph {et~al.}}]{qiu2023broadband}%
  \BibitemOpen
  \bibfield  {author} {\bibinfo {author} {\bibfnamefont {J.~Y.}\ \bibnamefont {Qiu}}, \bibinfo {author} {\bibfnamefont {A.}~\bibnamefont {Grimsmo}}, \bibinfo {author} {\bibfnamefont {K.}~\bibnamefont {Peng}}, \bibinfo {author} {\bibfnamefont {B.}~\bibnamefont {Kannan}}, \bibinfo {author} {\bibfnamefont {B.}~\bibnamefont {Lienhard}}, \bibinfo {author} {\bibfnamefont {Y.}~\bibnamefont {Sung}}, \bibinfo {author} {\bibfnamefont {P.}~\bibnamefont {Krantz}}, \bibinfo {author} {\bibfnamefont {V.}~\bibnamefont {Bolkhovsky}}, \bibinfo {author} {\bibfnamefont {G.}~\bibnamefont {Calusine}}, \bibinfo {author} {\bibfnamefont {D.}~\bibnamefont {Kim}}, \emph {et~al.},\ }\bibfield  {title} {\bibinfo {title} {Broadband squeezed microwaves and amplification with a josephson travelling-wave parametric amplifier},\ }\href {https://www.nature.com/articles/s41567-022-01929-w} {\bibfield  {journal} {\bibinfo  {journal} {Nat. Phys.}\ }\textbf {\bibinfo {volume} {19}},\ \bibinfo {pages} {706} (\bibinfo {year} {2023})}\BibitemShut
  {NoStop}%
\bibitem [{\citenamefont {Brecht}\ \emph {et~al.}(2015)\citenamefont {Brecht}, \citenamefont {Reddy}, \citenamefont {Silberhorn},\ and\ \citenamefont {Raymer}}]{brecht2015photon}%
  \BibitemOpen
  \bibfield  {author} {\bibinfo {author} {\bibfnamefont {B.}~\bibnamefont {Brecht}}, \bibinfo {author} {\bibfnamefont {D.~V.}\ \bibnamefont {Reddy}}, \bibinfo {author} {\bibfnamefont {C.}~\bibnamefont {Silberhorn}},\ and\ \bibinfo {author} {\bibfnamefont {M.~G.}\ \bibnamefont {Raymer}},\ }\bibfield  {title} {\bibinfo {title} {Photon temporal modes: a complete framework for quantum information science},\ }\href {https://journals.aps.org/prx/abstract/10.1103/PhysRevX.5.041017} {\bibfield  {journal} {\bibinfo  {journal} {Phys. Rev. X}\ }\textbf {\bibinfo {volume} {5}},\ \bibinfo {pages} {041017} (\bibinfo {year} {2015})}\BibitemShut {NoStop}%
\bibitem [{\citenamefont {Pechal}\ \emph {et~al.}(2014)\citenamefont {Pechal}, \citenamefont {Huthmacher}, \citenamefont {Eichler}, \citenamefont {Zeytino{\u{g}}lu}, \citenamefont {Abdumalikov~Jr}, \citenamefont {Berger}, \citenamefont {Wallraff},\ and\ \citenamefont {Filipp}}]{pechal2014microwave}%
  \BibitemOpen
  \bibfield  {author} {\bibinfo {author} {\bibfnamefont {M.}~\bibnamefont {Pechal}}, \bibinfo {author} {\bibfnamefont {L.}~\bibnamefont {Huthmacher}}, \bibinfo {author} {\bibfnamefont {C.}~\bibnamefont {Eichler}}, \bibinfo {author} {\bibfnamefont {S.}~\bibnamefont {Zeytino{\u{g}}lu}}, \bibinfo {author} {\bibfnamefont {A.}~\bibnamefont {Abdumalikov~Jr}}, \bibinfo {author} {\bibfnamefont {S.}~\bibnamefont {Berger}}, \bibinfo {author} {\bibfnamefont {A.}~\bibnamefont {Wallraff}},\ and\ \bibinfo {author} {\bibfnamefont {S.}~\bibnamefont {Filipp}},\ }\bibfield  {title} {\bibinfo {title} {Microwave-controlled generation of shaped single photons in circuit quantum electrodynamics},\ }\href {https://journals.aps.org/prx/abstract/10.1103/PhysRevX.4.041010} {\bibfield  {journal} {\bibinfo  {journal} {Phys. Rev. X}\ }\textbf {\bibinfo {volume} {4}},\ \bibinfo {pages} {041010} (\bibinfo {year} {2014})}\BibitemShut {NoStop}%
\bibitem [{\citenamefont {Szeg}(1939)}]{szeg1939orthogonal}%
  \BibitemOpen
  \bibfield  {author} {\bibinfo {author} {\bibfnamefont {G.}~\bibnamefont {Szeg}},\ }\href {https://bookstore.ams.org/coll-23} {\emph {\bibinfo {title} {Orthogonal polynomials}}},\ Vol.~\bibinfo {volume} {23}\ (\bibinfo  {publisher} {American Mathematical Soc.},\ \bibinfo {year} {1939})\BibitemShut {NoStop}%
\bibitem [{\citenamefont {Gottesman}(2024)}]{gottesman2024surviving}%
  \BibitemOpen
  \bibfield  {author} {\bibinfo {author} {\bibfnamefont {D.}~\bibnamefont {Gottesman}},\ }\href {https://www.cs.umd.edu/~dgottesm/QECCbook-2024.pdf} {\emph {\bibinfo {title} {Surviving as a quantum computer in a classical world}}}\ (\bibinfo  {publisher} {Textbook manuscript preprint},\ \bibinfo {year} {2024})\BibitemShut {NoStop}%
\bibitem [{\citenamefont {Shapiro}(2016)}]{ShapiroQOC_OCW}%
  \BibitemOpen
  \bibfield  {author} {\bibinfo {author} {\bibfnamefont {J.~H.}\ \bibnamefont {Shapiro}},\ }\href {https://ocw.mit.edu/courses/6-453-quantum-optical-communication-fall-2016/} {\bibinfo {title} {Quantum optical communication}},\ \bibinfo {howpublished} {MIT OpenCourseWare lecture notes} (\bibinfo {year} {2016})\BibitemShut {NoStop}%
\bibitem [{\citenamefont {Glancy}\ and\ \citenamefont {Knill}(2006)}]{glancy2006error}%
  \BibitemOpen
  \bibfield  {author} {\bibinfo {author} {\bibfnamefont {S.}~\bibnamefont {Glancy}}\ and\ \bibinfo {author} {\bibfnamefont {E.}~\bibnamefont {Knill}},\ }\bibfield  {title} {\bibinfo {title} {Error analysis for encoding a qubit in an oscillator},\ }\href {https://journals.aps.org/pra/abstract/10.1103/PhysRevA.73.012325} {\bibfield  {journal} {\bibinfo  {journal} {Phys. Rev. A}\ }\textbf {\bibinfo {volume} {73}},\ \bibinfo {pages} {012325} (\bibinfo {year} {2006})}\BibitemShut {NoStop}%
\end{thebibliography}
\providecommand{\noopsort}[1]{}\providecommand{\singleletter}[1]{#1}%

\appendix
\section{Photon temporal basis}
\label{app:PhotonTM}
Temporal modes (TMs) form a complete orthonormal basis for single-photon states \cite{brecht2015photon}. Although the concept of TMs can be extended to more complex photonic states, this study focuses primarily on single-photon temporal modes. Given a temporal mode shape, the wavepacket quantum state of a single-photon is defined as 
\begin{align}
    \lvert 1\rangle_{\Gamma_{n}} = \int dt \hspace*{0.2em}\Gamma_{n} (t) a^{\dagger}(t) \lvert 0 \rangle_{\Gamma},
\end{align}
where $\Gamma_{n}(t)$ is the complex amplitude of the temporal mode at time $t$, and $\lvert 0\rangle_{\Gamma}$ denotes the multi-mode vacuum state at different times. 

The action of the creation operator $a^{\dagger}$ on the multi-mode vacuum state creates a localized photon at a particular time slot $t$. Consequently, the single-photon wavepacket state is a coherent superposition over all possible creation times, weighted by the temporal mode function. A broadband mode operator that creates the wavepacket state $\lvert 1 \rangle_{\psi_{n}}$ can also be defined as 
\begin{align}
    a^{\dagger} [\Gamma_{n}] &= \int dt \hspace*{0.2em}\Gamma_{n} (t) a^{\dagger}(t), \nonumber \\ 
    \lvert 1 \rangle_{\Gamma_{n}} &= a^{\dagger} [\Gamma_{n}] \lvert 0 \rangle_{\Gamma}.
\end{align}
Furthermore, a finite orthonormal family of mode shapes $\{\Gamma\}_{m \in \mathbb{Z}}$ form a complete basis for single-photon TMs
\begin{align}
    {_{\Gamma_{n}}\langle 1 \lvert} 1 \rangle_{\Gamma_{m}}&= \iint dt dt' \hspace*{0.2em} \overline{\Gamma_{n}(t)} \Gamma_{m}(t') \langle 0 \lvert_{\Gamma} a(t)a^{\dagger}(t') \lvert 0 \rangle_{\Gamma}\nonumber \\ 
    &= \iint dt dt' \hspace*{0.2em} \overline{\Gamma_{n}(t)} \Gamma_{m}(t') \delta(t-t') \nonumber \\ 
    &= \int dt \hspace*{0.2em} \overline{\Gamma_{n}(t)} \Gamma_{m}(t) = \delta_{nm}, \hspace*{0.2em} \forall n, m \in \mathbb{Z},
\end{align}
where $[a(t),a^{\dagger}(t')] = \delta(t-t')$. One notable example of an orthonormal basis family is the set of \textit{Hermite-Gaussian} modes, which is widely used in the optical domain. 

By spanning two orthogonal mode shapes, a qubit TM can be defined as 
\begin{align}
    \lvert \varphi \rangle = c_{0} \lvert 1\rangle_{\Gamma_{0}} + c_{1} \lvert 1 \rangle_{\Gamma_{1}} 
\end{align}
where, $\Gamma_{0}, \Gamma_{1}$ can be taken as the two lowest-order Hermite-Gaussian functions, and $c_{0}, c_{1} \in \mathbb{C}$. In microwave cQED, single-photon TMs have been successfully demonstrated experimentally \cite{pechal2014microwave}, and proposed as viable information carriers for microwave quantum communications~\cite{kurpiers2019quantum}.

\section{Pure-loss channel description}
\label{app:Channel}
Losses due to information transfer and storage during entanglement generation and swapping are accurately described by a beamsplitter loss-channel model.
In our sequential remote entanglement  protocol, the degradation of the transmitted wavepacket is modeled by a unitary beamsplitter transformation that couples the transmission line input to an environment mode in its vacuum state. As a result, the output single-photon temporal mode can be expressed as
\begin{align}
{a[\Gamma]}^{\rm (out)}&=(U_{\Gamma,{\rm v}}^{\theta})a[\Gamma] {(U_{\Gamma,{\rm v}}^{\theta }}) ^{\dagger} \nonumber \\ 
     &= \sqrt{\eta} a[\Gamma]+\sqrt{1-\eta} \hspace*{0.2em}a_{\rm v}.
\end{align}
Here, $U_{\Gamma,{\rm v}}^{\theta} = e^{-i \theta H_{\texttt{BS}}}$ is the beamsplitter operator, where $\theta$ is its angle, $H_{\texttt{BS}}= i({a[\Gamma]}^{\dagger}a_{\rm v}+a_{\rm v} a[\Gamma])$ is the corresponding Hamiltonian. The transmissivity of the beamsplitter is given by $\cos{(\theta)}=\eta \in [0,1]$. In this expression, $a_{\rm v}$ denotes the annihilation operator for the environment vacuum mode. 

It is also worth noting that a beamsplitter loss channel can be represented as a $2 \times 2$ matrix . 
\begin{align}
    \begin{pmatrix}
        {a[\Gamma]}^{\rm (out1)} \\
        {a[\Gamma]}^{\rm (out2)}
    \end{pmatrix} = \begin{pmatrix}
        \sqrt{\eta} & \sqrt{1-\eta} \\
        \sqrt{1-\eta} & \sqrt{\eta}
    \end{pmatrix} = \begin{pmatrix}
        a[\Gamma] \\
        a_{v}.
    \end{pmatrix},
     \label{eq:BSMatrix}
\end{align}
where in this case $\eta$ denotes the transmissivity of the transmitting channel. 

We can also define losses due to stationary damping by recasting the transmissivity of the beamsplitter as $\gamma = e^{-t/\tau}$ where $t$ is the storage time in \SI{}{\milli\second}, and $\tau=\kappa^{-1}_{\rm damp}$ is the lifetime of the stored codeword in \SI{}{\milli\second}. We note that here we have changed $\eta$ to $\gamma$ to avoid confusion with the main text.

The effect of stationary losses on the stored codewords is to increase their variance. The added variance can be calculated from the codewords position quadrature after undergoing a beamsplitter channel, $Q_{\text{out}} = \sqrt{\gamma} Q_{\text{in}}+\sqrt{1-\gamma}Q_{v}$, and hence the variance of the output is calculated as  
\begin{align}
    &\langle \Delta Q_{\text{out}} \rangle^{2} = \langle Q^{2}_{\text{out}} \rangle - \langle Q_{\text{out}} \rangle^{2}, \nonumber \\ 
    &\langle Q_{\text{out}}^{2} \rangle = \langle (\sqrt{\gamma} Q_{\text{in}}+\sqrt{1-\gamma}Q_{v})(\sqrt{\gamma} Q_{\text{in}}+\sqrt{1-\gamma}Q_{v}) \rangle,\nonumber \\ 
    &\langle Q_{\text{out}}^{2} \rangle= \gamma \langle Q_{\text{in}}^{2} \rangle+\frac{1-\gamma}{2}, \nonumber \\
    &\langle Q_{\text{out}} \rangle^{2} = \gamma \langle Q_{\text{in}} \rangle^{2}, \nonumber \\ 
    &\langle \Delta Q_{\text{out}} \rangle^{2} = \gamma\langle \Delta Q_{\text{in}} \rangle^{2}+\frac{1-\gamma}{2}.
\end{align}
where $\langle Q_{v} \rangle = 0$.

For the stationary damping values considered in the main text we calculate their corresponding added variance as follows. When the stationary damping is ~\SI{40}{\milli\second}, and a storage time of ~\SI{1}{\milli\second}, the transmissivity becomes $\gamma= e^{\frac{-1}{40}}= 0.97$, and hence the added variance becomes $0.01$. The rest of the values can be calculated in a similar manner. 
The table below summarizes the results. 

\begin{table} [H]
\centering
\scalebox{0.7}{\begin{tabular}{|l|c|c|c|c|c|}
\hline
 \shortstack{\vspace*{-0.1em}\text{Device } \\ { \text{Parameter}} } & \shortstack{\\$\kappa_{\text{damp}}^{-1}=40\text{ms}$} & \shortstack{\\$\kappa_{\text{damp}}^{-1}= 25 \text{ms}$} &  \shortstack{ \\ $\kappa_{\text{damp}}^{-1}= 15 \text{ms}$ } &\shortstack{\\  $\kappa_{\text{damp}}^{-1}= 10 \text{ms}$ } & \shortstack{\\$\kappa_{\text{damp}}^{-1}=8\text{ms}$}\\ \hline
$\gamma$ & $0.97$ & $0.96$ &$ 0.93$ & $0.90$ &$0.88$ \\ \hline
$\text{Variance}$ & 0.010 & 0.020& 0.030 & 0.046 & 0.058\\ \hline

\end{tabular}}
\caption{ The added variance values due to different levels of stationary losses of GBMQR. Included in the the damping value and its corresponding channel transmissivity, $\gamma$.}
\label{Tab:AddedVariance}
\end{table}

\section{GKP grid qubit: single and joint operations }
\label{app:GridPauli}
Ideally, as discussed in the main text, we assume a square grid encoding. We further assume that the state of the oscillator is encoded in the Z-basis. Hence the qubit subspace is panned by the following basis states
\begin{align}
    \lvert \bar{0} \rangle &= \underset{k \in \mathbb{Z}}{\sum} \lvert 2k \sqrt{\pi} \rangle_{q}, \nonumber \\ 
    \lvert \bar{1} \rangle &= \underset{k \in \mathbb{Z}}{\sum} \lvert (2k+1)\sqrt{\pi} \rangle_{q}.
\end{align}
In the continuous variable momentum basis, the position eigenkets are replaced with momentum states and the same superposition structure is preserved. 

We define a logical Pauli-Z operation as $\bar{Z} = D(i\sqrt{\frac{\pi}{2}} ) = e^{i\sqrt{\pi} Q}$, where $Q= (a+a^{\dagger})/2$. This operator generates translations in the momentum space, whereas on a \texttt{GKP} basis states it acts as
\begin{align}
    &\bar{Z} \lvert \bar{0} \rangle \nonumber \\ 
    &= \underset{k \in \mathbb{Z}}{\sum} e^{i\sqrt{\pi} Q} \lvert 2k \sqrt{\pi} \rangle_{q}  = \underset{k \in \mathbb{Z}}{\sum} (e^{i 2 \pi})^{k} \lvert 2k \sqrt{\pi} \rangle_{q} = +1 \lvert \bar{0} \rangle, \nonumber \\ 
    &\bar{Z} \lvert \bar{1} \rangle \nonumber \\
    &= \underset{k \in \mathbb{Z}}{\sum} e^{i\sqrt{\pi} Q} \lvert (2k+1) \sqrt{\pi} \rangle_{q} = e^{i \pi} \underset{k \in \mathbb{Z}}{\sum} (e^{i 2 \pi})^{k} \lvert (2k+1) \sqrt{\pi} \rangle_{q} \nonumber \\ 
    &= -1 \lvert \bar{1} \rangle.
\end{align}
where $Q \lvert z \rangle_{q} = z \lvert z \rangle_{q}$.

On the other hand, a logical Pauli-X defined as $\bar{X} =D(\sqrt{\frac{\pi}{2}})= e^{-i \sqrt{\pi}P}$ generates translations in position space and its action on a \texttt{GKP} basis is shown as
\vspace*{-0em}
\begin{align}
    \bar{X} \lvert \bar{1} \rangle &= \underset{k \in \mathbb{Z}}{\sum} e^{-i \sqrt{\pi}P} \lvert (2k+1) \sqrt{\pi} \rangle_{q} \nonumber \\
    &=\underset{k \in \mathbb{Z}}{\sum} \int dx e^{-i \sqrt{\pi}p} \lvert x \rangle_{p} {_{p}\langle x \lvert (2k+1) \sqrt{\pi} \rangle_{q}} \nonumber \\
    &=\frac{1}{\sqrt{\pi}} \underset{k \in \mathbb{Z}}{\sum} \int dp  e^{-i \sqrt{\pi}x} e^{i\sqrt{\pi}x} e^{i 2k \sqrt{\pi}x} \lvert x \rangle_{p} \nonumber \\
    &=  \underset{k \in \mathbb{Z}}{\sum} \int dx  \frac{ e^{i 2k \sqrt{\pi}x}\lvert x \rangle_{p} }{\sqrt{\pi}} \nonumber \\ 
    &= \underset{k \in \mathbb{Z}}{\sum} \lvert 2k \sqrt{\pi} \rangle_{q} = \lvert \bar{0} \rangle, \nonumber \\ 
    \bar{X} \lvert \bar{0} \rangle &= \underset{k \in \mathbb{Z}}{\sum} \int dx e^{-i \sqrt{\pi}p} \lvert x \rangle_{p} {_{p}\langle x \lvert 2k \sqrt{\pi} \rangle_{q}} \nonumber \\
    &= \underset{k \in \mathbb{Z}}{\sum} \int dx  \frac{ e^{i (2k-1) \sqrt{\pi}x}\lvert x \rangle_{p} }{\sqrt{\pi}} \nonumber \\ 
    &= \underset{k \in \mathbb{Z}}{\sum} \lvert (2k+1) \sqrt{\pi} \rangle_{q} = \lvert \bar{1} \rangle,
\end{align}
where $\frac{e^{i xz}}{\sqrt{2\pi}} = {_{p}\langle x \lvert z\rangle_{q}}$, $\int dx \lvert{x} \rangle_{p} {_{p}\langle x \lvert} = \mathds{1}$, $\lvert z \rangle _{q} = \frac{1}{\sqrt{2 \pi}}\int dx e^{ixz }  \lvert x \rangle_{p}$, and we have exploited translation invariance of the infinite grid state. 

The anti-commutation relation between the logical qubit Pauli operations is straightforwardly derived as 
\begin{align}
    \bar{X}\bar{Z} &= e^{-i \sqrt{\pi}P}e^{i \sqrt{\pi} Q} \nonumber \\
    & = e^{i \sqrt{\pi} Q}e^{-i \sqrt{\pi}P} e^{[-i\sqrt{\pi}P,i\sqrt{\pi}Q]}\nonumber \\
    &= e^{i \pi} \bar{Z}\bar{X}=- \bar{Z}\bar{X},
\end{align}
where $e^{A+B}=e^{A}e^{B}e^{\frac{-1}{2}[A,B]}=e^{B}e^{A}e^{\frac{1}{2}[A,B]}$ was utilized.

\subsection{Controlled-Z joint operation}
\label{app:GridJoint}
A bosonic controlled-Z operation is defined as $e^{-i {\rm } Q_{c}Q_{t}}$. The action of the gate on logical states is shown as
\begin{align}
    &e^{-i {\rm } Q_{c}Q_{t}} \lvert \bar{0} \rangle_{c} \lvert \bar{0} \rangle_{t} \nonumber \\ 
    &= e^{-i Q_{c}Q_{t}} \underset{k \in \mathbb{Z}}{\sum} \lvert 2k \sqrt{\pi} \rangle_{q_{c}} \otimes \underset{k' \in \mathbb{Z}}{\sum} \lvert 2k' \sqrt{\pi} \rangle_{q_{t}} \nonumber \\
    &= \underset{k \in \mathbb{Z}}{\sum} \underset{k' \in \mathbb{Z}}{\sum} (e^{-i 4 \pi})^{kk'} \lvert 2k \sqrt{\pi} \rangle_{q_{c}} \otimes \lvert 2k' \sqrt{\pi} \rangle_{q_{t}} = \lvert \bar{0} \rangle \lvert \bar{0} \rangle, \nonumber\\[0.25cm] 
    &e^{-i {\rm } Q_{c}Q_{t}} \lvert \bar{0} \rangle \lvert \bar{1} \rangle \nonumber \\ 
    &= e^{-i Q_{c}Q_{t}} \underset{k \in \mathbb{Z}}{\sum} \lvert 2k \sqrt{\pi} \rangle_{q_{c}} \otimes \underset{k' \in \mathbb{Z}}{\sum} \lvert (2k'+1) \sqrt{\pi} \rangle_{q_{t}} \nonumber \\
    &= \underset{k \in \mathbb{Z}}{\sum} \underset{k' \in \mathbb{Z}}{\sum} (e^{-i 2 \pi})^{k}(e^{-i 4 \pi})^{kk'} \lvert 2k \sqrt{\pi} \rangle_{q_{c}} \otimes \lvert (2k'+1) \sqrt{\pi} \rangle_{q_{t}} \nonumber \\
    &= \lvert \bar{0} \rangle \lvert \bar{1} \rangle, \nonumber \\[0.2cm]
    &e^{-i {\rm } Q_{c}Q_{t}} \lvert \bar{1} \rangle \lvert \bar{0} \rangle = \lvert \bar{1} \rangle \lvert \bar{0} \rangle, \nonumber \\[0.25cm]
    &e^{-i {\rm } Q_{c}Q_{t}} \lvert \bar{1} \rangle \lvert \bar{1}\rangle \nonumber \\ 
    &= \underset{k \in \mathbb{Z}}{\sum} \underset{k' \in \mathbb{Z}}{\sum}e^{-i \pi} (e^{-i 4\pi})^{kk'}(e^{-i 2\pi})^{k}(e^{i 2 \pi})^{k'} \nonumber \\ & \times\lvert (2k+1) \sqrt{\pi} \rangle_{q_{c}}\otimes \lvert (2k'+1) \sqrt{\pi} \rangle_{q_{t}} = - \lvert \bar{1} \rangle \lvert \bar{1}\rangle,  
\end{align}
where the minus sign is only acquired when both of the inputs of the gate are in the logical-one state.
\section{Finite-energy \texttt{GKP} codewords analytical expressions}
\label{app:OVerlapExps}
As described in the main text, the finite-energy \texttt{GKP} code words are defined as 
\begin{align}
    \lvert \bar{0}\rangle_{\Delta} &=  e^{- \Delta^{2}\hat{n}} \lvert \bar{0}\rangle, \nonumber \\ 
     \lvert \bar{1}\rangle_{\Delta} &=  e^{- \Delta^{2}\hat{n}} \lvert \bar{1} \rangle,
\end{align}
where $\lvert \bar{0} \rangle \propto \underset{k \in \mathbb{Z}}{\sum} \lvert 2k \sqrt{\pi} \rangle_{q} $, and $\lvert \bar{1} \rangle \propto \underset{k \in \mathbb{Z}}{\sum} \lvert (2k+1) \sqrt{\pi} \rangle_{q}$. To show the effect of applying an envelope operator, $e^{-\Delta^{2}\hat{n}}$, the ideal code words are expanded in the complete orthonormal Fock-basis: 
    \begin{align}
    e^{- \Delta^{2}\hat{n}} \lvert \bar{0} \rangle&=   \underset{n \in \mathbb{Z}^{+}_{0}}{\sum} e^{- \Delta^{2}\hat{n}}\lvert n \rangle \langle n \lvert \bar{0} \rangle \nonumber \\ 
    &= \underset{n \in \mathbb{Z}^{+}_{0}}{\sum} e^{- \Delta^{2}n } \underset{k \in \mathbb{Z}^{}_{}}{\sum} \langle n \lvert 2k \sqrt{\pi} \rangle_{q} \lvert n \rangle \nonumber \\
    &= \underset{n \in \mathbb{Z}^{+}_{0}}{\sum} \underset{k \in \mathbb{Z}}{\sum} \frac{e^{- \Delta^{2}n } e^{-(2k \sqrt{\pi} )^{2}/2}   }{\pi^{1/4} \sqrt{2^{n} n !}} {\mathrm{H}}_{n}(2k \sqrt{\pi}) \lvert n \rangle, \\[0.25cm]
    e^{- \Delta^{2}\hat{n}} \lvert \bar{1} \rangle 
    &= \underset{n \in \mathbb{Z}^{+}_{0}}{\sum} e^{- \Delta^{2}\hat{n}}\lvert n \rangle \langle n \lvert \bar{1} \rangle \nonumber \\ 
    &= \underset{n \in \mathbb{Z}^{+}_{0}}{\sum} e^{- \Delta^{2}n } \underset{k \in \mathbb{Z}}{\sum} \langle n \lvert (2k+1) \sqrt{\pi} \rangle_{q} \lvert n \rangle \nonumber \\ 
    = \underset{n \in \mathbb{Z}^{+}_{0}}{\sum} &\underset{k \in \mathbb{Z}}{\sum} \frac{e^{- \Delta^{2}n } e^{-( [2k+1] \sqrt{\pi} )^{2}/2} }{\pi^{1/4} \sqrt{2^{n} n !}} {\mathrm{H}}_{n}([2k+1] \sqrt{\pi}) \lvert n \rangle,
\end{align}
where $\lvert q \rangle_{q}$ is a position eigenstate, $Q\lvert q \rangle_{q} = q \lvert q \rangle_{q}$, with a Fock-basis representation, $\langle q \lvert n \rangle = \frac{e^{-q^{2}/2}}{\pi^{1/4} \sqrt{2^{n}n!}} {\mathrm{H}}_{n}(q)$, such that, ${\mathrm{H}}_{n}(q)$ is a Hermite polynomial, and $\Delta^{2}$ is the variance of the Gaussian envelope introduced by the truncation process.

Because of the non-orthogonality of the truncated codewords, the overlap and cross-overlap between codewords can be calculated as 
    \begin{align}
    &\scalebox{1}{${_{\Delta}\langle} \bar{0}\lvert \bar{0}\rangle_{\Delta}$} \nonumber \\
    &= \scalebox{1}{$\underset{k,k' \in \mathbb{Z}}{\sum} \hspace{0.2em}\underset{n,m \in \mathbb{Z}^{+}_{0}}{\sum} \delta_{mn} e^{-\Delta^{2}(n+m)} {_{q}\langle} 2k' \sqrt{\pi} \lvert m \rangle \langle n \lvert 2k \sqrt{\pi} \rangle_{q}$}  \nonumber \\ 
    &= \scalebox{1}{$\underset{k,k' \in \mathbb{Z}^{}}{\sum} \hspace{0.2em}\underset{n \in \mathbb{Z}^{+}_{0}}{\sum} \Big(\frac{e^{-2\Delta^{2} }}{2 } \Big)^{n} \hspace*{0.3em}\frac{{\mathrm{H}}_{n}(2k' \sqrt{\pi}) {\mathrm{H}_{n}(2k \sqrt{\pi})}}{\sqrt{\pi} \hspace*{0.2em} n!}$} \nonumber \\ 
    & \scalebox{1}{$\times \exp{\Big(\frac{-( 2k' \sqrt{\pi} )^{2}}{2}}\Big) \exp{\Big( \frac{-(2k\sqrt{\pi} )^{2}} {2}\Big)}$}\nonumber \\ 
    &= \scalebox{1}{$\frac{1}{\sqrt{\pi(1-e^{- 4\Delta^{2}})}} \underset{k,k' \in \mathbb{Z}}{\sum}\exp{\Big(\frac{-( 2k'\sqrt{\pi}  )^{2}}{2}}\Big) \exp{\Big(\frac{-( 2k\sqrt{\pi}  )^{2}}{2}\Big)}$} \nonumber \\ 
    & \scalebox{1}{$\times \exp{\Big( \frac{8k'k\pi e^{-2 \Delta^{2}}-\big( 4k'^{2} \pi + 4k^{2} \pi\big)e^{-4 \Delta^{2}} }{1-e^{-4 \Delta^{2}}}\Big)}$} 
    \label{eq: Logical0NormOverlap}
\end{align}
where normalization factor is $\mathcal{N}_{\bar{0}}= \sqrt{{_{\Delta}\langle} \bar{0}\lvert \bar{0}\rangle_{\Delta}}$.  Similarly one can compute the other codeword overlaps
    \begin{align}
    &\scalebox{1}{${_{\Delta}\langle} \bar{1}\lvert \bar{1}\rangle_{\Delta} $}\nonumber \\
    &= \scalebox{0.88}{$\underset{k,k' \in \mathbb{Z}}{\sum} \hspace{0.2em}\underset{n,m \in \mathbb{Z}^{+}_{0}}{\sum} \delta_{mn}e^{-\Delta^{2}(n+m)} {_{q}\langle} (2k'+1) \sqrt{\pi} \lvert m \rangle \langle n \lvert (2k+1) \sqrt{\pi} \rangle_{q}$}  \nonumber \\ 
    &= \scalebox{1}{$\underset{k,k' \in \mathbb{Z}^{}}{\sum} \hspace{0.3em}\underset{n \in \mathbb{Z}^{+}_{0}}{\sum} \Big(\frac{e^{-2\Delta^{2} }}{2 } \Big)^{n} \hspace*{0.2em}\frac{{\mathrm{H}}_{n}([2k'+1] \sqrt{\pi}) {\mathrm{H}_{n}([2k+1] \sqrt{\pi})}}{\sqrt{\pi} \hspace*{0.2em} n!}$} \nonumber \\ 
    &\scalebox{1}{$ \times \exp{\Big(\frac{-( (2k'+1) \sqrt{\pi} )^{2}}{2}}\Big) \exp{\Big(\frac{-( (2k+1)\sqrt{\pi} )^{2}}{2}\Big)}$}\nonumber \\ 
    &= \scalebox{1}{$\frac{1}{\sqrt{\pi(1-e^{- 4\Delta^{2}})}}\underset{k,k' \in \mathbb{Z}}{\sum} \exp{\Big(\frac{-((2k'+1) \sqrt{\pi} )^{2}}{2}}\Big) $}\nonumber \\ & \times \scalebox{1}{$\exp{\Big(\frac{-((2k+1)\sqrt{\pi} )^{2}}{2}\Big)}$} \nonumber \\ 
    &\times \scalebox{1}{$\exp{\Big( \frac{2(2k'+1)(2k+1)\pi e^{-2 \Delta^{2}}-\big( (2k'+1)^{2} \pi + (2k+1)^{2} \pi\big) e^{-4 \Delta^{2}} }{1-e^{-4 \Delta^{2}}}\Big)}$},
    \label{eq: Logical1NormOverlap}
\end{align}
where normalization is $\mathcal{N}_{\bar{1}}= \sqrt{{_{\Delta}\langle} \bar{1}\lvert \bar{1}\rangle_{\Delta}}$.
    \begin{align}
     &{_{\Delta}\langle} \bar{1}\lvert \bar{0}\rangle_{\Delta} \nonumber \\ 
     &= \scalebox{0.95}{$\underset{k,k' \in \mathbb{Z}}{\sum} \hspace{0.2em}\underset{n,m \in \mathbb{Z}^{+}_{0}}{\sum} \delta_{mn}  e^{-\Delta^{2}(n+m)} {_{q}\langle} 2k' \sqrt{\pi} \lvert m \rangle \langle n \lvert (2k+1) \sqrt{\pi} \rangle_{q}$} \nonumber \\ 
      &= \scalebox{1}{$\underset{k,k' \in \mathbb{Z}^{}}{\sum} \hspace{0.2em}\underset{n \in \mathbb{Z}^{+}_{0}}{\sum} \Big(\frac{e^{-2\Delta^{2} }}{2 } \Big)^{n} \hspace*{0.2em}\frac{{\mathrm{H}}_{n}(2k' \sqrt{\pi}) {\mathrm{H}_{n}([2k+1] \sqrt{\pi})}}{\sqrt{\pi} \hspace*{0.2em} n!}$} \nonumber \\ 
      & \times \scalebox{1}{$\exp{\Big(-\frac{(2k'\sqrt{\pi}  )^{2}}{2}}\Big) \exp{\Big(\frac{-( (2k+1)\sqrt{\pi} )^{2}}{2}\Big)}$}\nonumber \\ 
     &= \scalebox{1}{$\frac{1}{\sqrt{\pi(1-e^{- 4\Delta^{2}})}}\underset{k,k' \in \mathbb{Z}}{\sum}\exp{\Big(\frac{-(2k' \sqrt{\pi} )^{2}} {2}}\Big) $}\nonumber \\ 
     & \times \scalebox{1}{$\exp{\Big( -\frac{((2k+1)\sqrt{\pi} )^{2}}{2}\Big)}$} \nonumber \\ 
     & \times \scalebox{1}{$\exp{\Big( \frac{4k'(2k+1)\pi e^{-2 \Delta^{2}}-\big( 4k'^{2} \pi + (2k+1)^{2} \pi\big)e^{-4 \Delta^{2}} }{1-e^{-4 \Delta^{2}}}\Big)}$} \nonumber \\
    &={_{\Delta}\langle} \bar{0}\lvert \bar{1}\rangle_{\Delta},
    \label{eq: Logical01NormCrossOverlap}
\end{align}    
where $\mathcal{N}_{\bar{0}\bar{1}}= \sqrt{{_{\Delta}\langle \bar{0}\lvert \bar{0}\rangle_{\Delta}} {_{\Delta}\langle} \bar{1}\lvert \bar{1}\rangle_{\Delta}}$ denotes normalization constant.

The previous derivations were performed routinely using the standard Mehler's formula \cite{szeg1939orthogonal}, $\underset{n \in \mathbb{Z}^{+}}{\sum} \frac{{\mathrm{H}_{n}(x)}{\mathrm{H}}_{n}(y)}{n!} \hspace*{0.2em} \Big(\frac{z}{2}\Big)^{n} = \frac{1}{\sqrt{1-z^{2}}} \exp{\Big(\frac{2xyz-(x^{2}+y^{2})z^{2}}{1-z^{2}} \Big)}$. Numerical values can be obtained by truncating the sums over $k, k'$ to a finite value. Alternatively, the double sum expressions can be recast interms of Jacobi-Theta functions \cite{royer2020stabilization}. However, in this paper a standard Hermite-Mehler representation, which is mathematically equivalent to the theta-function representation, was favored in order to highlight explicitly the evolution of the codewords in the Fock-basis.   
\subsection{Coupling and mode-mismatch errors}
\label{app:MisError}
In a beamsplitter-based Bell-state measurement, two identical \texttt{GKP} codewords are routed to a coupling balanced beamsplitter. However, losses due to couplings, detection inefficiency, and mode mismatch decrease the indistinguishability between the two interfering codewords. In order to quantify this error, the overlap between codewords is considered, when each is undergoing a different loss channel. For the case of two logical-zero states, the overlap becomes
    \begin{align}
& \scalebox{1}{${_{\Delta}\langle \bar{0} \lvert \otimes {\langle 0}_{c} \lvert (U^{\eta_{2} })^{\dagger} \lvert U^{ \eta_{1} } \lvert 0 \rangle_{b} \otimes \lvert \bar{0} \rangle_{\Delta}}$} \nonumber \\ 
&= \scalebox{1}{$\underset{k,k' \in \mathbb{Z}}{\sum} \hspace*{0.2em}\underset{n \in \mathbb{Z}^{+}_{0}}{\sum} (\eta_{1}\eta_{2} e^{- 2\tilde{\Delta}^{2}})^{n} \langle 2k'\sqrt{\pi} \lvert n \rangle \langle n \lvert 2k \sqrt{\pi} \rangle $} \nonumber\\
&= \scalebox{1}{$\underset{k,k' \in \mathbb{Z}}{\sum} \hspace*{0.2em}\underset{n \in \mathbb{Z}^{+}_{0}}{\sum} \Big( \frac{ e^{- 2\tilde{\Delta}^{2}} \sqrt{\eta_{1}\eta_{2}}   }{2 } \Big)^{n}\frac{{\mathrm{H}}_{n}(2k'\sqrt{\pi})  {\mathrm{H}_{n}(2k\sqrt{\pi})}}{\sqrt{\pi} \, n!} $}  \nonumber \\
& \times \scalebox{1}{$\exp{\!\Big(-\frac{(2 k' \sqrt{\pi})^{2}}{2}\Big)} \exp{\!\Big(-\frac{(2 k \sqrt{\pi})^{2}}{2}\Big)} $}\nonumber \\
&= \scalebox{1}{$\frac{1}{{\sqrt{ \pi(1-( e^{- 2\tilde{\Delta}^{2}} \sqrt{\eta_{1}\eta_{2}}  )^{2}) } }}  \underset{k,k' \in \mathbb{Z}}{\sum} \exp{\!\Big(-\frac{(2k' \sqrt{\pi})^{2}}{2}\Big)} $} \nonumber \\ 
& \times \scalebox{1}{$\exp{\!\Big(-\frac{(2k \sqrt{\pi})^{2}}{2}\Big)} $}  \nonumber \\
& \times \scalebox{1}{$\exp{\!\Bigg(\frac{8 \pi kk' \Big( e^{- 2\tilde{\Delta}^{2}} \sqrt{\eta_{1}\eta_{2}}  \Big)-(4\pi k^{2} + 4 \pi k'^{2})\Big( e^{- 2\tilde{\Delta}^{2}} \sqrt{\eta_{1}\eta_{2}}  \Big)^{2}}{1-( e^{- 2\tilde{\Delta}^{2}} \sqrt{\eta_{1}\eta_{2}} )^{2}}\Bigg) }$},
\end{align}
 where $\tilde{\Delta} = \Delta + \frac{1-\eta_{1}}{2}+\frac{1-\eta_{2}}{2}$, $U^{ \eta_{1} }$, $U^{ \eta_{2} }$ are two different beamsplitter channels defined in Eq.~\eqref{eq:BSMatrix}. As for logical-1 states,
\begin{align}
& \scalebox{0.8}{${_{\Delta}\langle \bar{1} \lvert \otimes {\langle 0}_{c} \lvert (U^{\eta_{2} })^{\dagger} \lvert U^{ \eta_{1} } \lvert 0 \rangle_{b} \otimes \lvert \bar{1} \rangle_{\Delta}}$} \nonumber \\ 
&= \scalebox{0.95}{$\underset{k,k' \in \mathbb{Z}}{\sum} \hspace*{0.2em}\underset{n \in \mathbb{Z}^{+}_{0}}{\sum} ( e^{- 2\tilde{\Delta}^{2}}\sqrt{\eta_{1}\eta_{2}})^{n} \langle (2k'+1)\sqrt{\pi} \lvert n \rangle \langle n \lvert (2k+1)\sqrt{\pi} \rangle$}  \nonumber  \\
&= \scalebox{0.8}{$\underset{k,k' \in \mathbb{Z}}{\sum} \hspace*{0.2em}\underset{n \in \mathbb{Z}^{+}_{0}}{\sum}  \Big(\frac{ e^{- 2\tilde{\Delta}^{2}} \sqrt{\eta_{1}\eta_{2}}  }{2 } \Big)^{n} \frac{{\mathrm{H}}_{n}((2k'+1)\sqrt{\pi})  {\mathrm{H}_{n}((2k+1)\sqrt{\pi})}}{\sqrt{\pi} \, n!} $} \nonumber \\ 
& \times \scalebox{0.8}{$\exp{\Big( -\frac{\big((2k'+1)\sqrt{\pi}\big)^{2}}{2}\Big)} \exp{\Big( -\frac{\big((2k+1)\sqrt{\pi}\big)^{2}}{2}\Big)}$}  \nonumber 
\end{align}
\begin{align}
&= \scalebox{0.8}{$\frac{1}{\sqrt{ \pi(1-( e^{- 2\tilde{\Delta}^{2}} \sqrt{\eta_{1}\eta_{2}}  )^{2}) } } \underset{k,k' \in \mathbb{Z}}{\sum}  \exp{\Big( -\frac{\big((2k'+1)\sqrt{\pi}\big)^{2}}{2}\Big)}$} \nonumber \\
& \times \scalebox{0.8}{$\exp{\Big( -\frac{\big((2k+1)\sqrt{\pi}\big)^{2}}{2}\Big) } $}   \nonumber \\
&\times  \scalebox{0.8}{$\exp{ \!\Bigg(\frac{8 \pi (k+\tfrac{1}{2})(k'+\tfrac{1}{2}) \Big( e^{- 2\tilde{\Delta}^{2}} \sqrt{\eta_{1}\eta_{2}} \Big)-(4\pi (k+\tfrac{1}{2})^{2} + 4 \pi (k'+\tfrac{1}{2})^{2})\Big( e^{- 2\tilde{\Delta}^{2}} \sqrt{\eta_{1}\eta_{2}}  \Big)^{2}}{1-\Big( e^{- 2\tilde{\Delta}^{2}} \sqrt{\eta_{1}\eta_{2}}  \Big)^{2}}\Bigg) } $},
\end{align}
where $\tilde{\Delta} = \Delta + \frac{1-\eta_{1}}{2}+\frac{1-\eta_{2}}{2}$, $U^{ \eta_{1} }$, $U^{ \eta_{2} }$ .

It is important to note here that for the purpose of this article we choose to normalize the lossy overlaps of the codewords by the lossless normalization factors defined earlier, $\mathcal{N}_{\bar{0}}$, and $\mathcal{N}_{\bar{1}}$. This approach is standard when stationary damping is in the range of ~\SIrange{25}{40}{\milli\second} as treated in this manuscript \cite{grimsmo2021quantum}. However, this is not the case when losses are high as in the case of BSMQR operating at $\eta_{1}=\eta_{2}=0.6$. In this case, the codewords become a statistical mixture and the environment captures too much information, rendering any recovery operation impossible \cite{gottesman2024surviving}. 
 \section{Entanglement generation exact success probability}
\label{app:EntanglementGeneration}

This section provides the exact expression of the state shared between two nodes that undergo the entanglement generation step described in the main text. Because of transmission losses, the state evolves as 
\begin{align}
\lvert \varphi \rangle_{\Delta,AB} \nonumber \\
 =\frac{1}{2} \Big(&
- \alpha^2 \eta \lvert gg \rangle_{AB} \otimes \lvert \bar{0} \bar{0} \rangle_{\Delta, AB}
- \alpha \beta \eta \lvert gg \rangle_{AB} \otimes \lvert \bar{0} \bar{1} \rangle_{\Delta, AB} \nonumber \\ &- \alpha \beta \eta \lvert gg \rangle_{AB} \otimes \lvert \bar{1} \bar{0} \rangle_{\Delta, AB} 
- \beta^2 \eta\lvert gg \rangle_{AB} \otimes \lvert \bar{1} \bar{1} \rangle_{\Delta, AB}\nonumber \\ & + \alpha \eta \lvert ge \rangle_{AB} \otimes \lvert \bar{0} \bar{0} \rangle_{\Delta, AB}+ \beta \lvert eg \rangle_{AB} \otimes \lvert \bar{0} \bar{1} \rangle_{\Delta, AB} \nonumber \\
&+ \beta \eta \lvert ge \rangle_{AB} \otimes \lvert \bar{1} \bar{0} \rangle_{\Delta, AB} 
+ \lvert ee \rangle_{AB} \otimes \lvert \bar{0} \bar{0} \rangle_{\Delta, AB} \nonumber \\ &+ \alpha \lvert e g\rangle \otimes\lvert {\bar{0}\bar{0}} \rangle_{\Delta, AB}
\Big).
\label{eq:node-node EG}
\end{align}
After projective measurements on the transmons, followed by local operations and feedforward on the bosonic parts, the state becomes
    \begin{align}
    \lvert \varphi^{S} \rangle_{\Delta, AB}
    = \frac{1}{2 \mathcal{N}_{S}}\big( &(1-\eta\alpha^{2}) \lvert \bar{0}\bar{0} \rangle_{\Delta, AB} - \beta^{2} \eta \lvert \bar{1}\bar{1} \rangle_{\Delta, AB} \nonumber \\ 
    &- \alpha \beta \eta \sqrt{2} \lvert \bar{\Psi}^{+} \rangle_{\Delta, AB} \big),
\end{align}
where the normalization factor $\mathcal{N}_{S} = \frac{\sqrt{(1-\eta\alpha^{2})^{2} + \eta^{2} \beta^{4} + 2\alpha^{2}\beta^{2} \eta^{2}}}{2}$ and the overlaps ${_{\Delta}\langle \bar{0}} \lvert \bar{0} \rangle_{\Delta}={_{\Delta}\langle \bar{1}} \lvert \bar{1} \rangle_{\Delta}=\beta$, and ${_{\Delta}\langle \bar{0}} \lvert \bar{1} \rangle_{\Delta}={_{\Delta}\langle \bar{1}} \lvert \bar{0} \rangle_{\Delta}=\alpha$.
Accordingly, the density matrix of this state is 
    \begin{align}
    &\rho^{S}_{\Delta,AB} \nonumber \\
    &= \frac{1}{4\mathcal{N}^{2}_{S}} \Big(  (1-\eta\alpha^{2})^{2} \lvert \bar{0}\bar{0} \rangle_{\Delta, AB} {_{\Delta, AB}\langle \bar{0}\bar{0} \lvert} \nonumber \\ 
    &\qquad\qquad-(1-\eta\alpha^{2})\beta^{2}\eta \lvert \bar{0}\bar{0} \rangle_{\Delta, AB} {_{\Delta, AB}\langle \bar{1}\bar{1} \lvert} \nonumber \\
    &\qquad\qquad+ (1-\eta\alpha^{2}) \alpha \beta \eta\sqrt{2 } \lvert \bar{0}\bar{0} \rangle_{\Delta, AB} {_{\Delta, AB}\langle \bar{\Psi}^{+} \lvert}  
    \nonumber \\ 
    &\qquad\qquad- (1-\eta\alpha^{2})\beta^{2}\eta \lvert \bar{1}\bar{1} \rangle_{\Delta, AB} {_{\Delta, AB}\langle \bar{0}\bar{0} \lvert} \nonumber \\
    &\qquad\qquad-(1-\eta\alpha^{2})^{2}\alpha \beta \eta\sqrt{2 } \lvert \Psi^{+}\rangle_{\Delta, AB}{_{\Delta, AB}\langle \bar{0}\bar{0} \lvert}\nonumber \\
    &+ \beta^{4}\eta^{2}  \lvert \bar{1}\bar{1} \rangle_{\Delta, AB} {_{\Delta, AB}\langle \bar{1}\bar{1} \lvert} +\alpha \beta^{3} \eta^{2} \sqrt{2} \lvert \bar{1}\bar{1} \rangle_{\Delta, AB} {_{\Delta, AB}\langle \bar{\Psi}^{+} \lvert} \nonumber \\ 
    &+ \scalebox{0.9}{$\alpha \beta^{3} \eta^{2} \sqrt{2} \lvert \Psi^{+}\rangle_{\Delta, AB}{_{\Delta, AB}\langle \bar{1}\bar{1} \lvert}
    + \alpha^{2}\beta^{2}2\eta^{2}\lvert \Psi^{+}\rangle_{\Delta, AB}{_{\Delta, AB}\langle \bar{\Psi}^{+} \lvert}$}
\Big).
\end{align}
The success probability of the entanglement generation protocol is quantified by calculating the fidelity $\mathcal{F}$ of the previous state against the target finite-energy Bell state $\lvert \bar{\Phi}^{-}  \rangle_{\Delta, AB}= \frac{1}{\sqrt{2}}(\lvert \bar{0}\bar{0}\rangle_{\Delta, AB}- \lvert \bar{1}\bar{1}\rangle_{\Delta, AB}$ 
\begin{align}
    \mathcal{F} = \mathcal{P}^{\rm succ}_{\rm EG}=  \text{Tr} \hspace*{0.2em} \{{_{\Delta, AB}\langle \bar{\Phi}^{-}  \lvert} \rho^{S}_{\Delta,AB} \lvert \bar{\Phi}^{-}  \rangle_{\Delta, AB} \}.
\end{align}
 Thus, the entanglement success probability becomes 
    \begin{align}
   \ \mathcal{P}^{\rm succ}_{\rm EG} 
    &= \frac{1}{4 \mathcal{N}_{S}^{2}} \Big[
     (1-\eta\alpha^{2})^{2} \frac{(\beta^{2}-\alpha^{2}) }{2}  \nonumber \\
     & \qquad+\beta^{2}\eta (1-\eta\alpha^{2}) (\alpha^{2}-\beta^{2})^{2}+\frac{\beta^{4}\eta^{2}}{2} (\alpha^{2}-\beta^{2})^{2} \Big].
\end{align}
As can be seen, the success probability expression is a function of the transmissivity of the channel and the overlap between the codewords. 
\section{Entanglement swapping practical success probability}
\label{app:EGSWAP}
After establishing entanglement successfully, each of the two repeater segments is in a similar bipartite state. In the finite-energy \texttt{GKP} basis, the overall state can be written as
    \begin{align}
    \lvert \bar{\Psi} \rangle_{\Delta, ABCD}= \frac{1}{2} \Big[& \lvert \bar{0}\bar{0}\bar{0}\bar{0} \rangle_{\Delta, ABCD}- \bar{0}\bar{0}\bar{1}\bar{1} \rangle_{\Delta, ABCD} \nonumber \\
    &- \bar{1}\bar{1}\bar{0}\bar{0} \rangle_{\Delta, ABCD}+\bar{1}\bar{1}\bar{1}\bar{1} \rangle_{\Delta, ABCD}\Big].
\end{align}
The swapping operation proceeds by first applying a controlled-Z operation between nodes B and C, followed by a projective measurement on the two nodes in the X-basis. Device imperfections and losses are considered by recasting the involved gates in the non-orthogonal finite-energy \texttt{GKP} basis. 

An erroneous controlled-Z gate can be written as: 
\begin{align}
    \textbf{CZ}_{\Delta} &= {\lvert \bar{0}\rangle_{\Delta}}{_{\Delta}\langle \bar{0} \lvert} \otimes \mathds{1} + {\lvert \bar{1}\rangle_{\Delta}}{_{\Delta}\langle \bar{1} \lvert} \otimes Z_{\Delta}
\end{align}
where $Z_{\Delta} = {\lvert \bar{0}\rangle_{\Delta}}{_{\Delta}\langle \bar{0} \lvert} -{\lvert \bar{1}\rangle_{\Delta}}{_{\Delta}\langle \bar{1} \lvert}$. Subsequently, the state becomes
\begin{align}
     &\lvert \bar{\Psi} \rangle_{\Delta, ABCD} \nonumber \\
     &= \frac{1}{2} \Big [ \beta \big(\lvert \bar{0}\bar{0}\bar{0}\bar{0} \rangle_{\Delta, ABCD}-  \lvert \bar{0}\bar{0}\bar{1}\bar{1} \rangle_{\Delta, ABCD} - \lvert \bar{1}\bar{1}\bar{0}\bar{0} \rangle_{\Delta, ABCD} \nonumber \\
     &\quad-  \lvert \bar{1}\bar{1}\bar{1}\bar{1} \rangle_{\Delta, ABCD}\big) +\alpha \big(\lvert \bar{0}\bar{1}\bar{0}\bar{0} \rangle_{\Delta, ABCD}  +  \lvert \bar{0}\bar{1}\bar{1}\bar{1} \rangle_{\Delta, ABCD} \nonumber \\ 
     &\quad-  \lvert \bar{1}\bar{1}\bar{0}\bar{0} \rangle_{\Delta, ABCD} +  \lvert \bar{1}\bar{0}\bar{1}\bar{1} \rangle_{\Delta, ABCD} \big)\Big],
     \label{eq: ErrorSwapState}
\end{align}
where as before $\beta = \langle \bar{0} \lvert \bar{0} \rangle = \langle \bar{1} \lvert \bar{1} \rangle $, and $\alpha = \langle \bar{0} \lvert \bar{1} \rangle = \langle \bar{1} \lvert \bar{0} \rangle $. The gate errors resulting in flipping the states of both the control and target qubits are second order in the cross overlap of the logical codewords $\alpha$, where $\alpha \lll 1$.

\subsection{Imperfect projective measurement in \texttt{GKP} basis}
\label{app:HOM}
The next step in the swapping process is to perform two projective homodyne measurements on the middle modes. For projective homodyne detection, an operator description of this process is captured by the continuous variable  positive operator-valued measure (POVM) formalism. As defined earlier, the \texttt{GKP} logical states adopted in this article are superpositions of the position eigenstates of the oscillator. Completeness and orthonormality in the continuous-variable setting take the following form
\begin{align}
    \int dk \hspace*{0.2em} \lvert k \rangle_{q} {_{q}\langle k \lvert} = \mathds{1}, \hspace*{0.5em} {_{q}\langle k' \lvert} k \rangle_{q} = \delta(k'-k).
\end{align}
In this representation, the position quadrature operator can be expanded in these eigenstates as
\begin{align}
    Q = \int dk \hspace*{0.2em} k \lvert k \rangle_{q} {_{q}\langle k \lvert}. 
\end{align}
Accordingly, a set of POVMs can be defined as the set of projectors on the position eigenstates \cite{ShapiroQOC_OCW}
\begin{align}
    \hat{\Pi}(k) = \lvert k \rangle_{q} {_{q}\langle k \lvert}.
\end{align}
Following directly from their definitions, it is evident that the POVM elements $\{\hat{\Pi}(k) \}_{k \in \mathbb{R}}$ are Hermitian, $\hat{\Pi}(k) = \hat{\Pi}^{\dagger}(k)$. Furthermore, POVMs are positive-semidefinite operators, $\langle \psi \lvert \hat{\Pi}(k) \lvert \psi \rangle \geq 0 \hspace*{0.2em} \forall \psi$, which can be verified in a straight forward manner by recalling the Hermite representation of the harmonic oscillator states and their orthongonality property. Finally, the POVM elements resolve the identity $\int dk \hspace*{0.2em} \hat{\Pi}(k) = \mathds{1}$.

With this mathematical machinery, a projective measurement of a logical codeword and its success probability can now be feasibly defined. By considering inefficiencies of the detector, the probability of successfully detecting a logical zero state $\lvert \bar{0} \rangle_{\Delta}$ can be defined as:
\begin{align}
    \mathcal{P}^{\text{succ}}_{\bar{0} } (k)&= \text{Tr} \{\hat{\Pi}_{k} U^{\tilde{\eta}}\rho _{\bar{0},\Delta}[U^{\tilde{\eta}}]^{\dagger} \},
\end{align}
where $U^{\tilde{\eta}}$ defines a beamsplitter transformation describing losses due to detection imperfections, such that $\tilde{\eta}$ is its transmissivity. 

Practically, due to the \textit{Gaussianity} of both the initial energy truncation of the \texttt{GKP} codewords and losses during detection, the decision interval of a homodyne detector for the logical-zero state is centered around the zero, i.e., $k \in [-\pi/2, \pi/2]$, and the probability of successful detection is a gaussian distribution \cite{glancy2006error}
\begin{align}
    \mathcal{P}^{\text{succ}}_{\bar{0} } (k)&= \frac{1}{\sqrt{2\pi  \Delta}^{2}} \underset{-\sqrt{\pi}/2}{\overset{\sqrt{\pi}/2}{\int}} dk \hspace*{0.2em} e^{-k^{2}/2 \Delta^{2}} \nonumber \\
    &=  \text{erf} (\sqrt{\pi}/2 / \sqrt{2} \Delta),
\end{align}
where $\text{erf}(k) = 2/\sqrt{\pi} \underset{0}{\overset{k}{\int}} dk' \hspace*{0.2em}e^{-k'^{2}}$.
Similarly, the success probability of detecting a logical-one \texttt{GKP} codeword can be written as
\begin{align}
    \mathcal{P}^{\text{succ}}_{\bar{1} } (k)&= \frac{1}{\sqrt{2\pi  \Delta^{2}}} \underset{\sqrt{\pi}/2}{\overset{3\sqrt{\pi}/2}{\int}} dk \hspace*{0.2em} e^{-(k-\sqrt{\pi})^{2}/2  \Delta^{2}} \nonumber \\ 
    &=  \text{erf}(\sqrt{\pi}/2/\sqrt{2}\Delta),
\end{align}
where the decision is centered in this case around the peak value of logical-one \texttt{GKP} state. 
As described in the main text these success probabilities can be improved with the assistance of digital optimization methods to reach values as high as $0.95\text{--}0.99$, and hence for our calculations we chose the success probability of detecting either a logical-zero or a logical-1 to be $\mathcal{P}^{\text{succ}}_{\bar{0}/\bar{1} }\approx 0.95$.
\pagebreak

\end{document}